\documentclass[twocolumn]{aastex701}
\usepackage{placeins}
\usepackage{textcomp}
\usepackage{amsmath}
\usepackage{adjustbox}

\begin{document}

\title{HETDEX [\ion{O}{2}] galaxies at $z \le 0.48$: Growth of structures from power spectrum multipoles}

\author[0009-0009-2135-7061]{Jeongin Moon}
\affiliation{Max-Planck-Institut f\"{u}r Astrophysik, Karl-Schwarzschild-Str. 1, 85741 Garching, Germany}
\affiliation{Ludwig-Maximilians-Universit\"{a}t München, Schellingstr. 4, 80799 München, Germany}
\email{jmoon@mpa-garching.mpg.de}

\author[0000-0002-0136-2404]{Eiichiro Komatsu}
\affiliation{Max-Planck-Institut f\"{u}r Astrophysik, Karl-Schwarzschild-Str. 1, 85741 Garching, Germany}
\affiliation{Ludwig-Maximilians-Universit\"{a}t München, Schellingstr. 4, 80799 München, Germany}
\affiliation{Kavli Institute for the Physics and Mathematics of the Universe (WPI), The University of Tokyo Institutes for Advanced Study (UTIAS), The University of Tokyo, Chiba 277-8583, Japan}
\email{komatsu@mpa-garching.mpg.de}

\author[0000-0002-1328-0211]{Robin Ciardullo}
\affiliation{Department of Astronomy \& Astrophysics, The Pennsylvania State University, University Park, PA 16802, USA}
\affiliation{Institute for Gravitation and the Cosmos, The Pennsylvania State University, University Park, PA 16802, USA}
\email{rbc3@psu.edu}

\author[0000-0003-2575-0652]{Daniel J. Farrow}
\affiliation{E. A. Milne Centre for Astrophysics
University of Hull, Cottingham Road, Hull, HU6 7RX, UK}
\affiliation{Centre of Excellence for Data Science,
Artificial Intelligence \& Modelling (DAIM),
University of Hull, Cottingham Road, Hull, HU6 7RX, UK}
\email{D.J.Farrow@hull.ac.uk}

\author[0000-0002-8433-8185]{Karl Gebhardt}
\affiliation{Department of Astronomy, The University of Texas at Austin, 2515 Speedway Boulevard, Austin, TX 78712, USA}
\email{gebhardt@astro.as.utexas.edu}

\author[0000-0001-6842-2371]{Caryl Gronwall}
\affiliation{Department of Astronomy \& Astrophysics, The Pennsylvania
State University, University Park, PA 16802, USA}
\affiliation{Institute for Gravitation and the Cosmos, The Pennsylvania State University, University Park, PA 16802, USA}
\email{cag18@psu.edu}

\author[0000-0001-6717-7685]{Gary J. Hill}
\affiliation{Department of Astronomy, The University of Texas at Austin, 2515 Speedway Boulevard, Austin, TX 78712, USA}
\affiliation{McDonald Observatory, The University of Texas at Austin, 2515 Speedway Boulevard, Austin, TX 78712, USA}
\email{hill@astro.as.utexas.edu}

\author[0000-0002-8434-979X]{Donghui Jeong}
\affiliation{Department of Astronomy \& Astrophysics, The Pennsylvania State University, University Park, PA 16802, USA}
\affiliation{Institute for Gravitation and the Cosmos, The Pennsylvania State University, University Park, PA 16802, USA}
\affiliation{School of Physics, Korea Institute for Advanced Study, 85 Heogiro, Dongdaemun-gu, Seoul, 02455, Republic of Korea}
\email{djeong@psu.edu}

\author[0000-0002-2307-0146]{Erin {Mentuch Cooper}}
\affiliation{Department of Astronomy, The University of Texas at Austin, 2515 Speedway Boulevard, Austin, TX 78712, USA}
\affiliation{FirstPrinciples Foundation, 181 Bay Street, Toronto, ON, M5J 2T3, Canada}
\email[]{erin.hetdex@gmail.com}

\author[0000-0002-6186-5476]{Shun Saito}
\affiliation{Kavli Institute for the Physics and Mathematics of the Universe (WPI), The University of Tokyo Institutes for Advanced Study (UTIAS), The University of Tokyo, Chiba 277-8583, Japan}
\affiliation{Institute for Multi-messenger Astrophysics and Cosmology, Department of Physics, Missouri University of Science and Technology, 1315 N. Pine St., Rolla MO 65409, USA}
\email{saitos@mst.edu}

\author[0000-0003-1198-831X]{Ariel G. S\'anchez}
\affiliation{Max-Planck-Institut f\"ur extraterrestrische Physik, Postfach 1312, Giessenbachstr., 85748 Garching, Germany}
\email{arielsan@mpe.mpg.de}

\author[0000-0001-7240-7449]{Donald P. Schneider}
\affiliation{Department of Astronomy \& Astrophysics, The Pennsylvania State University, University Park, PA 16802, USA}
\affiliation{Institute for Gravitation and the Cosmos, The Pennsylvania State University, University Park, PA 16802, USA}
\email{dps7@psu.edu}


\begin{abstract}
We use the monopole and quadrupole power spectra of [\ion{O}{2}]-emitting galaxies in the HETDEX Public Data Release 1 to constrain the growth of cosmic structures at $z\le0.48$. We interpret the power spectra using a simulation-based model built from the Mucho-Uchuu-1G $N$-body simulation. The best-fit model reproduces the observed power spectra remarkably well up to $k_\mathrm{max}=0.7~h~\mathrm{Mpc}^{-1}$, and goodness-of-fit tests confirm that the HETDEX [\ion{O}{2}] clustering is consistent with the \textit{Planck} flat $\Lambda$CDM cosmology. We test the general relativity (GR) prediction for the linear growth index, $\gamma\simeq 0.55$. When the Alcock-Paczy\'nski parameters are held fixed at their fiducial values, we find consistency with the GR prediction, $\gamma=0.55\pm 0.05$. And, when the AP parameters are allowed to vary, $\gamma=0.60\pm 0.07$, which is consistent with both the GR prediction and the suppressed growth rate reported by \citet{Nguyen_2023}. These results establish HETDEX [\ion{O}{2}] galaxies as a competitive, independent low-redshift probe of the growth of structures and motivate future full cosmological inference.
\end{abstract}


\section{Introduction} 
\label{sec:intro}
The Hobby-Eberly Telescope Dark Energy Experiment~\citep[HETDEX;][]{Gebhardt_2021} has produced new data on the cosmic web---the large-scale structure (LSS) of the universe~\citep{peebles:1980}---over a wide range of cosmic look-back times. These data were obtained through an untargeted integral field unit spectroscopic survey using the Visible Integral-field Replicable Unit Spectrograph~\citep[VIRUS;][]{Hill_2021}, on the 10m Hobby-Eberly Telescope~\citep[HET;][]{Ramsey_1998,Hill_2021}. 

HETDEX's catalog of emission-line-selected sources contains half a million [\ion{O}{2}]-emitting galaxies at redshifts of $z\le 0.48$ and 1.6 million Ly$\alpha$ emitter candidates at $1.88\le z\le 3.52$~\citep{Mentuch_Cooper_2023,HETDEX_PDR1}. This paper builds upon \textit{volume-limited samples} of HETDEX [\ion{O}{2}] galaxies constructed in \citet{Moon_2026} (Paper I).

By measuring the clustering statistics of the LSS, we can study the expansion history of the universe, $H(z)$~\citep[e.g.,][]{Seo_Eisenstein_2003}, and the growth history of cosmic structures, $f(z)\equiv\mathrm{d}\ln D(a)/\mathrm{d}\ln a$~\citep[e.g.,][]{Song_Percival_2009}. Here, $D(a)$ is the growth factor for linear matter density fluctuations, and $a\equiv(1+z)^{-1}$ is the cosmological scale factor. LSS statistics can also be used to determine the constituents of the universe and their proportions, including the absolute mass scale of neutrinos and the physical nature of dark energy~\citep[see][for reviews]{neutrino_review_2006,dark_energy_review_2008}. 

The late-time LSS measurements are highly complementary to the early-universe constraints from the cosmic microwave background (CMB), helping break parameter degeneracies and test the standard $\Lambda$ cold dark matter ($\Lambda$CDM) model~\citep{Huterer_2023}.

As discussed in~Paper I, the low-redshift universe has become increasingly important for LSS studies. For example, \citet{Nguyen_2023} reported evidence for a suppressed LSS growth rate, $f(z)$, at low redshifts (i.e., the dark-energy-dominated era) compared to the prediction of general relativity in a flat $\Lambda$CDM cosmology,
\begin{equation}
    f(z)\simeq \left[\Omega_\mathrm{m}(z)\right]^\gamma,
\end{equation}
with $\gamma\simeq 0.55$~\citep{peebles:1980,Lahav_1991,Bouchet_1995}.
Here,
\begin{equation}
    \Omega_\mathrm{m}(z)\equiv\Omega_\mathrm{m,0}\frac{(1+z)^3H^2(z)}{H^2_0},\quad \Omega_\mathrm{m,0}\equiv\Omega_\mathrm{m}(z=0),
\end{equation}
and $H_0\equiv H(z=0)=100~h~\mathrm{km~s^{-1}~Mpc^{-1}}$ is the Hubble constant.
\citet{Nguyen_2023} found $\gamma=0.633^{+0.025}_{-0.024}$~(68\%~credible intervals) using various LSS probes. Since $\Omega_\mathrm{m}<1$, $\gamma>0.55$ implies a suppressed growth rate.

There is also an ongoing debate surrounding the so-called ``$S_8$ tension,'' a discrepancy between the clumpiness of the present-day universe 
inferred from the CMB and weak gravitational lensing surveys. Here, $S_8\equiv \sigma_8\sqrt{\Omega_\mathrm{m,0}/0.3}$ characterizes the present-day amplitude of linear matter density fluctuations~\citep{Jain_1997}, where $\sigma_8$ is the r.m.s.\ mass fluctuation enclosed within a sphere of $8~h^{-1}~\mathrm{Mpc}$. 
The Subaru Hyper Suprime-Cam Year 3~\citep[HSC;][]{HSCY3_shear} and the Dark Energy Survey Year 6 cosmic shear analyses~\citep[DES;][]{DESY6_shear} reported $S_8$ values that are in a $\simeq 2\sigma$ tension with the CMB value, while the Kilo-Degree Survey's Legacy cosmic shear analysis~\citep[KiDS;][]{KIDSLegacy_shear} and the cross-correlation between galaxies found in the Dark Energy Spectroscopic Instrument (DESI) Legacy Imaging Survey~\citep{DESIlegacy_2019} and CMB lensing measurements from the Atacama Cosmology Telescope Data Release 6 (ACT DR6) and \textit{Planck} Public Release 4~\citep{ACT_2025} found them to be in good agreement. However, \citet{ACT_2025} also found a hint of suppressed $S_8$ at $\simeq 2\sigma$ in the lowest redshift bin. 

The DESI DR2 Baryon Acoustic Oscillation (BAO) analysis \citep{DESIDR2_BAO} presented evidence for dynamical dark energy, i.e., redshift evolution in the dark energy equation of state, by combining DESI DR2 BAO measurements with Type Ia supernovae data. Since time-evolving dark energy affects the growth of structures, this dynamical dark energy scenario can also be tested through measurements of $f(z)$.

Testing these findings requires independent LSS probes in the low-redshift universe ($z\lesssim 0.5$) that can provide precise and accurate measurements of $f(z)$. Several spectroscopic surveys have probed the growth of low-$z$ structures using photometrically selected galaxies, such as the Galaxy And Mass Assembly survey~\citep[$z\lesssim 0.5$;][]{GAMADR2,GAMA_2021}, the DESI Bright Galaxy Survey~\citep[$z\lesssim 0.6$;][]{DESIBGS_2023}, and the DESI Peculiar Velocity Survey~\citep[$z\lesssim 0.1$;][]{DESIPV_2025}. In this context, the HETDEX [\ion{O}{2}] sample offers a complementary and independent probe of the growth of structures, using its unprecedented sample of $z \leq 0.48$ emission-line galaxies with no photometric pre-selection. By comparing the results of these surveys, which differ in their target selection criteria and systematics, we can obtain robust constraints on the cosmological parameters.

In this work, we use the HETDEX volume-limited [\ion{O}{2}] sample, as described in Paper I, to study the growth history of cosmic structures and estimate $\gamma$. 
The rest of this paper is organized as follows.
Section~\ref{sec:hetdex_data} presents the HETDEX [\ion{O}{2}] galaxies from the HETDEX Public Data Release 1~\citep[PDR1;][]{HETDEX_PDR1} and their volume-limited subsamples~(Paper I). 
Section~\ref{sec:methodology} reviews our methodology for constraining the growth of structures using galaxy power spectrum multipoles. Section~\ref{sec:pk_multipoles} describes the power spectrum multipoles of the volume-limited [\ion{O}{2}] galaxies from data and mocks with varying peculiar velocities and Alcock-Paczy\'{n}ski~\citep{AP_1979} parameters. In Section~\ref{sec:interpretation}, we infer the growth history of cosmic structures characterized by $\gamma$ from the best-fit monopole and quadrupole power spectra and we conclude in Section~\ref{sec:conclusions}. 

Throughout this paper, we assume a flat $\Lambda$CDM model following the \textit{Planck} 2018 CMB measurement~\citep{Planck18} when converting redshfit $z$ to comoving distance in units of $h^{-1}~\mathrm{Mpc}$, and applying galaxy’s peculiar velocity along the line of sight (LOS) to the galaxy position in simulations.

\section{HETDEX Data}
\label{sec:hetdex_data}

\subsection{HETDEX Public Data Release 1}
\label{subsec:hetdex_PDR1}
HETDEX's PDR1~\citep{HETDEX_PDR1} reports $\simeq 86$~deg$^2$ of spectroscopic observations that span roughly 540~deg$^2$ of sky.  
The HETDEX Public Source Catalog 2
(HPSC2) of PDR1 contains approximately $4 \times 10^5$ candidates of Ly$\alpha$ emitters (LAEs) at $1.88 \le z \le 3.52$ and $5 \times 10^5$ [\ion{O}{2}] emitting galaxies at $z \le 0.48$, along with stars, low-redshift continuum-selected galaxies, and active galactic nuclei.  Robust classifications and redshifts for the full catalog have been obtained through several independent approaches~\citep{Dustin_2023, Mentuch_Cooper_2023}, with 94.1$\%$ of sources agreeing to within $|\Delta z|<0.02$ when compared to external spectroscopic measurements \citep{HETDEX_PDR1}. 

HETDEX uses a sparse-sampling approach to measure LSS~\citep{Chiang_2013}. Due to the gaps between adjacent VIRUS integral field units (IFUs) on the focal plane, the survey footprint is sampled sparsely rather than being fully covered by the IFUs. This format results in a fill factor of approximately $1/4.6$ per observation. As described in Section~\ref{subsec:mock}, we use simulations to forward-model the window functions induced by this observing pattern.

Following~Paper I, we use data from the two primary HETDEX fields: a high-declination ``Spring'' field, which spans $390~\mathrm{deg}^2$, and an equatorial ``Fall'' field, which extends over $150~\mathrm{deg}^2$ \citep{HETDEX_PDR1}. 
We use the PDR1 v1.3 catalog (\texttt{hetdex\_sc2\_v1.3.fits}), which, for the Spring and Fall fields, contains 286,992 and 140,122 [\ion{O}{2}]-emitting galaxies within comoving volumes of $4.7\times10^7$ and $2.6\times10^7~h^{-3}~\mathrm{Mpc}^3$, respectively. The full [\ion{O}{2}] catalog has a number density of about $6\times10^{-3}~h^3~\mathrm{Mpc}^{-3}$.

\subsection{HETDEX PDR1 \texorpdfstring{[\ion{O}{2}]}{[O II]} Volume-limited Catalog}
\label{subsec:hetdex_pdr1_oii}

In Paper I, we constructed volume-limited [\ion{O}{2}] samples from the HETDEX HPSC2 in three luminosity bins: $40.1 < \log L \leq 40.5$ (Bin 2),
$40.5 <\log L \leq40.9$ (Bin 3), and $40.9 <\log L \leq 41.3$ (Bin 4), where $L$ is in units of erg~s$^{-1}$. These subsamples provide number densities, $\bar n$, five to ten times higher than those of typical spectroscopic surveys of emission-line galaxies. See Table \ref{tab:table_oii_sub} for their properties. The effective redshift, $z_\mathrm{eff}=N_\mathrm{gal}^{-1}\int \mathrm{d}z\frac{\mathrm{d}V}{\mathrm{d}z}\bar n(z)$, in the table is computed as the number-weighted mean redshift of the galaxy sample. 
The DESI collaboration defines $z_\mathrm{eff}$ as the $\bar{n}^2$-weighted mean redshift~\citep{Adame_2025}. Since our volume-limited samples have nearly constant $\bar{n}(z)$, the two approaches yield similar results, with differences smaller than $10^{-3}$.

\begin{deluxetable*}{cccccccc}
\tabletypesize{\small}
\tablecaption{Key properties of the [\ion{O}{2}] volume-limited samples in the 
Spring and Fall fields. \label{tab:table_oii_sub}}
\tablehead{
\colhead{Samples} & \colhead{Field} & \colhead{Luminosity range} & \colhead{$z$ range} & 
\colhead{$z_\mathrm{eff}$} & \colhead{$N_\mathrm{gal}$} & \colhead{$10^3\bar{n}$} & \colhead{Volume} \\
\colhead{} & \colhead{} & \colhead{$\log(L~[\mathrm{erg~s}^{-1}])$} & \colhead{} & 
\colhead{} & \colhead{} & \colhead{[$h^3~\mathrm{Mpc}^{-3}$]} & \colhead{[$h^{-3}~\mathrm{Mpc}^3$]}
}
\startdata
Bin 2 & Spring & 40.1 $<$ $\log L$ $\leq$ 40.5 & 0.091 $\leq z \leq$ 0.231 & 0.176 & 23347 & 3.9 & $6.0\times 10^6$\\
Bin 3 & Spring & 40.5 $<$ $\log L$ $\leq$ 40.9 & 0.183 $\leq z \leq$ 0.273 & 0.235 & 21573 & 3.1 & $6.9\times 10^6$ \\
Bin 4 & Spring & 40.9 $<$ $\log L$ $\leq$ 41.3 & 0.258 $\leq z \leq$ 0.468 & 0.378 & 64794 & 1.9 & $3.5\times 10^7$ \\
\hline
Bin 2 & Fall & 40.1 $<$ $\log L$ $\leq$ 40.5 & 0.102 $\leq z \leq$ 0.212 & 0.168 & 11354 & 4.6 & $2.5\times 10^6$ \\
Bin 3 & Fall & 40.5 $<$ $\log L$ $\leq$ 40.9 & 0.146 $\leq z \leq$ 0.316 & 0.247 & 24536 & 3.2 & $7.6\times 10^6$ \\
Bin 4 & Fall & 40.9 $<$ $\log L$ $\leq$ 41.3 & 0.313 $\leq z \leq$ 0.453 & 0.389 & 27494 & 1.9 & $1.4\times 10^7$ \\
\enddata
\end{deluxetable*}

The monopole and quadrupole power spectra of these samples have shown excellent agreement with the mocks generated from the Uchuu simulation~\citep{Uchuu_dr1}, based on \textit{Planck}'s flat $\Lambda$CDM cosmology~\citep{Planck15}. We created the mock catalogs by populating dark matter halos, including subhalos, of mass $M_\mathrm{h}$ with [\ion{O}{2}] galaxies using a lognormal function for the expectation value (see Equation~\eqref{eq:lognormal} and Section~\ref{subsec:mock} for details). 

The best-fit mock reveals characteristic dark matter halo masses of $\log(M_0~[h^{-1}M_\sun]) \simeq 11.9$--$12.3$, which increase weakly with [\ion{O}{2}] luminosity ($M_0 \propto L^{0.37}$), with approximately 13\% of galaxies residing in subhalos.

In Paper I, only the monopole power spectra were used for inference. In this paper, we also include the quadrupole power spectra to constrain the growth history of cosmic structures from the HETDEX volume-limited [\ion{O}{2}] galaxies.

\section{Methodology for constraining the growth of structures}
\label{sec:methodology}

We use the effect of redshift-space distortions~\citep[RSD;][]{Sargest_Turner_1977,Kaiser_1987,Hamilton_1998,Peacock_2001,Guzzo_2008} on the galaxy power spectrum multipoles to constrain the growth history of cosmic structures. The peculiar velocity of galaxies, $\mathbf{v}$, shifts their observed positions in comoving coordinates, $\mathbf{s}$, with respect to their true positions, $\mathbf{r}$, along the LOS as $\mathbf{s}=\mathbf{r}-f\left(\mathbf{u}\cdot\hat n\right)\hat n$. Here, $\hat n$ is a LOS unit vector and $\mathbf{u}\equiv -\mathbf{v}/(faH)$ is a scaled velocity~\citep{Scoccimarro_2004}. As a result, the observed galaxy distribution becomes statistically anisotropic, with $\hat n$ being a preferred direction; thus, the power spectrum depends on $\mu\equiv \hat{k}\cdot\hat{n}$ where $\hat k\equiv \mathbf{k}/|\mathbf{k}|$ is a unit vector and $\mathbf{k}$ is a wave vector. 
We write $k=|\mathbf{k}|$ in this paper. 
Expanding the power spectrum, $P(k,\mu)$, in terms of the Legendre polynomials, ${\cal L}_\ell(\mu)$, RSD modifies the monopole ($\ell=0$) and generates a non-vanishing quadrupole ($\ell=2)$ and higher, even multipole power spectra~\citep{Cole_Fisher_Weinberg_1994}. 
Due to symmetry, odd multipoles vanish for a single tracer~\citep{McDonald_2009,Yoo_2012,Bonvin_2014,Gaztanaga_2017}, however, we find non-zero dipole measurements in the HETDEX [\ion{O}{2}] samples, and these results are well reproduced by the mocks. Further details of the dipole measurements are presented in Appendix~\ref{appendix:dipole}.

In linear theory, the continuity equation gives $\nabla\cdot\mathbf{u}=\delta$, where $\delta$ is the matter density contrast. Therefore, RSD is sensitive to $f\sigma_R$, where $\sigma_R$ is the amplitude of the matter density contrast at a reference scale $R$~\citep{Song_Percival_2009}. 
Often, $R$ is taken to be $8~h^{-1}\,\mathrm{Mpc}$~\citep{Peebles_1982}. However, \citet{Sanchez_2020} proposes eliminating the dependence of $R$ on $h$, as this approach introduces a spurious dependence on the Hubble constant in cosmological inference.
Different values of $h$ imply the comparison of different scales, which can be alleviated by setting $R$ in physical units, e.g., 12 Mpc (transforming $8~h^{-1}\,\mathrm{Mpc}$ assuming $h \simeq 0.67$). 

As described in Section~\ref{subsec:mock}, we use cosmological $N$-body simulations to forward-model the power spectrum multipoles. 
In this analysis, we rescale the velocities of all halos, including subhalos, in the simulation by a factor of $\alpha_v$, so that $\mathbf{v}=\alpha_v\mathbf{v}_\mathrm{fid}$, where $\mathbf{v}_\mathrm{fid}$ is the original velocity. This parameter is similar to $\gamma_\mathrm{HV}$~\citep{Reid_2014,GAMA_2021} and $\gamma_f$~\citep{Zhai_2019} used in the literature. However, since subhalos were not resolved in their simulations, the correspondence between $\alpha_v$ and their parameters is not exact. Since we use the same $\alpha_v$ for all halos, including subhalos, and do not include any additional contributions, we do not account for a potential velocity bias for subhalos for this work.

Previous studies have shown that the linear growth parameter can be inferred accurately from $\alpha_v$ using both linear and nonlinear scales~\citep{Reid_2014,Zhai_2019,GAMA_2021}. We will discuss the effect of $\alpha_v$ on the power spectrum multipoles and the interpretation of this parameter in detail in Sections~\ref{subsec:pk_results} and~\ref{sec:interpretation}.

RSD is not the only source of apparent statistical anisotropy.
The Alcock-Paczy\'nski (AP) effect arises when an incorrect cosmological model is used to convert redshifts into comoving radial distances~\citep{AP_1979,Ballinger_1996,Matsubara_Suto_1996}. This effect reflects the different cosmological dependence of comoving length scales observed along and across the LOS:
\begin{equation}
    \Delta r_{\parallel}=\frac{c\, \Delta z}{H(z)}, \qquad \Delta r_{\perp}=(1+z)D_\mathrm{A}(z)\Delta \theta,
    \label{eq:AP_effect}
\end{equation}
where $D_\mathrm{A}(z)$ is the angular diameter distance, and $\Delta z$ and $\Delta\theta$ denote redshift and angular separations, respectively.

We incorporate the AP effect into our analysis as follows.
The galaxy distribution is expected to remain isotropic as the universe expands. Using the correct cosmological model to convert galaxy redshifts into comoving radial distances recovers this isotropy in comoving space, $\mathbf{r}$. However, using an incorrect cosmology alters $D_A$ and $H(z)$ by different amount, producing an anisotropic distribution that appears squashed or stretched along the LOS\null. This effect is similar to RSD anisotropy in the linear regime~\citep{Ballinger_1996}. We therefore also estimate the AP parameters defined by
\begin{equation}
    q_{\parallel}=\frac{H_\mathrm{fid}(z)}{H_\mathrm{true}(z)}, \qquad q_{\perp}=\frac{D_{A,\mathrm{true}}(z)}{D_{A,\mathrm{fid}}(z)},
    \label{eq:AP_parameters}
\end{equation}
where ``true'' and ``fid'' denote the true and fiducial values, respectively.

Instead of directly rescaling the comoving positions of every galaxy in the mock catalogs to apply a given choice of $q_\perp$ and $q_\parallel$, we measure the power spectrum multipoles once at the fiducial AP values and then rescale them analytically using the $(k,\mu)\to(k_\mathrm{true},\mu_\mathrm{true})$ mapping of \citet{Ballinger_1996},
\begin{equation}
\begin{split}
    k_\mathrm{true} = \frac{k}{\alpha_\perp}\left[1+\mu^2\left(\frac{1}{F^2}-1\right)\right]^{1/2},\\
    \qquad
    \mu_\mathrm{true} = \frac{\mu}{F}\left[1+\mu^2\left(\frac{1}{F^2}-1\right)\right]^{-1/2},
    \label{eq:ap_mapping}
\end{split}
\end{equation}
where $F\equiv \alpha_\parallel/ \alpha_\perp$, and following the convention of \citet{Brieden_2021}, $\alpha_\parallel$ and $\alpha_\perp$ are related to $q_\parallel$ and $q_\perp$ by $\alpha_\parallel = (r_\mathrm{d,fid}/r_\mathrm{d,true})\,q_\parallel$ and $\alpha_\perp = (r_\mathrm{d,fid}/r_\mathrm{d,true})\,q_\perp$, where $r_\mathrm{d}$ is the size of the sound horizon at the baryon drag epoch. In this work, we fix $r_\mathrm{d}$ to its fiducial value, so that $\alpha_\parallel = q_\parallel$ and $\alpha_\perp = q_\perp$.
We therefore refer to these as the AP parameters, $\alpha_\parallel$ and $\alpha_\perp$, in the sections that follow.
This analytic ``$P(k)$-level'' rescaling is significantly less computationally expensive than repeatedly rescaling and re-measuring the power spectra from mock catalogs directly (``catalog-level'' rescaling) over our full grid of AP parameters. In Appendix~\ref{appendix:AP_param_test}, we validate that this approximation accurately reproduces the catalog-level power spectrum multipoles.

The goal of this paper is to constrain $\alpha_v$, $\alpha_\parallel$, and $\alpha_\perp$ from the power spectrum multipoles of the volume-limited samples of HETDEX [\ion{O}{2}] galaxies.

\section{Power Spectrum Multipoles}
\label{sec:pk_multipoles}

We use the same power spectrum measurements of the HETDEX data as in Paper I, to which we refer the reader for further details on the estimator and measurements. A brief summary of the power spectrum estimation procedure is provided in Section~\ref{subsec:pk_estimator}.

However, the mocks used to interpret the observed clustering measurements are different from those of Paper I. Instead of using the Uchuu simulation~\citep{Uchuu_dr1}, we adopt halo catalogs from the Mucho-Uchuu-1G (MU1) simulation suite~\citep{Fernandez_2026,MuchoUchuu}\footnote{\url{https://hpci-web01.r-ccs.riken.jp/nextcloud/s/Uchuu}}.  The rest of the mock production procedure remains the same as in the previous paper, and is summarized in Section~\ref{subsec:mock}.
Section~\ref{subsec:pk_results} provides a comparison of the data with mocks of various values of $\alpha_v$, $\alpha_\perp$, and $\alpha_\parallel$.

\subsection{Estimator}
\label{subsec:pk_estimator}

We estimate the power spectrum multipoles using an FFT-based approach following the Feldman-Kaiser-Peacock formalism~\citep{FKP_1994}. We use the Yamamoto estimator~\citep{Yamamoto_2006} to account for the varying LOS directions, implemented via multiple FFTs~\citep{Hand_2017}. 

Galaxies are assigned to the grid using a Cloud-in-Cell (CIC) scheme, with each object weighted by $w(r)=N_\mathrm{overlap}^{-1}$ to account for repeated observational shots overlapping at the same sky position~(see Appendix A of Paper I). The shot noise term is subtracted from the monopole power spectrum and its contribution is negligible for higher-order multipoles. Then, the CIC kernel is deconvolved following~\citet{Wang_2024}. 

Random catalogs, with number densities $\alpha^{-1}$ times the data density (ranging from $\alpha^{-1}\simeq 54$ to 164 depending on the luminosity bin and field), are constructed using the fiber and sky emission line masks~\citep{HETDEX_PDR1}. We adopt a grid-cell resolution of $H = 2.2~h^{-1}~\mathrm{Mpc}$ for all bins to minimize window function effects from sparse IFU sampling~\citep{Chiang_2013}. We compute power spectra up to $k_\mathrm{max}=0.7~h~\mathrm{Mpc}^{-1}$, approximately half the Nyquist wavenumber, $\pi/H$. See Section 4 of Paper I for further details on the estimator and measurements.

\subsection{Mock: Simulation-based Model}
\label{subsec:mock}

To interpret the measured power spectrum multipoles and investigate the [\ion{O}{2}] galaxy--halo connection, we model the clustering signal using mock catalogs constructed with a hybrid framework. Namely, we combine a halo occupation distribution~\citep[HOD; e.g.,][]{Zheng_2005,Geach_2012} approach with subhalo abundance matching~\citep[SHAM; e.g.,][]{Kravtsov_2004,Behroozi_2019} approach, following the same methodology as in Paper I. 

In this framework, we treat all halos, including subhalos, equally. The probability of a halo hosting a galaxy is parameterized as a function of halo mass, $M_\mathrm{h}$,
\begin{equation}
\label{eq:lognormal}
    N_\mathrm{g}(M_\mathrm{h})=F_\mathrm{g}\exp\left[-\frac{\log^2(M_\mathrm{h}/M_0)}{2\sigma^2_{\log M}}\right]\,,
\end{equation}
where $F_\mathrm{g}$ is the mean fraction of [\ion{O}{2}] emitting galaxies hosted by dark matter halos with a characteristic mass of $M_0$, and $\sigma_{\log M}$ is the standard deviation of the Gaussian distribution. 
We fix $\sigma_{\log M}=0.6$ and set $F_\mathrm{g}$ to match the number density of each sample, leaving $M_0$ as the only free parameter in this model. 
We refer to this approach as the Subhalo-based HOD (SHOD) model throughout the paper. 
Further details of this model are provided in Section 4.2 of Paper I.

We regard this model as a combination of HOD and SHAM because it shares key features of both approaches. It resembles SHAM in that we use simulations that resolve subhalos and populate each subhalo with either one or no galaxy. It also resembles HOD in that galaxies are assigned to halos via a binomial distribution based solely on the halo mass, without explicitly incorporating galaxy properties such as stellar mass or luminosity. 
Since the model does not strictly extend either HOD or SHAM but rather combines elements of both, we refer to it as the SHOD model. 
Nevertheless, because we apply this model to volume-limited samples within narrow luminosity bins, it can also be understood as SHAM that implicitly accounts for galaxy luminosity.

In our previous work, we applied the SHOD model to halo catalogs from the Uchuu $N$-body simulation~\citep{Uchuu_dr1, Uchuu_nu2GCC, Uchuu_UM, Uchuu_RMG, Uchuu_SDSS}, in which halos and subhalos are identified using the \textsc{Rockstar} halo finder~\citep{Behroozi_2013}. We use $M_\mathrm{h}=M_\mathrm{200b}~[h^{-1}M_{\sun}]$ which is the halo mass enclosed within the mean overdensity of $200\rho_\mathrm{m}$ in the halo catalog. Here, $\rho_\mathrm{m}$ is the mean background mass density of the universe.
Uchuu adopts cosmological parameters from the \textit{Planck} 2015 CMB data~\citep{Planck15}, with an $N$-body particle mass of $m_p=3.27\times10^8~h^{-1}M_\sun$ and a corresponding minimum resolved halo mass of $1.3\times10^{10}~h^{-1}M_\sun$.

Since only one realization of a $2~h^{-1}~\mathrm{Gpc}$ cubic box is available for Uchuu, in this work we adopt halo catalogs from MU1~\citep{Fernandez_2026,MuchoUchuu}, which will ultimately comprise 1000 realizations of a $1~h^{-1}~\mathrm{Gpc}$ cubic box.
MU1 adopts cosmological parameters from the 2018 \textit{Planck} CMB release~\citep{Planck18} for a flat $\Lambda$CDM model: $\Omega_\mathrm{m} = 0.3111$, $\Omega_\Lambda = 0.6889$, $h = 0.6766$, $\sigma_8 = 0.8102$, $\Omega_\mathrm{b} = 0.048975$, and $n_\mathrm{s} = 0.9665$. 
Its minimum resolved dark matter particle mass, $m_p=1.26\times 10^9~h^{-1}M_\sun$, is approximately four times larger than that of Uchuu~\citep[see Table A.1 of][]{Fernandez_2026}, resulting in a correspondingly higher minimum halo mass of $5\times 10^{10}~h^{-1}~M_\sun$. 
In Appendix~\ref{appendix:sim_comparison}, we demonstrate that the use of halo catalogs from MU1 instead of Uchuu to populate SHOD galaxies with the same best-fit $M_0$ remains reasonable for this analysis, despite the differences in cosmology, halo mass resolution, and box size between the two simulations.

In this paper, we use the 100 realizations of MU1 that are currently publicly available. 
Each box can hold two volumes of Spring Bin 4 and more than four volumes of the other bins. 
Aside from using halo catalogs from a different simulation, we follow the same mock production procedure as in our previous work.
We refer the reader to Section 4.2 of Paper I for further details.

\subsection{Power Spectrum Multipoles of HETDEX Volume-limited  \texorpdfstring{[\ion{O}{2}]}{[O II]} Samples}
\begin{figure*}
    \centering
    \begin{minipage}{0.32\textwidth}
        \centering
        \includegraphics[width=\textwidth]{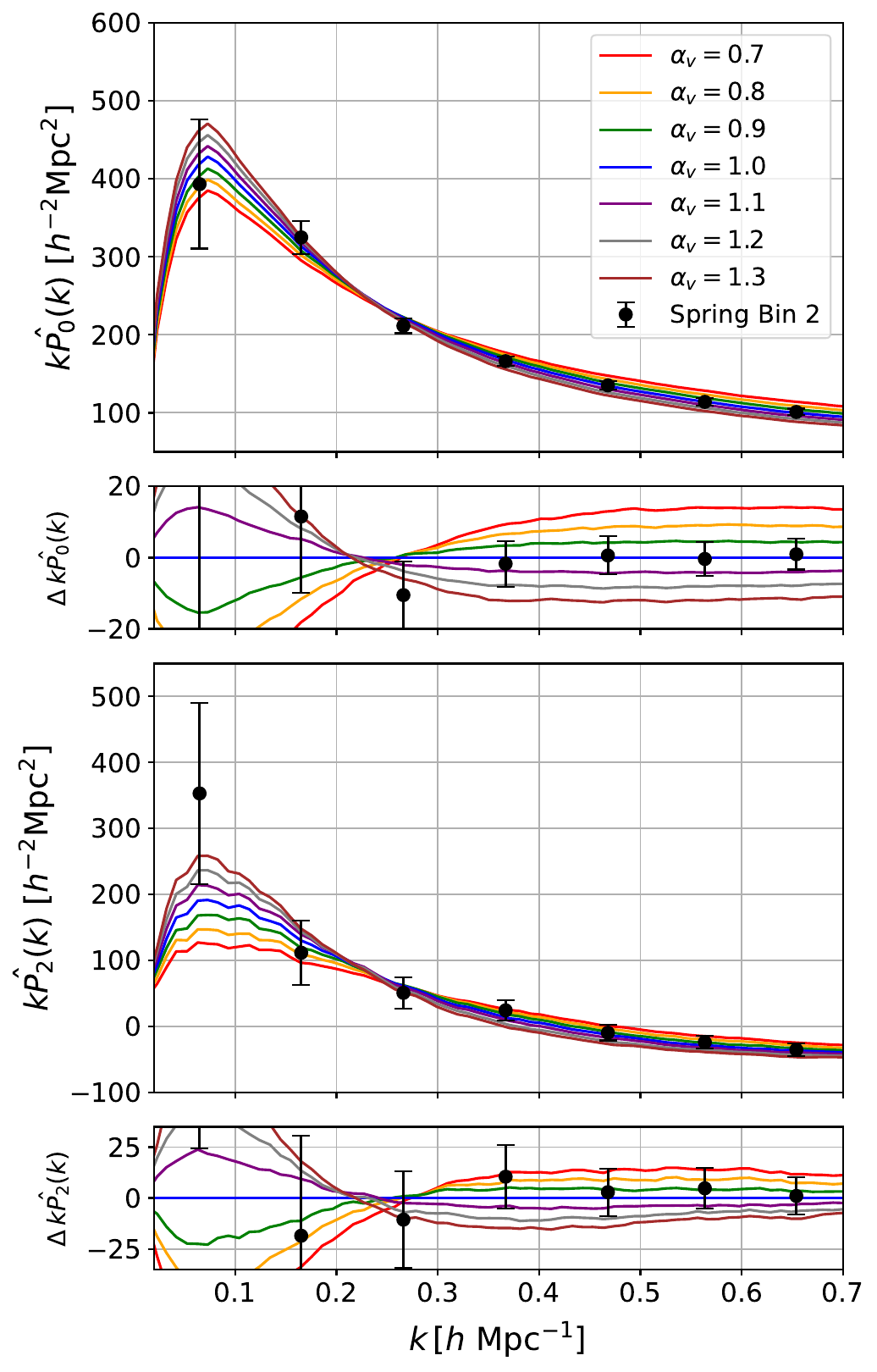}
    \end{minipage}
    \begin{minipage}{0.32\textwidth}
        \centering
        \includegraphics[width=\textwidth]{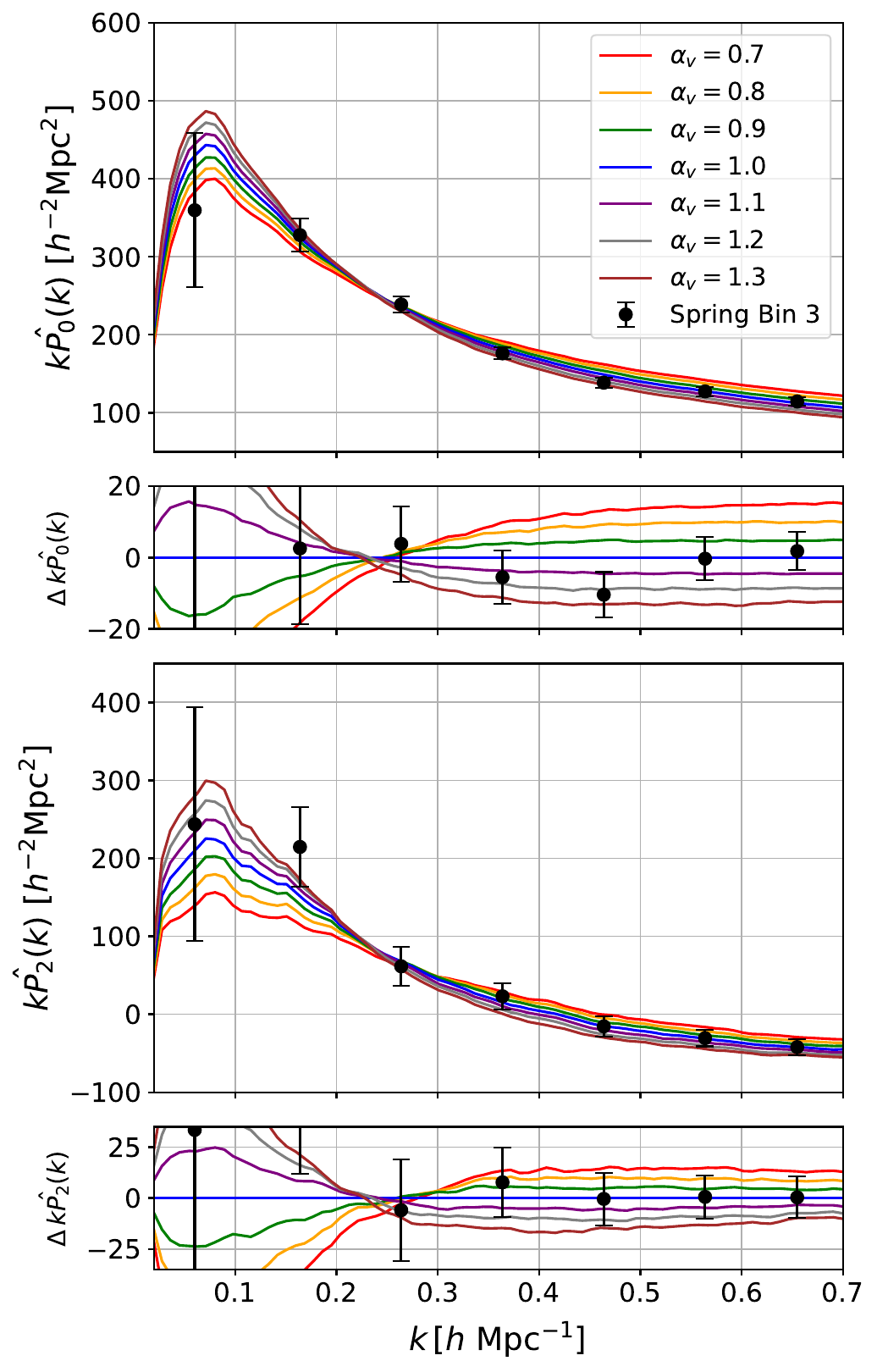}
    \end{minipage}
    \begin{minipage}{0.32\textwidth}
        \centering
        \includegraphics[width=\textwidth]{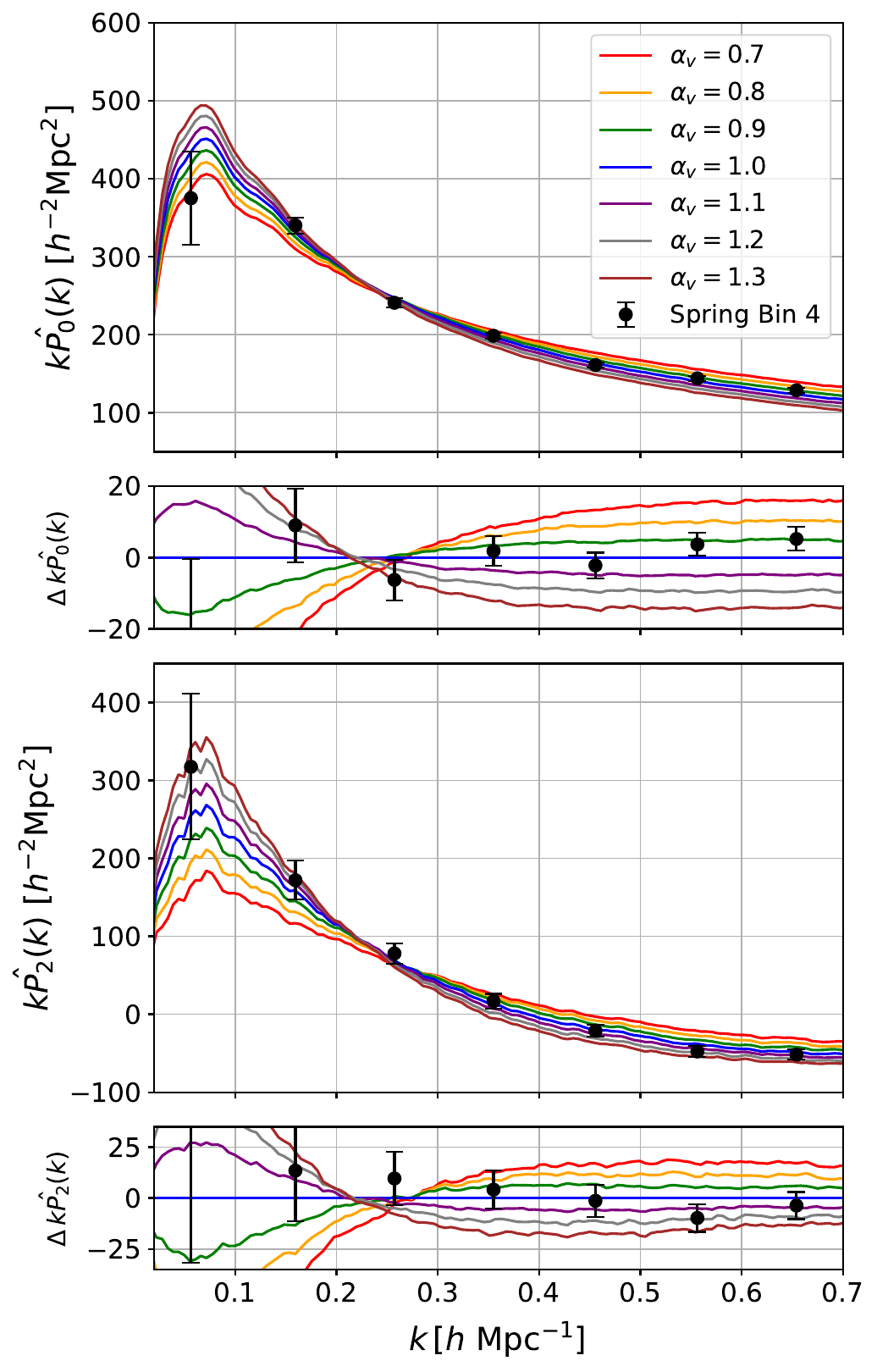}
    \end{minipage}
    \caption{Monopole (top panels) and quadrupole (third-row panels) power spectra of the HETDEX [\ion{O}{2}] volume-limited samples for Bins 2 (left), 3 (middle), and 4 (right) in the Spring field. The panels directly below each power spectrum show the difference, $\Delta\,k\hat{P}_\ell(k)$, relative to the $\alpha_v=1.0$ curve. The points with the error bars show the HETDEX data binned with a width $\Delta k=0.1~h~\mathrm{Mpc}^{-1}$. The $1\sigma$ uncertainties are estimated from the best-fit 400 mock realizations, except for Bin 4, which was estimated from 200 mocks. The colored solid lines represent the mean power spectra from the mocks for different values of $\alpha_v$ in finer bins of $\Delta k\simeq0.01~h~\mathrm{Mpc}^{-1}$, with the AP parameters held fixed at $\alpha_{\perp}=\alpha_{\parallel}=1$.}
    \label{fig:plot_pk_alpha_v_spring}
\end{figure*}
\begin{figure*}
    \centering
    \begin{minipage}{0.32\textwidth}
        \centering
        \includegraphics[width=\textwidth]{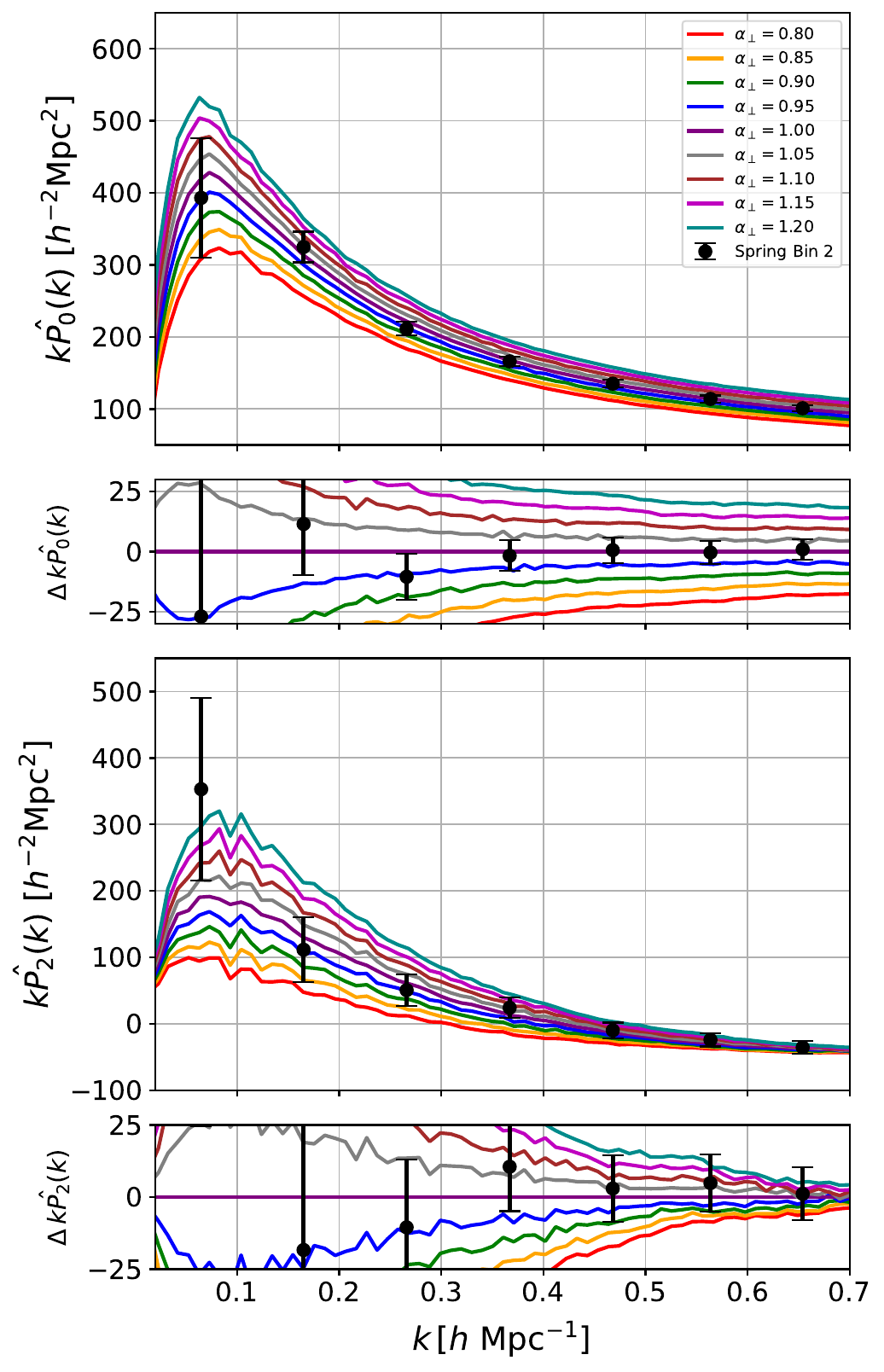}
    \end{minipage}
    \begin{minipage}{0.32\textwidth}
        \centering
        \includegraphics[width=\textwidth]{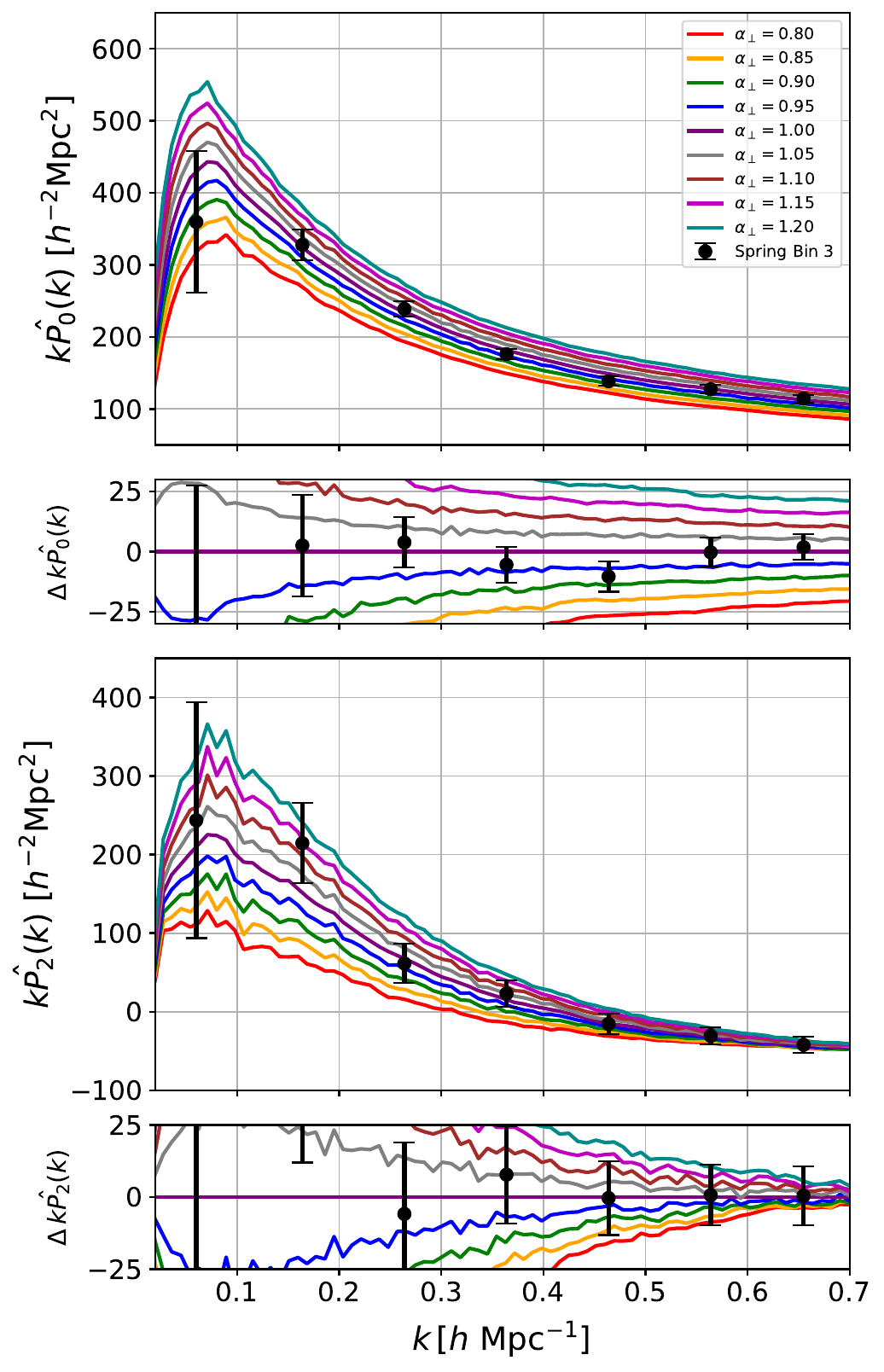}
    \end{minipage}
    \begin{minipage}{0.32\textwidth}
        \centering
        \includegraphics[width=\textwidth]{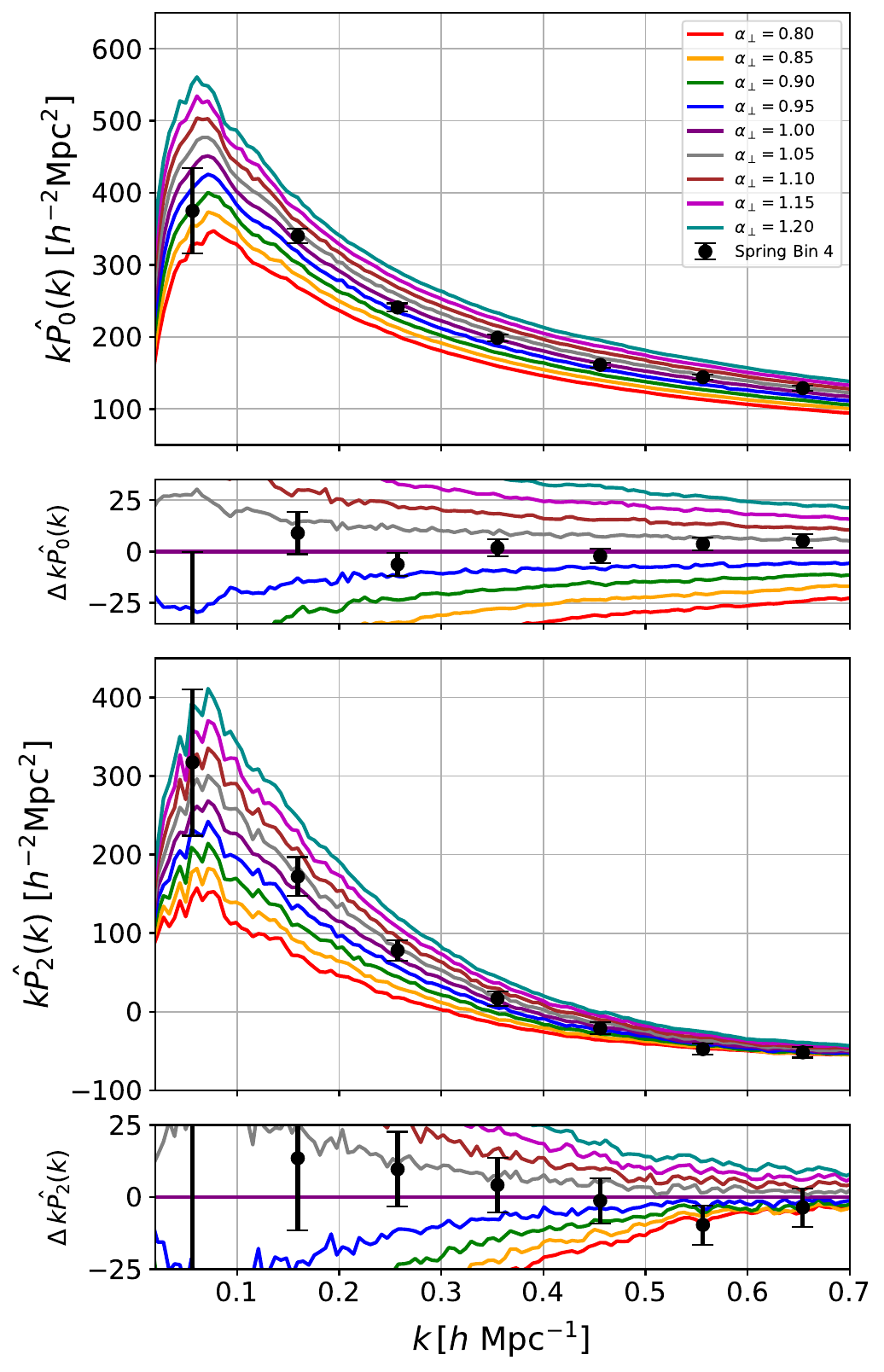}
    \end{minipage}
    \caption{Same format as Figure~\ref{fig:plot_pk_alpha_v_spring}, but for different values of one of the AP parameters, $\alpha_{\perp}$, with $\alpha_v$ and $\alpha_{\parallel}$ held fixed at unity.}
    \label{fig:plot_pk_alpha_perp_spring}
\end{figure*}
\label{subsec:pk_results}
\begin{figure*}
    \centering
    \begin{minipage}{0.32\textwidth}
        \centering
        \includegraphics[width=\textwidth]{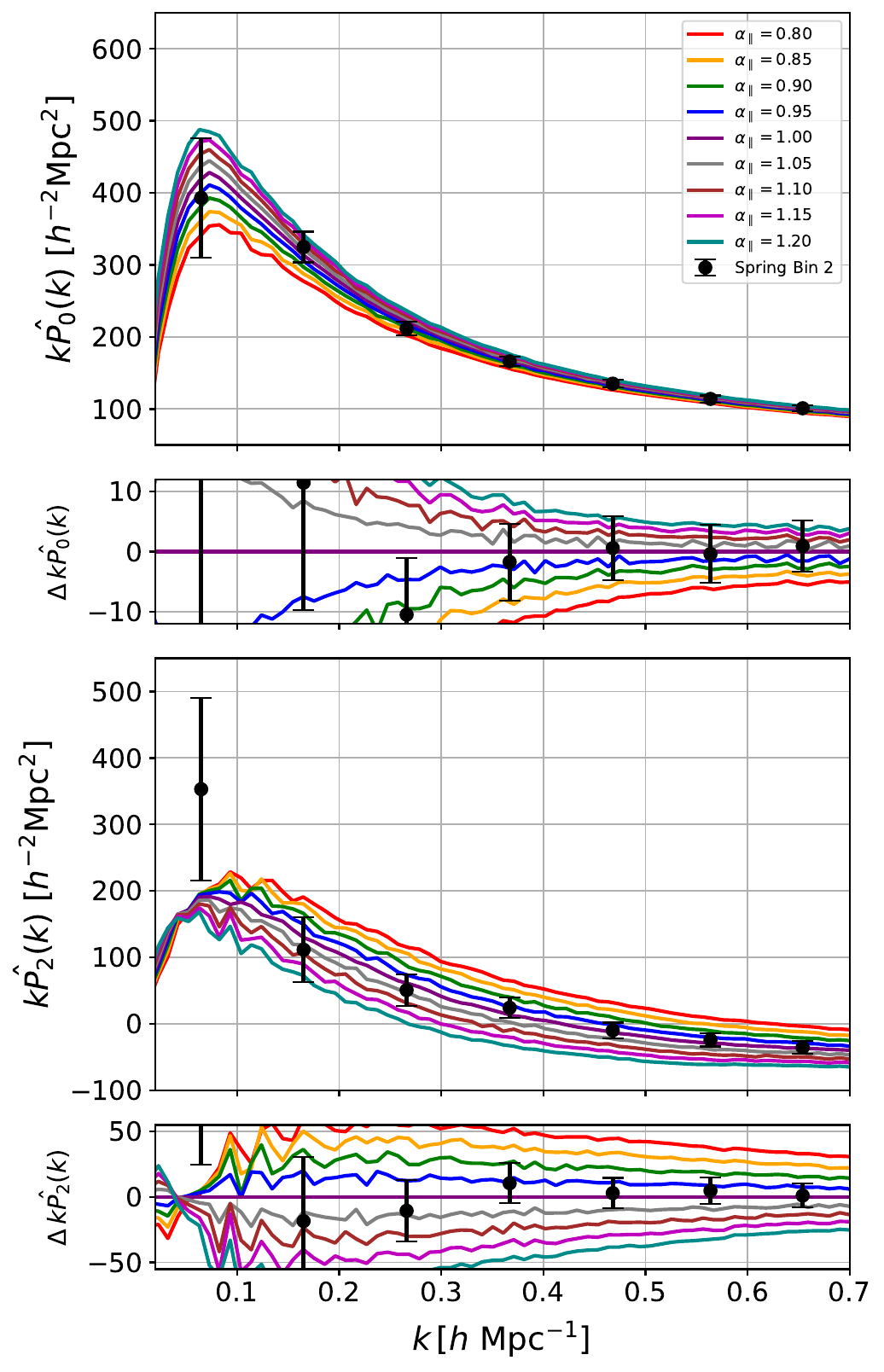}
    \end{minipage}
    \begin{minipage}{0.32\textwidth}
        \centering
        \includegraphics[width=\textwidth]{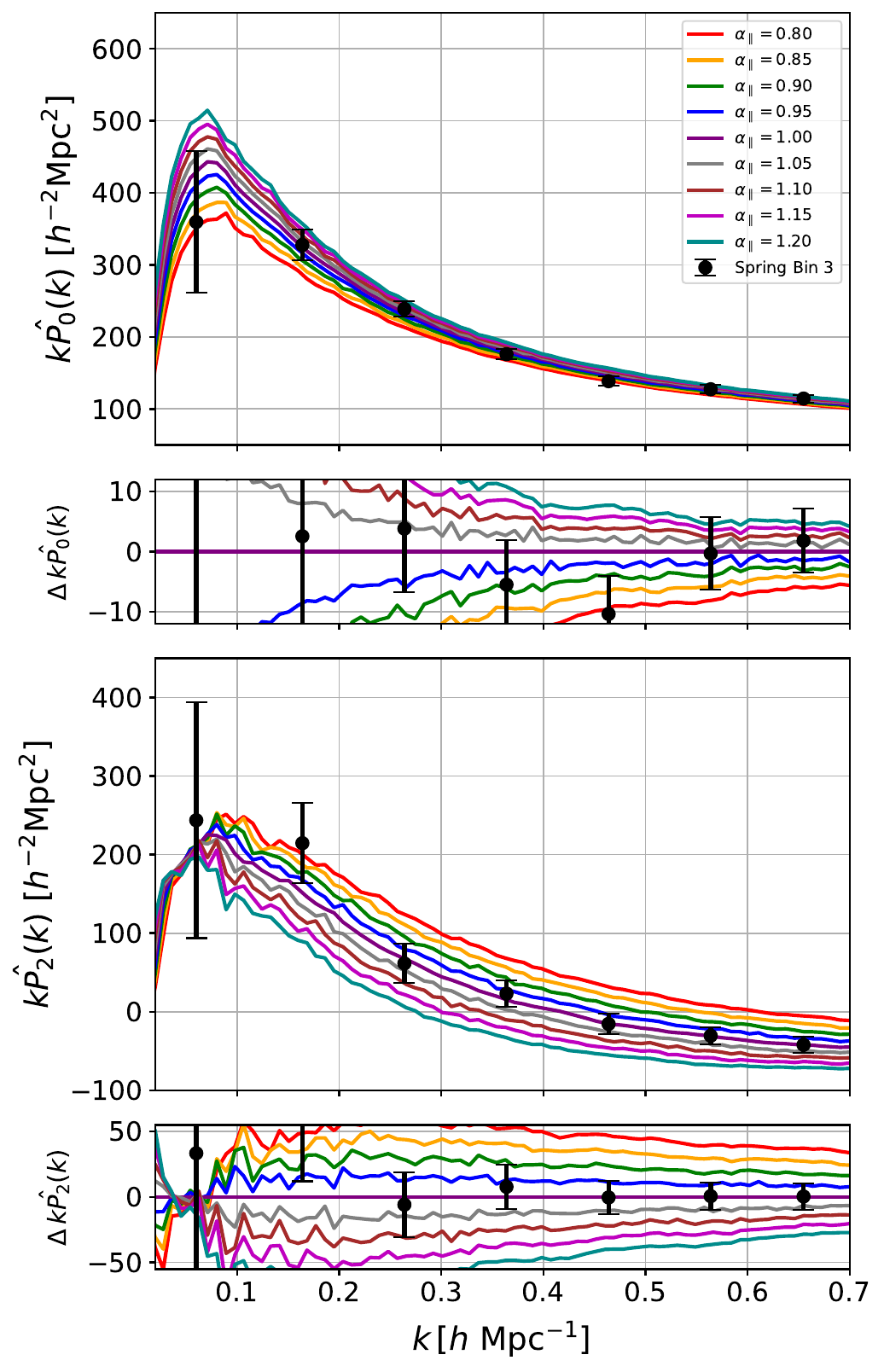}
    \end{minipage}
    \begin{minipage}{0.32\textwidth}
        \centering
        \includegraphics[width=\textwidth]{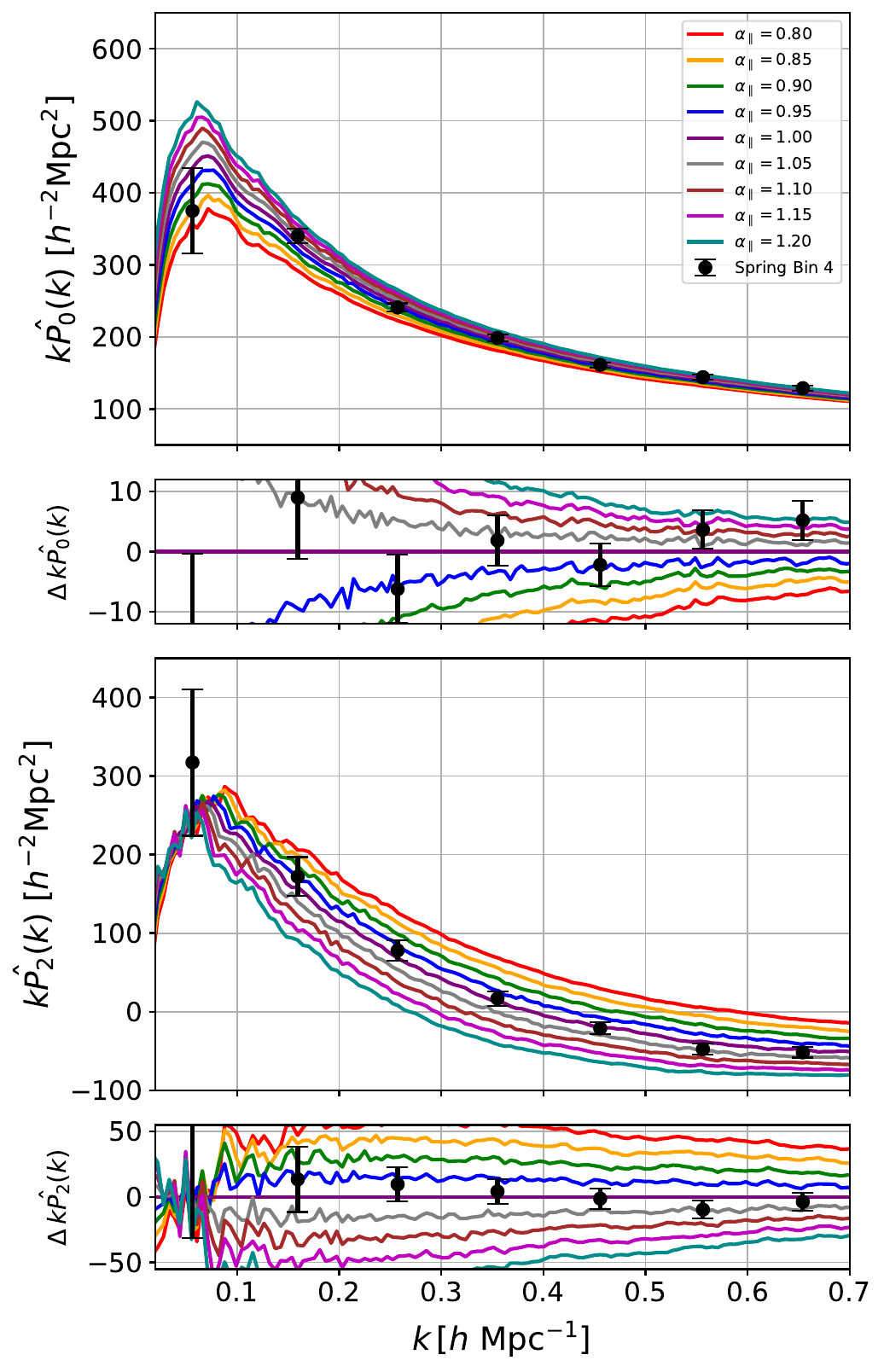}
    \end{minipage}
    \caption{Same format as Figure~\ref{fig:plot_pk_alpha_perp_spring}, but for different values of the other AP parameter, $\alpha_{\parallel}$, with $\alpha_v$ and $\alpha_{\perp}$ held fixed at unity.}
    \label{fig:plot_pk_alpha_par_spring}
\end{figure*}

Figure~\ref{fig:plot_pk_alpha_v_spring} 
presents the monopole and quadrupole power spectra measured from the HETDEX [\ion{O}{2}] volume-limited samples in the Spring field and compares them with mocks for different values of $\alpha_v$. The values of $M_0$ are taken from the best-fit values found in Paper I.
Corresponding figures for the Fall field are shown in Appendix~\ref{appendix:pk_fall}.
As predicted by the linear RSD effect~\citep{Kaiser_1987}, the monopole and quadrupole power spectra increase with increasing $\alpha_v$ at low $k\lesssim 0.25~h~\mathrm{Mpc}^{-1}$. However, they decrease at higher $k$ due to nonlinear RSD effect. This behavior differs greatly from that of $M_0$, as shown in Figures 6 and 7 of Paper I, and from that of the AP parameters, as shown in Figures~\ref{fig:plot_pk_alpha_perp_spring} and \ref{fig:plot_pk_alpha_par_spring}. This difference allows us to easily distinguish between the parameters during inference.

Figures~\ref{fig:plot_pk_alpha_perp_spring} and \ref{fig:plot_pk_alpha_par_spring} show the same comparison for varying $\alpha_{\perp}$ and $\alpha_{\parallel}$, respectively. The corresponding figures for the Fall field are shown in Appendix~\ref{appendix:pk_fall}.
Higher $\alpha_{\perp}$ increases the overall amplitude of the monopole and quadrupole power spectra at all $k$. However, for $\alpha_{\parallel}$, the trend differs by multipole: the monopole power spectrum increases, while the quadrupole power spectrum decreases at all $k$. Their dependence on $k$ is also different for $\alpha_\perp$ and $\alpha_\parallel$.

Because the three parameters affect the power spectrum multipoles in distinct ways, we can determine the optimal values of $\alpha_v$ and the AP parameters that provide an excellent description of the data at all wavenumbers up to $k_\mathrm{max}=0.7\,h\,\mathrm{Mpc}^{-1}$, as discussed in Section~\ref{sec:interpretation}. 
Also, an excellent quality of fit supports our choice of a lognormal SHOD model.

\section{Interpretation of the Monopole and Quadrupole Power Spectra}
\label{sec:interpretation}

In this section, we construct the likelihood function using the monopole and quadrupole power spectra of the observed data and mocks to estimate the best-fit parameters. The best-fit $\alpha_v$ values are obtained through a profile-likelihood analysis~\citep{Herold_2025}, combining the Sellentin-Heavens likelihood~\citep{Sellentin_2016} with parabolic fitting. We then report the corresponding best-fit linear growth index, $\gamma$,  derived from the $\alpha_v$ parabolic fits for different combinations of the luminosity bins.

\begin{figure*}
    \centering 
    \includegraphics[width=0.95\textwidth]{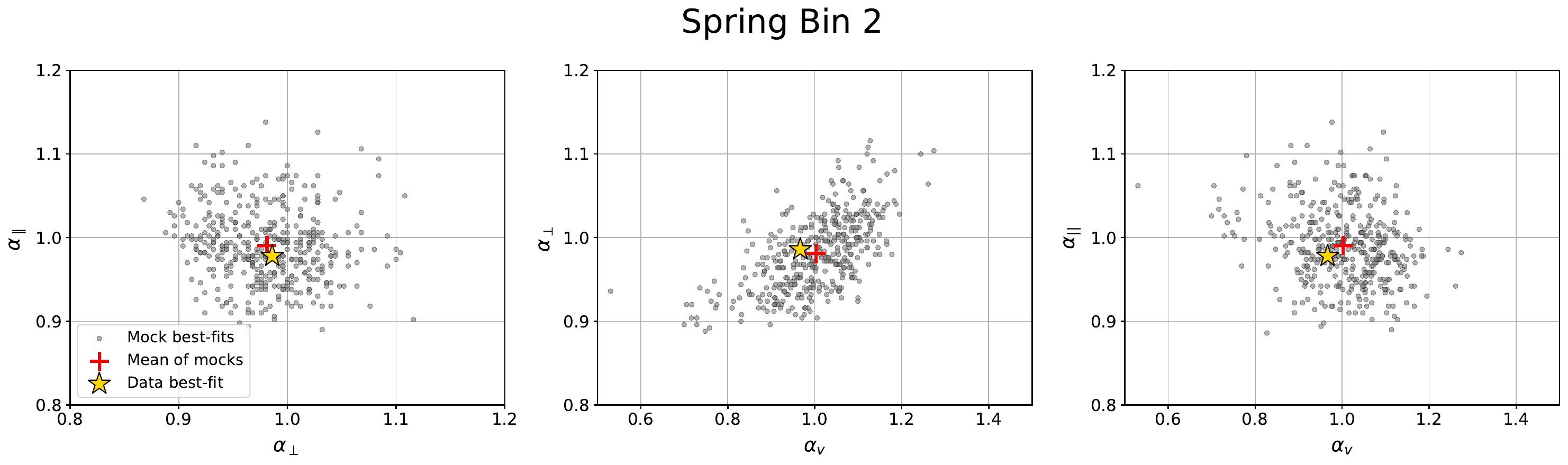}
    \includegraphics[width=0.95\textwidth]{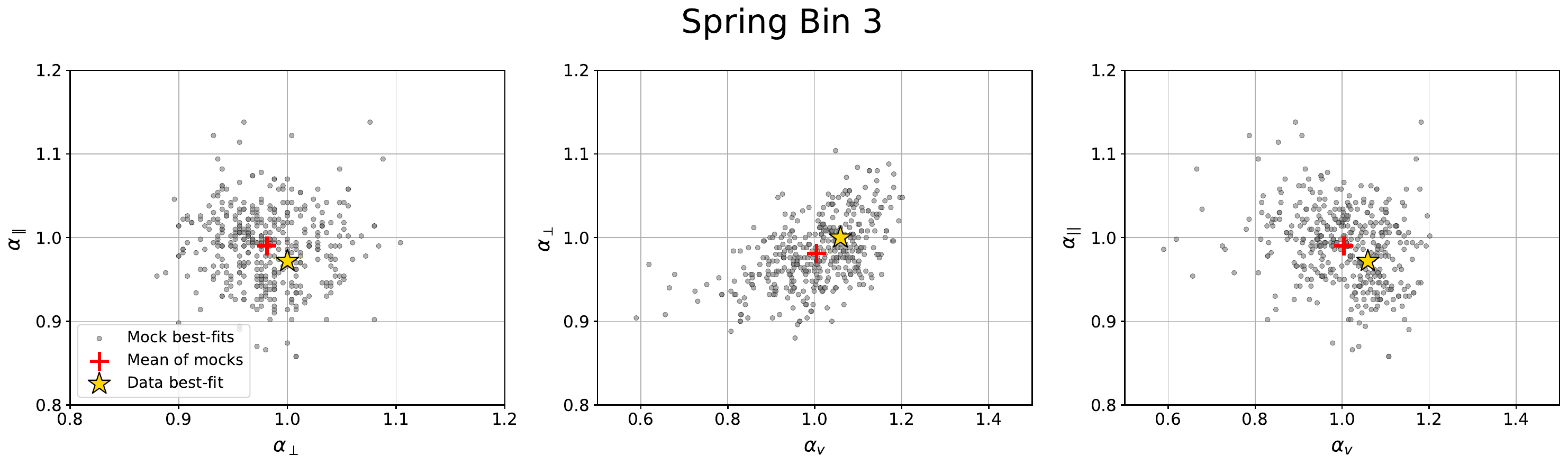}
    \includegraphics[width=0.95\textwidth]{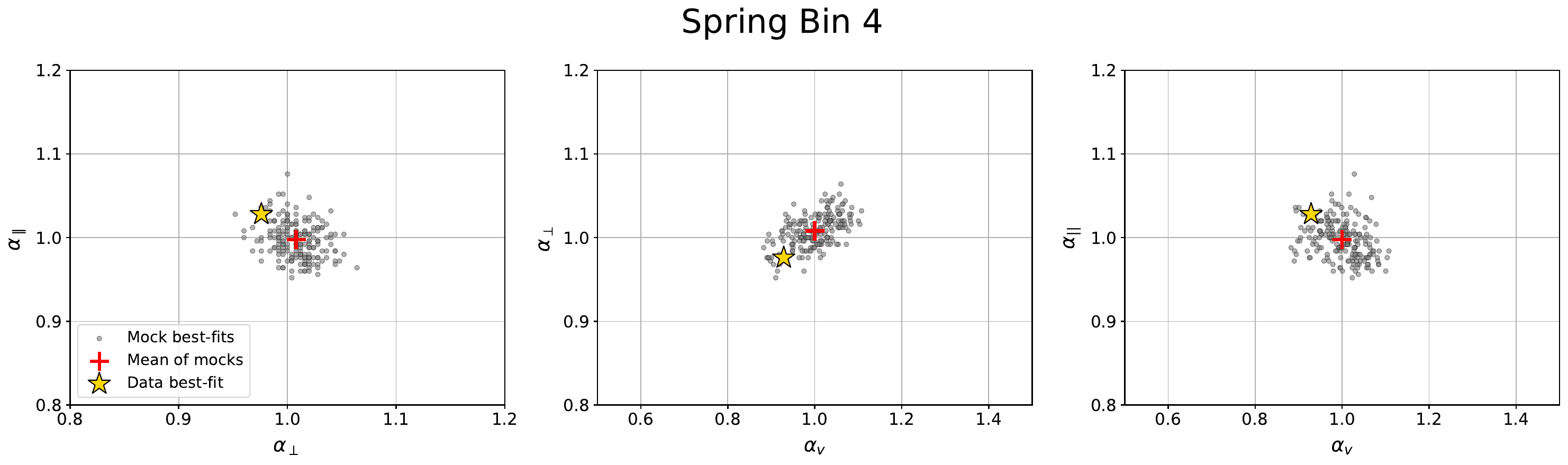}
    \caption{Distribution of the best-fit $\boldsymbol{\alpha}$ values, shown for different parameter pairs ($\alpha_{\perp}$-$\alpha_{\parallel}$, $\alpha_v$-$\alpha_{\perp}$, and $\alpha_v$-$\alpha_{\parallel}$), derived from the data (yellow star) and the mocks (gray dots) for Bins 2 (top), 3 (middle), and 4 (bottom) in the Spring field. The true $\boldsymbol{\alpha}$ for the mocks is set to unity, and the mean best-fit $\bar{\boldsymbol{\alpha}}\simeq \mathbf{1}$ (red cross) is well recovered.}
    \label{fig:plot_mock_recovery_spring}
\end{figure*}

\subsection{Likelihood Analysis}
\label{subsec:likelihood}
To estimate the best-fit parameters, we first combine the monopole and quadrupole power spectra into a single data vector, $\mathbf{D}$.
For the observed data with $N_k$ wavenumber bins $k_1,\dots,k_{N_k}$, the data vector is given by

\begin{equation}
    \mathbf{D}_{\rm data} =
    \begin{pmatrix}
        P_0^{\rm data}(k_1) \\
        \vdots \\
        P_0^{\rm data}(k_{N_k}) \\
        P_2^{\rm data}(k_1) \\
        \vdots \\
        P_2^{\rm data}(k_{N_k})
    \end{pmatrix},
    \qquad
\label{eq:data_vector}
\end{equation}
where $P_0(k)$ and $P_2(k)$ are the monopole and quadrupole power spectra, respectively.
The same construction is applied to each of the $N_{\rm mock}$ mock realizations,
$n = 1,\dots,N_{\rm mock}$, giving

\begin{equation}
    \mathbf{D}_{\rm mock}^{(n)}(\boldsymbol{\alpha}) =
    \begin{pmatrix}
        P_0^{(n)}(k_1,\boldsymbol{\alpha}) \\
        \vdots \\
        P_0^{(n)}(k_{N_k},\boldsymbol{\alpha}) \\
        P_2^{(n)}(k_1,\boldsymbol{\alpha}) \\
        \vdots \\
        P_2^{(n)}(k_{N_k},\boldsymbol{\alpha})
    \end{pmatrix} \, ,
\label{eq:mock_vector}
\end{equation}
where $\boldsymbol{\alpha} \equiv (\alpha_v, \alpha_{\perp}, \alpha_{\parallel})$ for brevity.

We exclude $M_0$ from inference because its impact on the monopole and quadrupole power spectra is not degenerate with that of $\boldsymbol{\alpha}$. Figure~7 of Paper I shows that, unlike $\boldsymbol{\alpha}$, $M_0$ hardly affects the quadrupole power spectra and only changes the overall amplitude of the monopole power spectra. Since $M_0$ is much more precisely determined than $\boldsymbol{\alpha}$, varying all four parameters simultaneously 
does not significantly change the results on $\boldsymbol{\alpha}$. Therefore, we fix $M_0$ to the best-fit values found in Paper I.

We compute $\chi^2(\boldsymbol{\alpha})$ across a three-dimensional grid of $\boldsymbol{\alpha}$ values,
\begin{equation}
\begin{split}
    \chi^2(\boldsymbol{\alpha}) = &\big[\mathbf{D}_\mathrm{data} - 
    \bar{\mathbf{D}}_{\rm mock}(\boldsymbol{\alpha})\big]^{\rm T}
    \mathbf{C}^{-1}(\boldsymbol{\alpha}) \\
    &\times\big[\mathbf{D}_\mathrm{data} - 
    \bar{\mathbf{D}}_{\rm mock}(\boldsymbol{\alpha})\big],
\end{split}
\label{eq:chi2_full}
\end{equation}
where $\mathbf{D}_\mathrm{data}$ is the data vector defined in Equation~\eqref{eq:data_vector}, and $\bar{\mathbf{D}}_{\rm mock}(\boldsymbol{\alpha})$ is the mean mock data vector, averaged over $N_\mathrm{mock}$ realizations at a given value of $\boldsymbol{\alpha}$.

A convergence test has demonstrated that $N_\mathrm{mock} = 400$ mock realizations are sufficient for the best-fit values to converge. However, since only 100 realizations are currently available for MU1, this limits our analysis to $N_\mathrm{mock} = 200$ mocks for Spring Bin 4, compared to 400 mocks for the remaining bins.
To assess the impact of this limitation, we compare the best-fit $\alpha_v$ values obtained from 200 and 400 mocks for Spring Bins 2 and 3, for which both covariance estimates are available. 
The best-fit values of $\alpha_v$ move by $3.3\%$ and $0.85\%$ for Spring Bins 2 and 3, respectively, when increasing the number of mock realizations from 200 to 400. These movements correspond only to $\simeq 0.3\sigma$ and $0.09\sigma$ compared to the statistical uncertainties. Thus, using 200 mocks for Spring Bin 4 is expected to introduce only a small shift that is well within its quoted statistical uncertainty.
We therefore utilize 200 mocks for Spring Bin 4 and 400 mocks for the remaining bins in this analysis.

The inverse covariance matrix, $\mathbf{C}^{-1}$, is obtained from inverting
\begin{equation}
\begin{split}
    \mathbf{C}(\boldsymbol{\alpha}) = \frac{1}{N_\mathrm{mock} - 1}\sum_{n=1}^{N_\mathrm{mock}}
    &\big[\mathbf{D}_\mathrm{mock}^{(n)}(\boldsymbol{\alpha}) - \bar{\mathbf{D}}_{\rm mock}(\boldsymbol{\alpha})\big] \\
    &\times\big[\mathbf{D}_\mathrm{mock}^{(n)}(\boldsymbol{\alpha}) - \bar{\mathbf{D}}_{\rm mock}(\boldsymbol{\alpha})\big]^{\rm T}.
\end{split}
\label{eq:cov}
\end{equation}

For each value of $\alpha_v$ on the discrete grid, we profile the $\chi^2$ statistic over the AP parameters by taking the minimum $\chi^2$ across the fully sampled grid of $(\alpha_{\perp},\alpha_{\parallel})$ at fixed $\alpha_v$~\citep{Herold_2025},
\begin{equation}
    \chi^2_p\left(\alpha_v\right)
    = \min_{\alpha_{\perp},\,\alpha_{\parallel}}\;
      \chi^2\left(\alpha_v,\,\alpha_{\perp},\,\alpha_{\parallel}\right).
    \label{eq:profile_chi2}
\end{equation}
We then construct the \citet{Sellentin_2016} likelihood from $\chi^2_p$,
\begin{equation}    \chi_\mathrm{SH,P}^2(\alpha_v)\equiv-2\ln\mathcal{L}_{\mathrm{SH,P}}(\alpha_v) = m \ln\left(1 + 
    \frac{\chi^2_p(\alpha_v)}{N_\mathrm{\rm mock} - 1}\right),
\label{eq:chisqSH}
\end{equation}
to account for the effect of the finite number of mock realizations used to estimate the covariance matrix. 
Rather than adopting the Jeffreys prior as described in \citet{Sellentin_2016}, we instead use the frequentist-matching prior~\citep{Percival_2022,Heavens_2026} because it yields slightly more conservative error bars, typically by a few percent~(see Paper I).
The factor $m$ multiplying the logarithm in Equation~\eqref{eq:chisqSH} reduces to $N_{\rm mock}$ under the Jeffreys prior, but becomes more complex under the frequentist-matching prior:
\begin{equation}
    m = 2+N_\mathrm{p}+\frac{N_\mathrm{mock}-1+B(2N_k-N_\mathrm{p})}{1+B(2N_k-N_\mathrm{p})},
\label{eq:percival_factor_m}
\end{equation}
with
\begin{equation}
    B = \frac{N_\mathrm{\rm mock}-2N_k-2}{(N_\mathrm{\rm mock}-2N_k-1)(N_\mathrm{\rm mock}-2N_k-4)},
\label{eq:percival_factor_B}
\end{equation}
where $N_\mathrm{p}$ is the number of parameters, which is $N_\mathrm{p}=1$, corresponding to $\alpha_v$ in this analysis.

We fit a parabola to $\chi_\mathrm{SH,P}^2(\alpha_v)$,
\begin{equation}
    \chi_\mathrm{SH,P}^2(\alpha_v) = a_{\mathrm{SH,P}}\,{\alpha_v}^2 + b_{\mathrm{SH,P}}\,{\alpha_v} + c_{\mathrm{SH,P}},
\label{eq:parabola_fit}
\end{equation}
which yields the best-fit value at the minimum, $\alpha_v^{\rm best} = -b_{\mathrm{SH,P}}/(2a_{\mathrm{SH,P}})$, with a $1\sigma$ uncertainty of $\sigma_{\alpha_v} = 1/\sqrt{a_{\mathrm{SH,P}}}$, derived from the condition $\Delta \chi^2_\mathrm{SH,P} \equiv \chi_\mathrm{SH,P}^2-\chi_\mathrm{SH,P,min}^2 = 1$.

Figure~\ref{fig:plot_mock_recovery_spring} validates our pipeline, showing how individual mocks recover the fiducial values of $\boldsymbol{\alpha}$.
The mocks in these figures were generated with $\boldsymbol{\alpha}$ set to unity, and the mean best-fit, $\bar{\boldsymbol{\alpha}}\simeq\mathbf{1}$, is well recovered within the distribution. The corresponding mock recovery test for the Fall field is shown in Figure~\ref{fig:plot_mock_recovery_fall} in Appendix~\ref{appendix:mock_recovery_fall}.

\subsection{Constraints on \texorpdfstring{$\alpha_v$}{alpha\_v} from the HETDEX data}
\begin{figure}
    \centering 
    \includegraphics[width=0.47\textwidth]{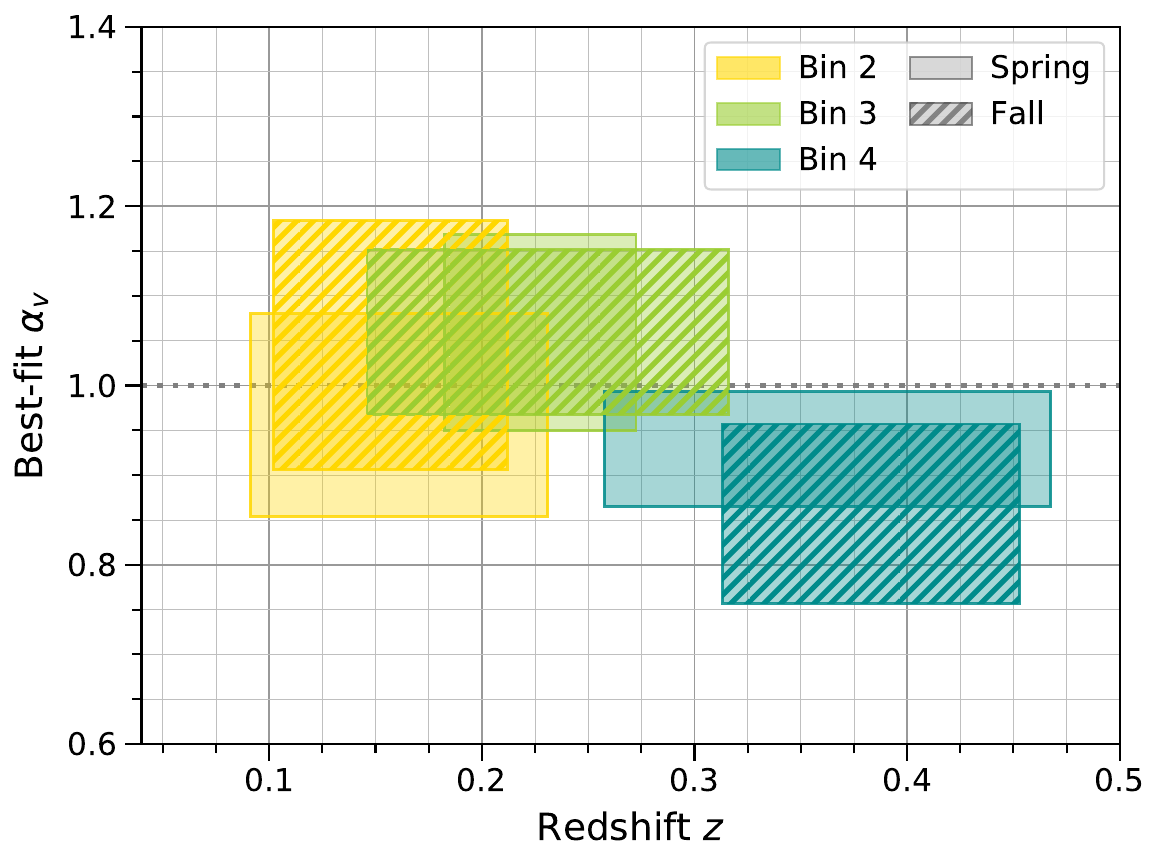}
    \caption{Best-fit $\alpha_v$ values as a function of redshift for Bins 2 (yellow), 3 (light green), and 4 (light blue), shown separately for the Spring (plain squares) and Fall (hatched squares) fields. The boxes indicate the redshift ranges of the bins, (see Table~\ref{tab:table_oii_sub}), horizontally, and the $1\sigma$ uncertainties of $\alpha_v$, vertically. The dotted horizontal line marks $\alpha_v = 1$.}
    \label{fig:plot_bestfit_alpha_v_redshift}
\end{figure}
\begin{figure*}
    \centering 
    \includegraphics[width=0.69\columnwidth]{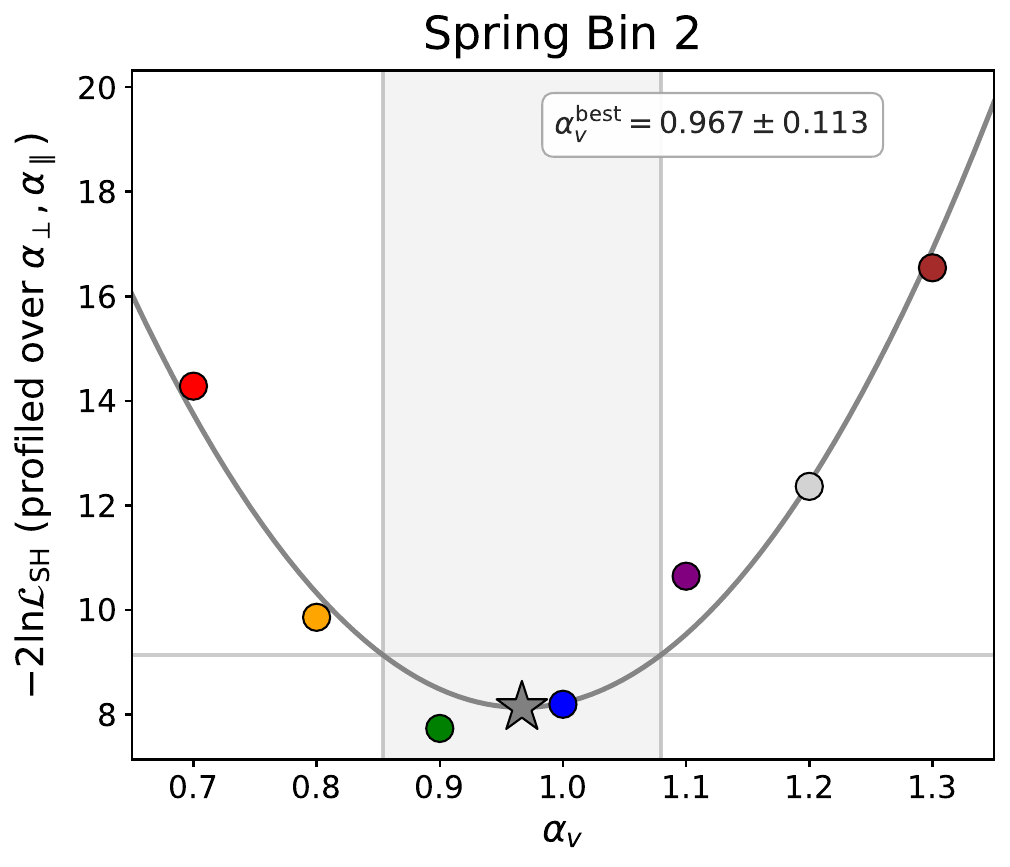}
    \includegraphics[width=0.69\columnwidth]{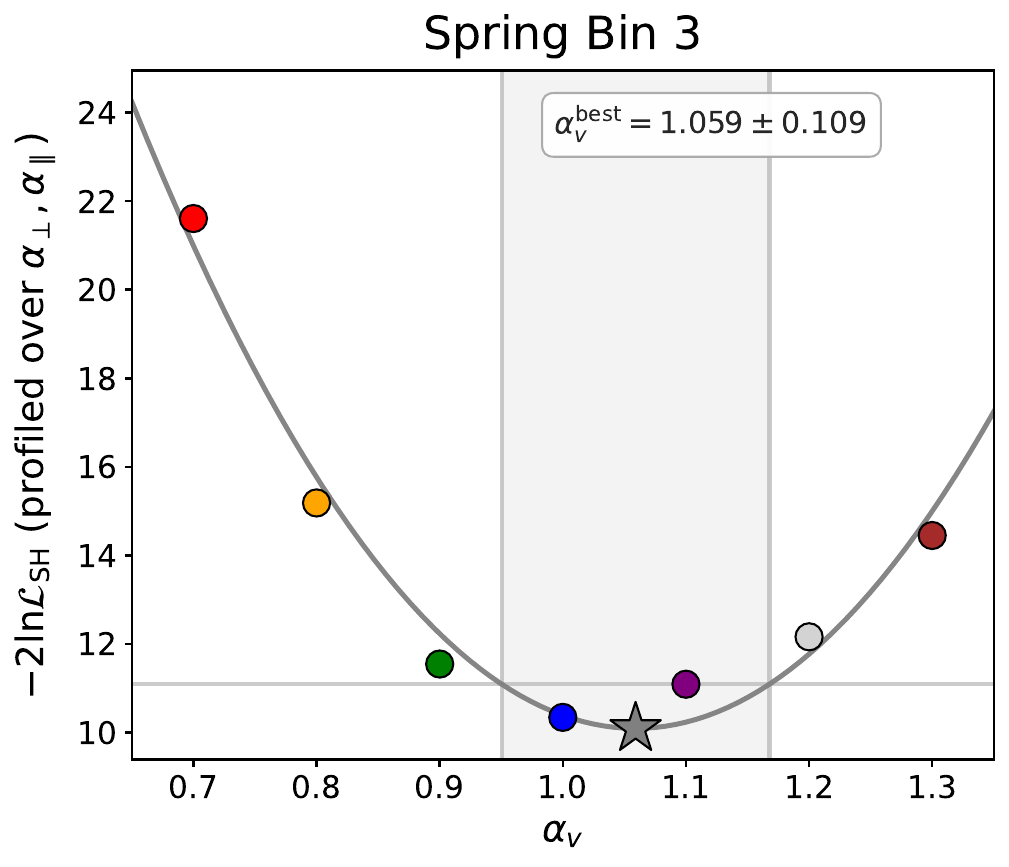}
    \includegraphics[width=0.69\columnwidth]{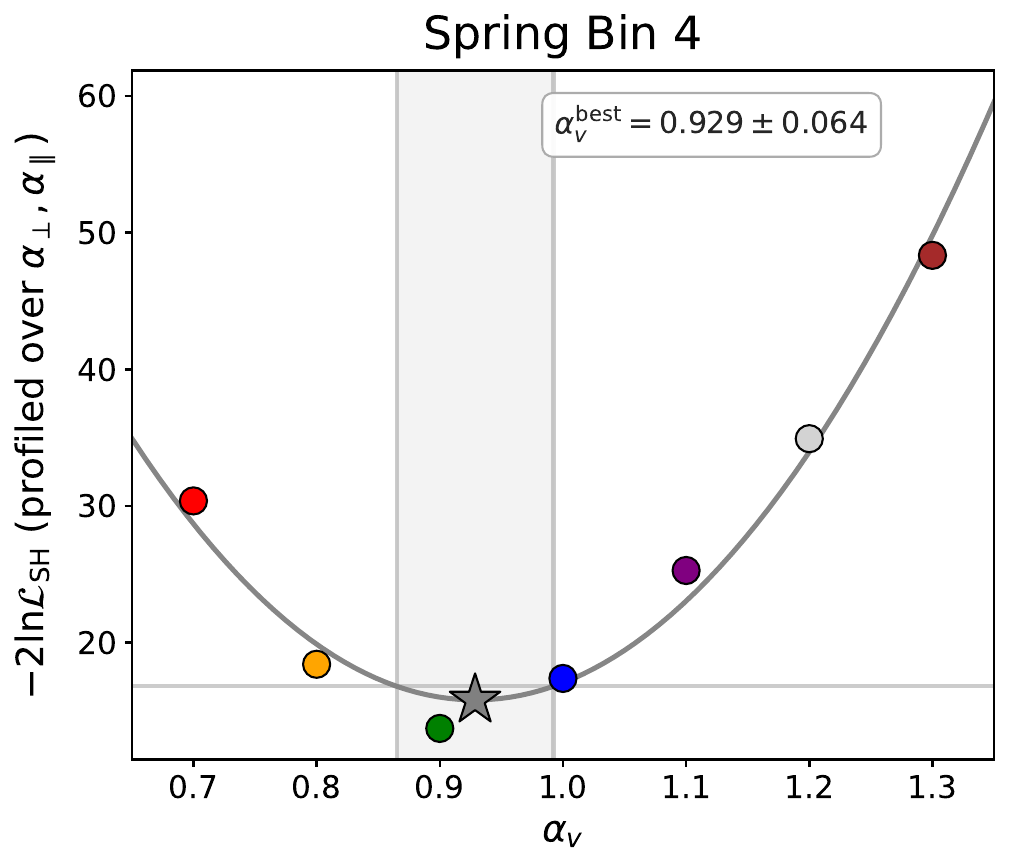}
    \includegraphics[width=0.69\columnwidth]{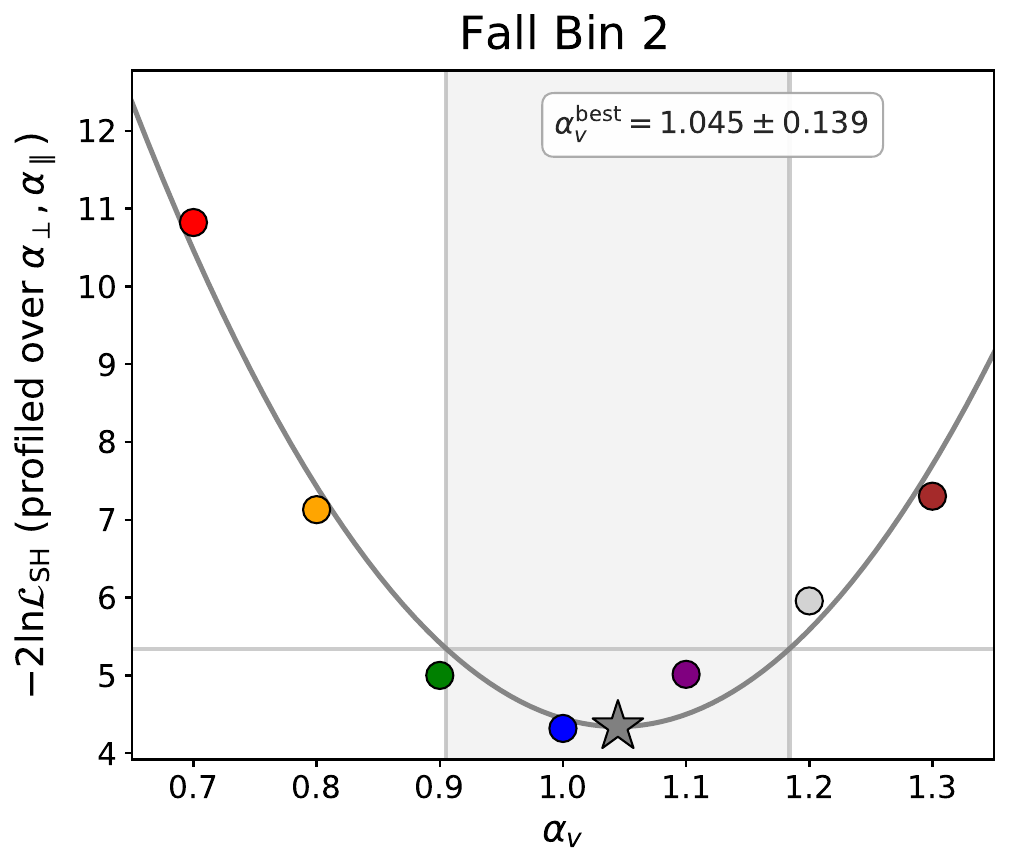}
    \includegraphics[width=0.69\columnwidth]{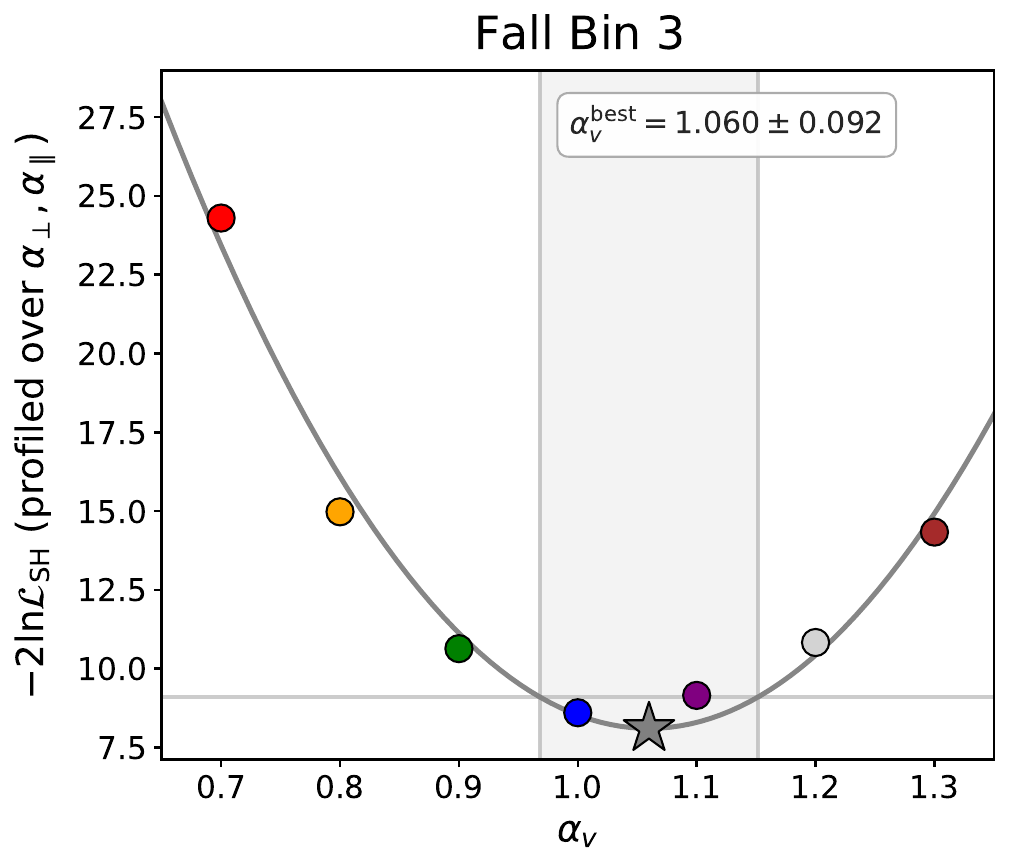}
    \includegraphics[width=0.69\columnwidth]{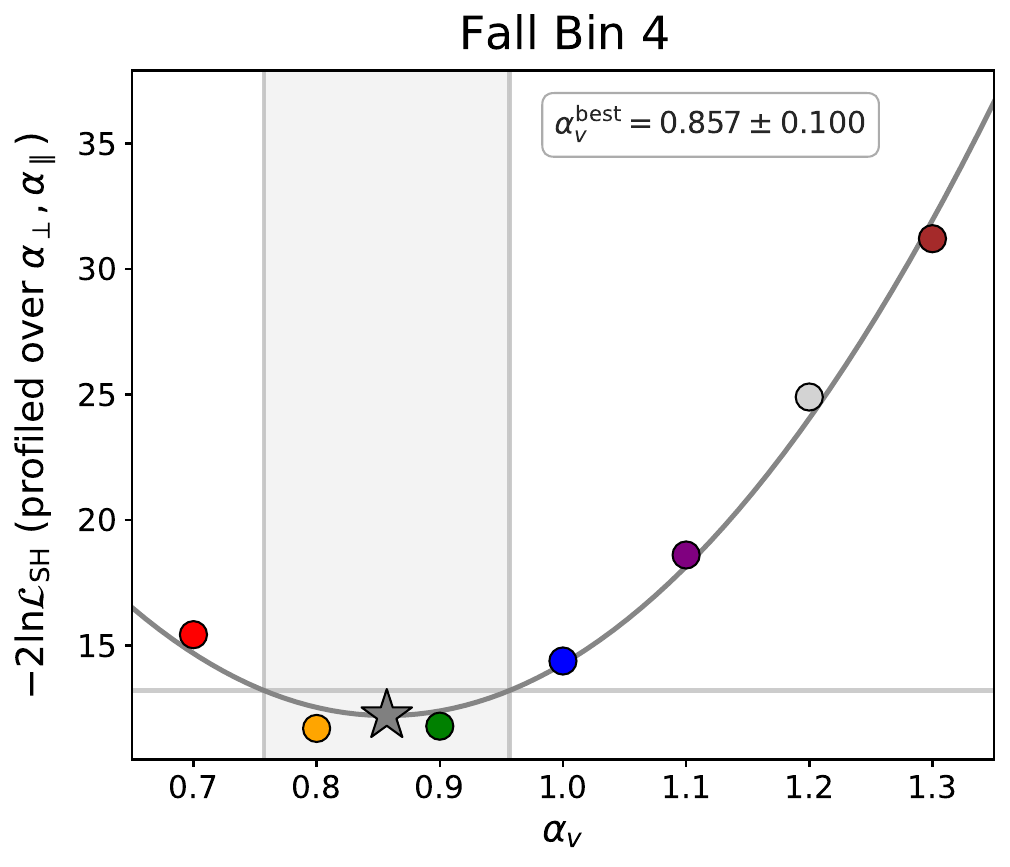}
    \caption{Parabolic fits to $\chi_\mathrm{SH,P}^2(\alpha_v)=-2\ln{\cal L}_\mathrm{SH,P}$ [Equation~\eqref{eq:chisqSH}], used to estimate the best-fit $\alpha_v$, for Bins 2 (left panels), 3 (middle panels), and 4 (right panels) in the Spring (top panels) and Fall (bottom panels) fields. The filled circles show $\chi_\mathrm{SH,P}^2(\alpha_v)$ evaluated at different $\alpha_v$ values, the stars mark the best-fit $\alpha_v$, and the horizontal and vertical lines indicate the $1\sigma$ uncertainty, defined by $\Delta \chi^2_\mathrm{SH,P} = 1$. The best-fit $\alpha_v$ values are given in the upper-right corner of each panel.}
    \label{fig:plot_chi2_interpolation}
\end{figure*}

Figure~\ref{fig:plot_bestfit_alpha_v_redshift} displays the constraints on $\alpha_v$ as a function of $z$, which are derived from parabolic fits to $\chi_\mathrm{SH,P}^2(\alpha_v)$ shown in Figure~\ref{fig:plot_chi2_interpolation}.

The stars in Figures~\ref{fig:plot_mock_recovery_spring} and \ref{fig:plot_mock_recovery_fall} indicate the best-fit values of $\boldsymbol{\alpha}$ derived from HETDEX and the gray dots show the distribution derived from the mocks with the fiducial values of $\boldsymbol{\alpha}=\mathbf{1}$. All values of $\boldsymbol{\alpha}$ derived from HETDEX fall comfortably within the distribution. Therefore, the monopole and quadrupole power spectra of the HETDEX [\ion{O}{2}] galaxies are consistent with the \textit{Planck} flat $\Lambda$CDM cosmology used for MU1.

\begin{figure*}
    \centering 
    \includegraphics[width=0.69\columnwidth]{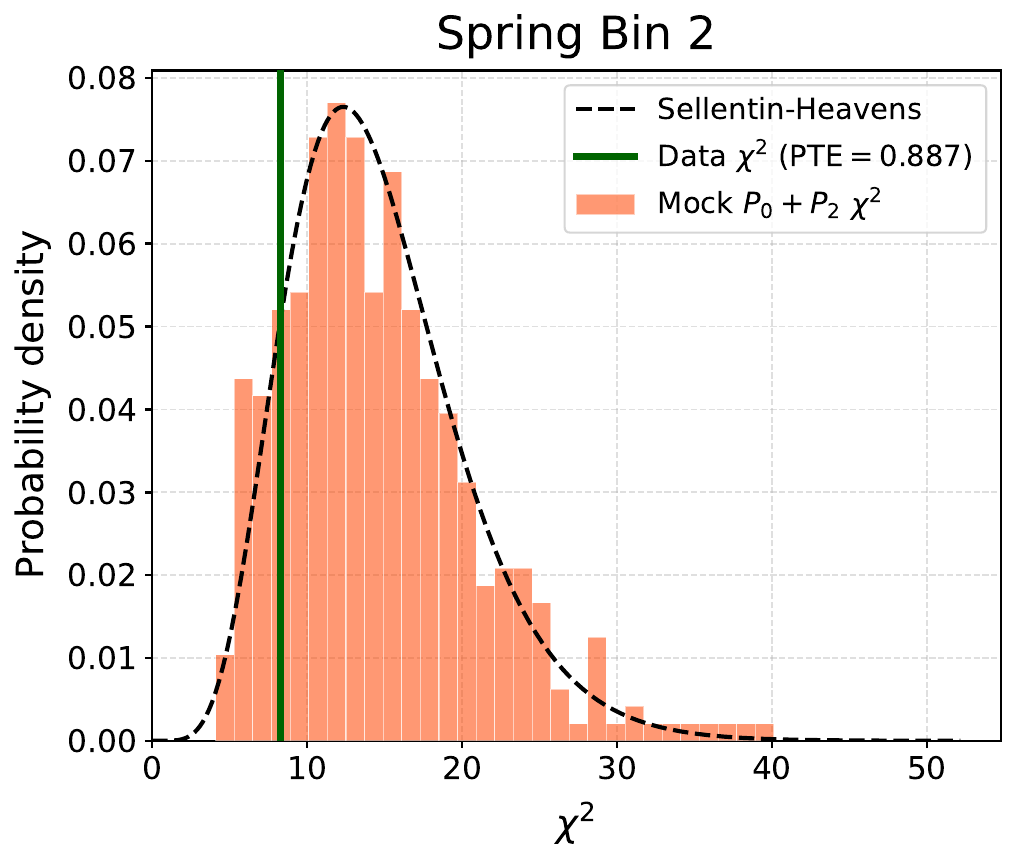}
    \includegraphics[width=0.69\columnwidth]{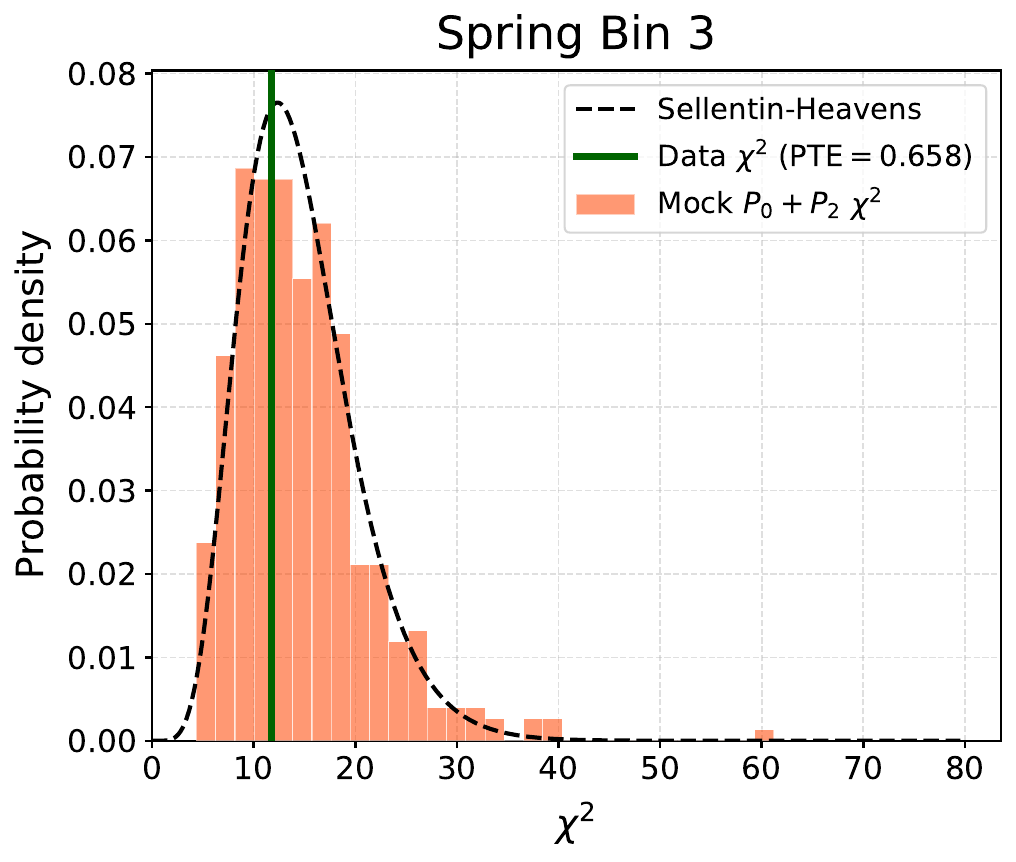}
    \includegraphics[width=0.69\columnwidth]{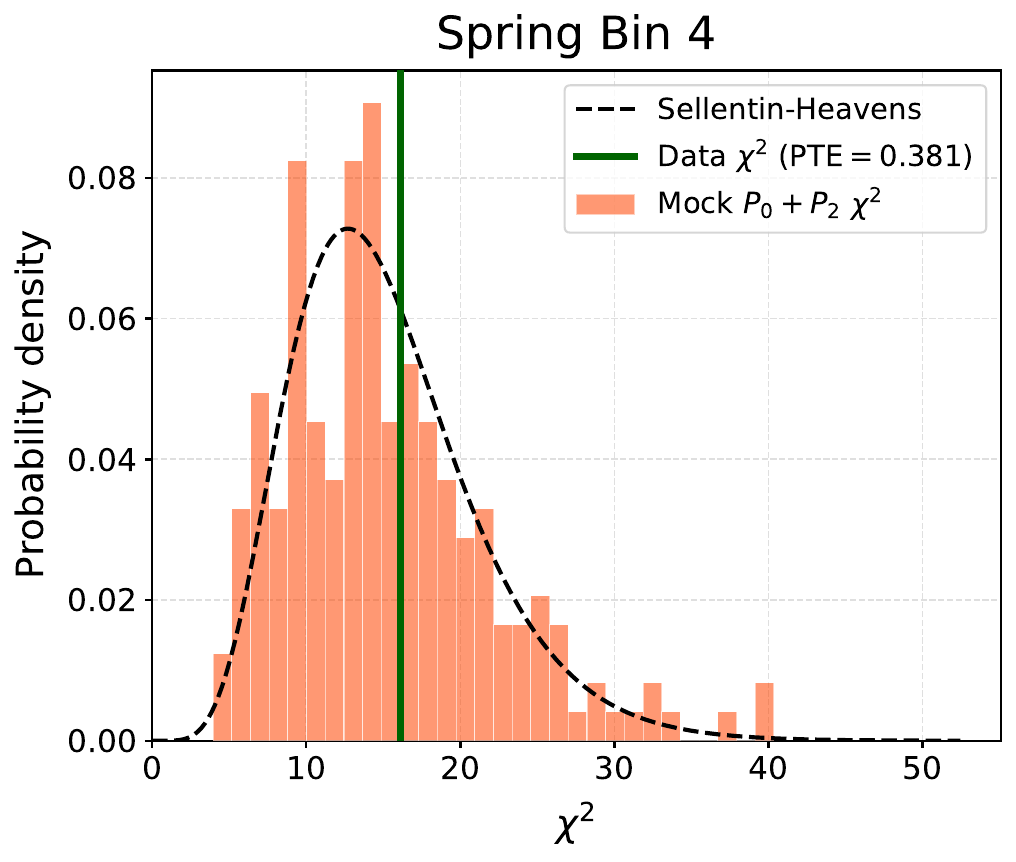}
    \includegraphics[width=0.69\columnwidth]{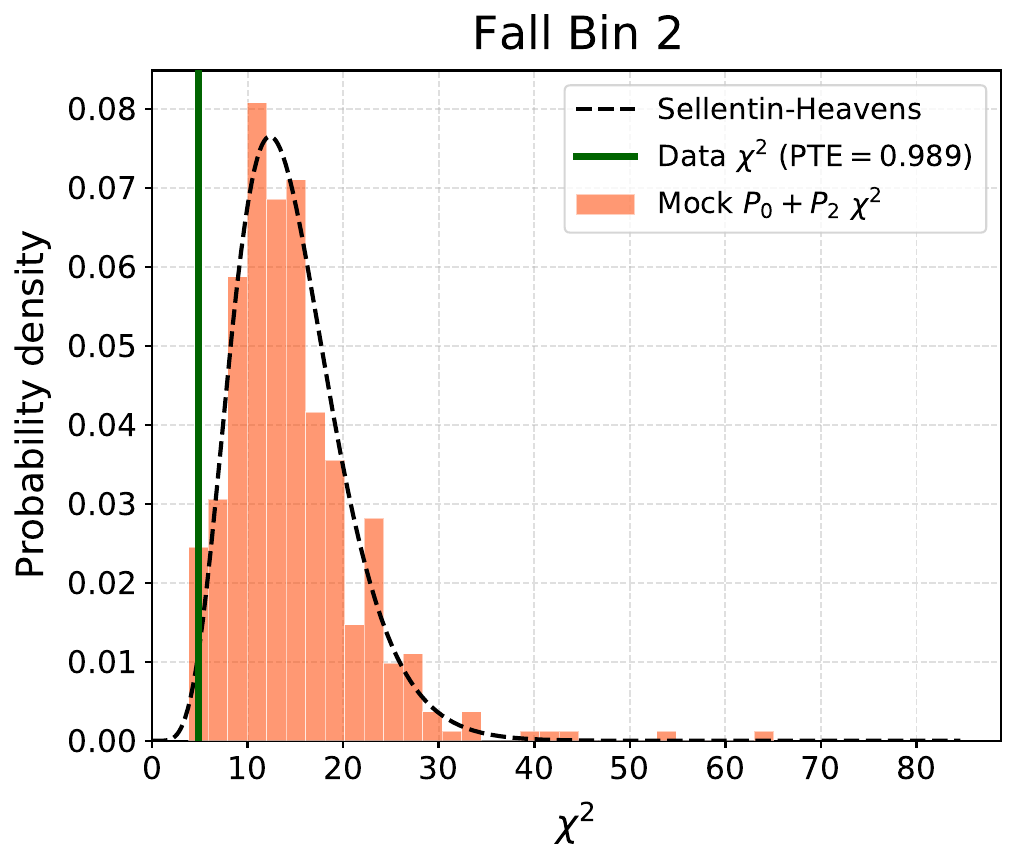}
    \includegraphics[width=0.69\columnwidth]{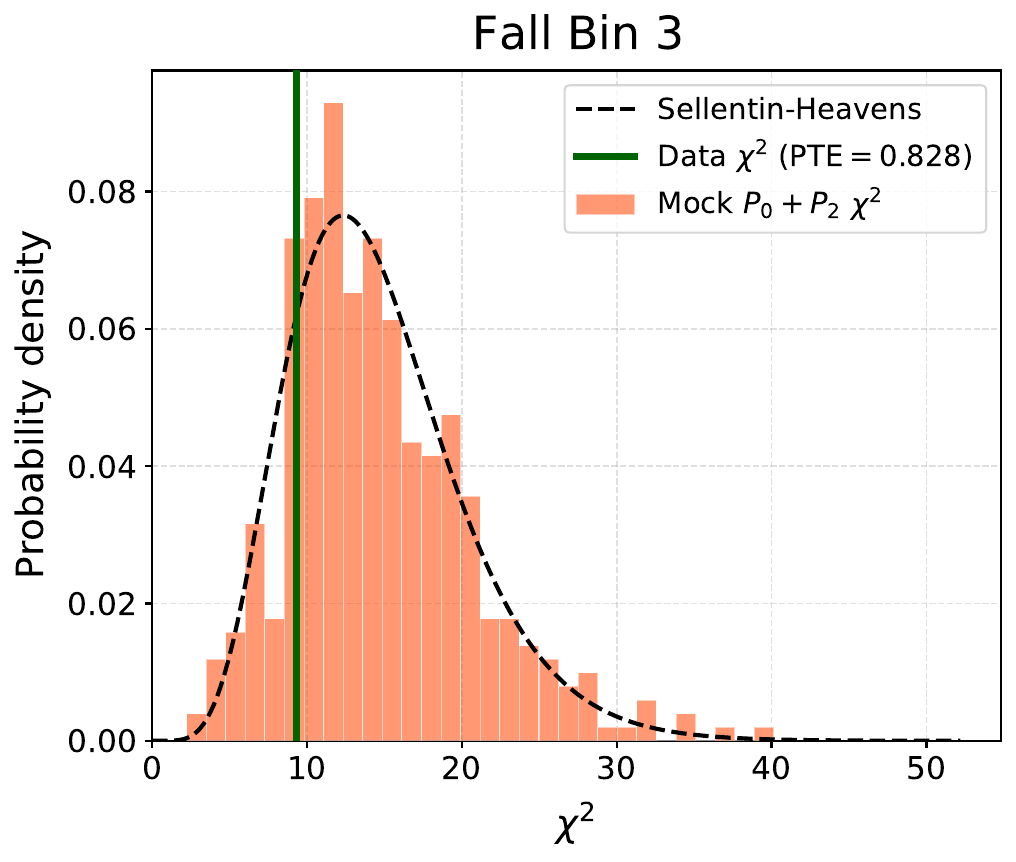}
    \includegraphics[width=0.69\columnwidth]{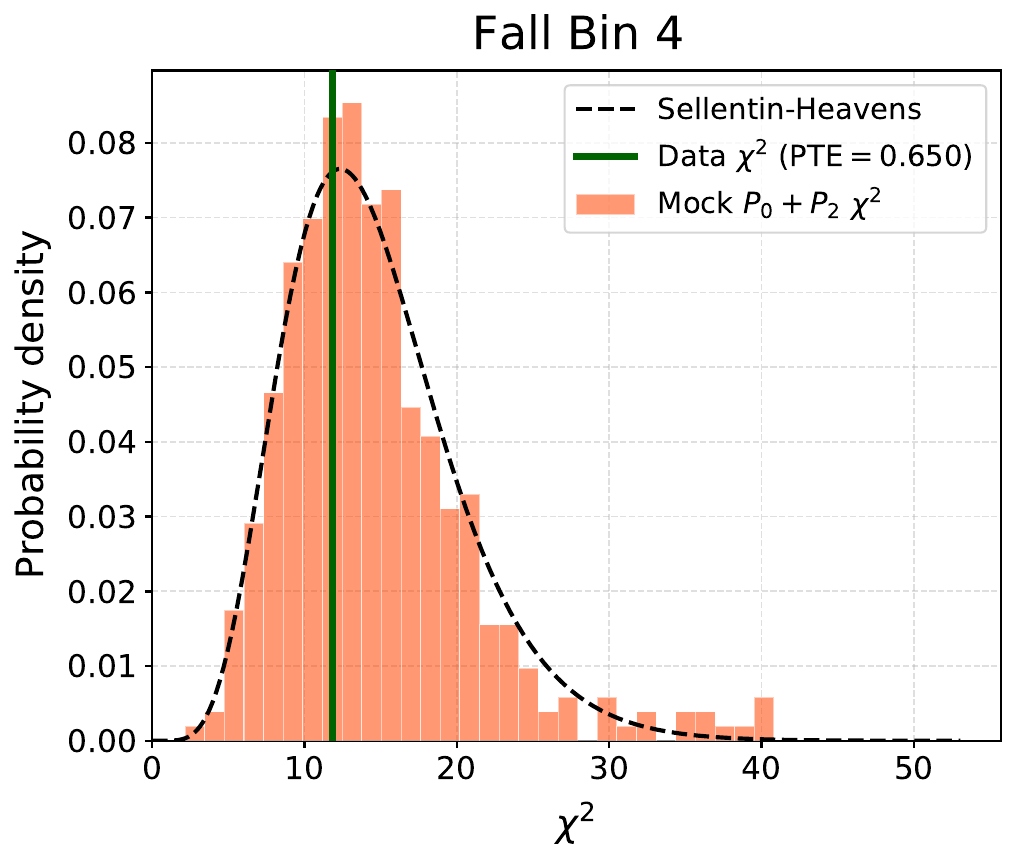}
    \caption{Distribution of $\chi^2$ [Equation~\eqref{eq:chi2_full}] for the joint monopole and quadrupole power spectra, computed from mock realizations at the best-fit value of $\boldsymbol{\alpha}$, shown separately for Bins 2 (left),  3 (middle), and 4 (right) in the Spring (top) and Fall (bottom) fields. In each panel, the histogram shows the $\chi^2$ values obtained from the mocks, the vertical line indicates the $\chi^2$ value measured from the data, and the dashed curve shows the corresponding Sellentin-Heavens PDF\null. The mock $\chi^2$ distributions agree well with theoretical expectations, and in every case, the data's $\chi^2$ falls comfortably within the bulk of the mock distribution, yielding acceptable PTE values.}
    \label{fig:plot_chi2_distribution}
\end{figure*}

To quantify the agreement between the data and the mocks with the best-fit values of $\boldsymbol{\alpha}$, we assess the goodness-of-fit, following Equation~(17) of \citet{Sellentin_2016}, in the same manner as described in Section 5.1 of Paper I.

Figure~\ref{fig:plot_chi2_distribution} presents the distribution of $\chi^2$ for the joint monopole and quadrupole power spectra. The mock $\chi^2$ values are computed using a leave-one-out (jackknife) approach, in which each best-fit mock realization serves as pseudo-data while the covariance matrix is estimated from the remaining realizations: 399 mocks for all bins except Spring Bin 4, for which 199 mocks are used.
The theoretical $\chi^2$ distribution under the Sellentin-Heavens framework, shown as dashed lines, follows a scaled $F$-distribution with $n_\mathrm{dof} = 14$ degrees of freedom, and $N_\mathrm{mock} = 400$ for all bins except Spring Bin 4, for which $N_\mathrm{mock} = 200$.
We compare the empirical $\chi^2$ distribution from the mock realizations against this theoretical prediction and check whether the data $\chi^2$ fall within the expected range. In all cases, the resulting probability-to-exceed (PTE) values are acceptable, confirming that the measured $\chi^2$ is fully consistent with expectations from the mock ensemble.

\begin{figure*}
    \centering 
    \includegraphics[width=0.69\columnwidth]{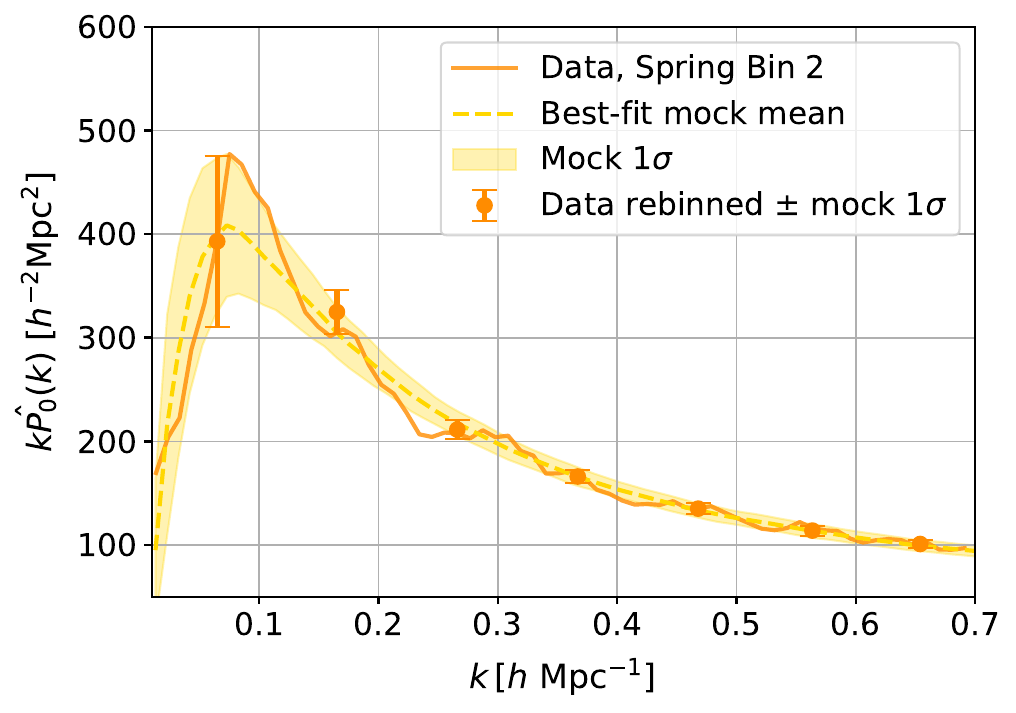}
    \includegraphics[width=0.69\columnwidth]{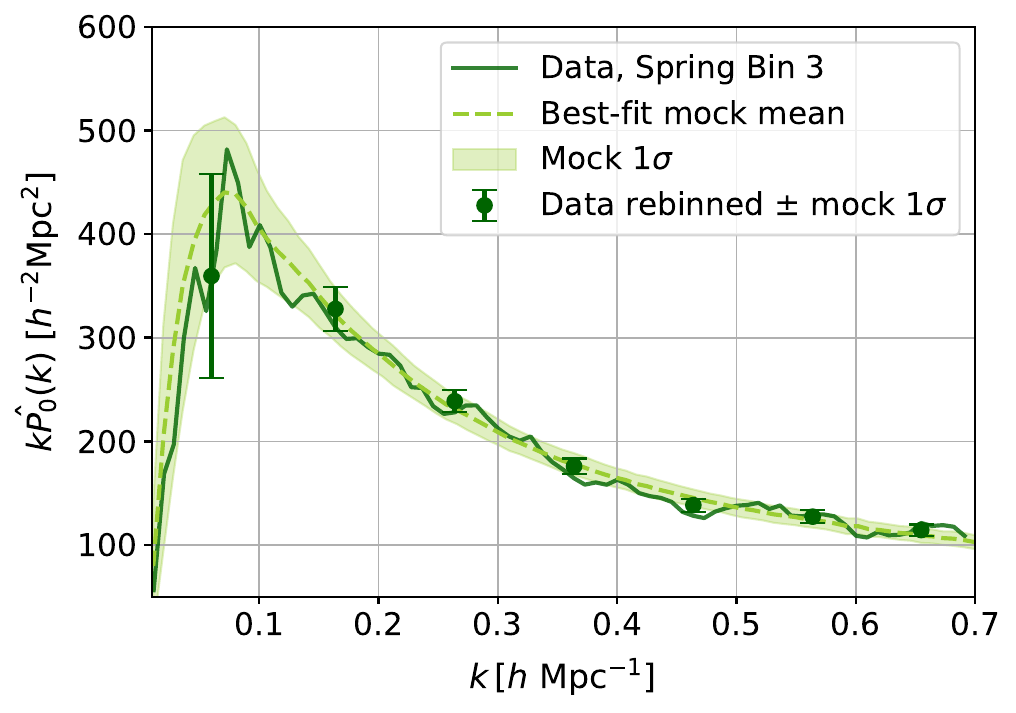}
    \includegraphics[width=0.69\columnwidth]{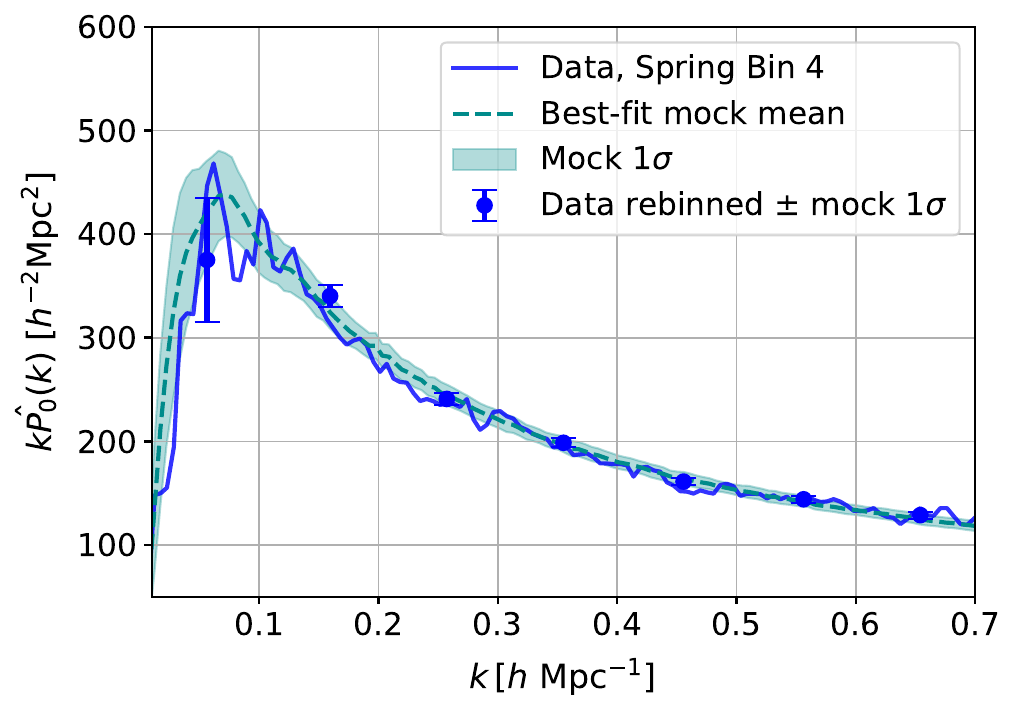}
    \includegraphics[width=0.695\columnwidth]{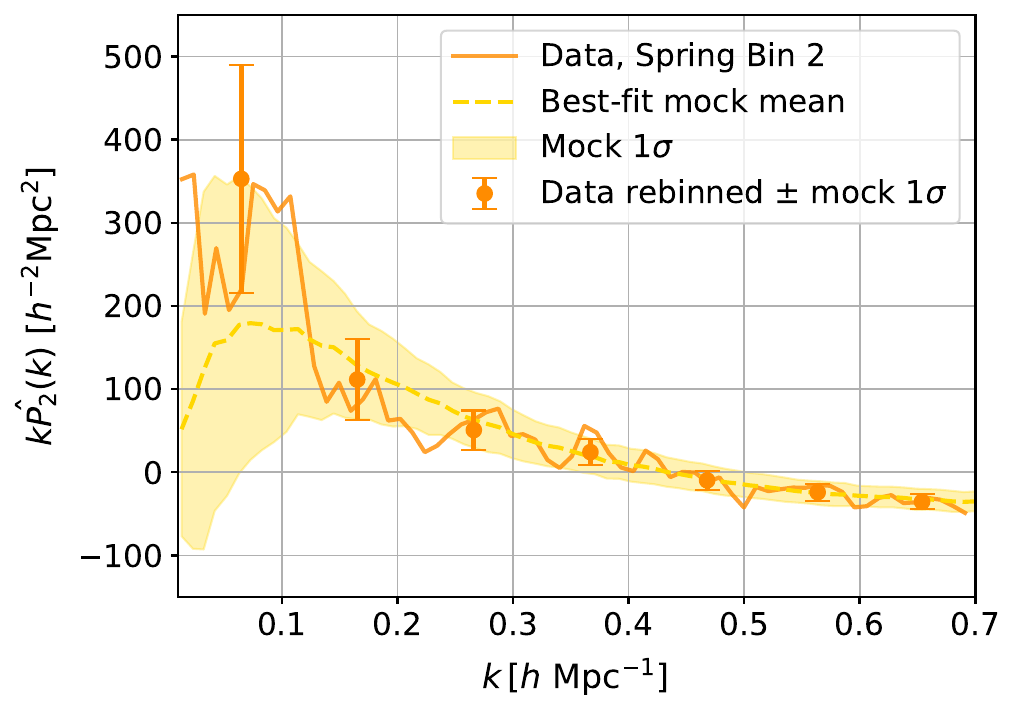}
    \includegraphics[width=0.695\columnwidth]{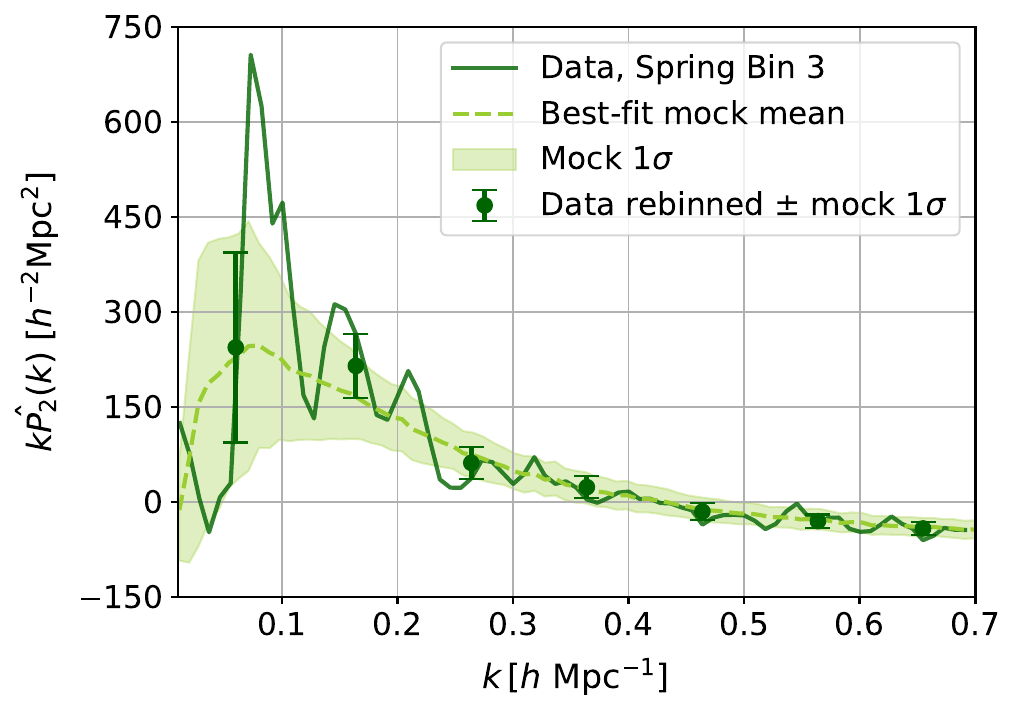}
    \includegraphics[width=0.695\columnwidth]{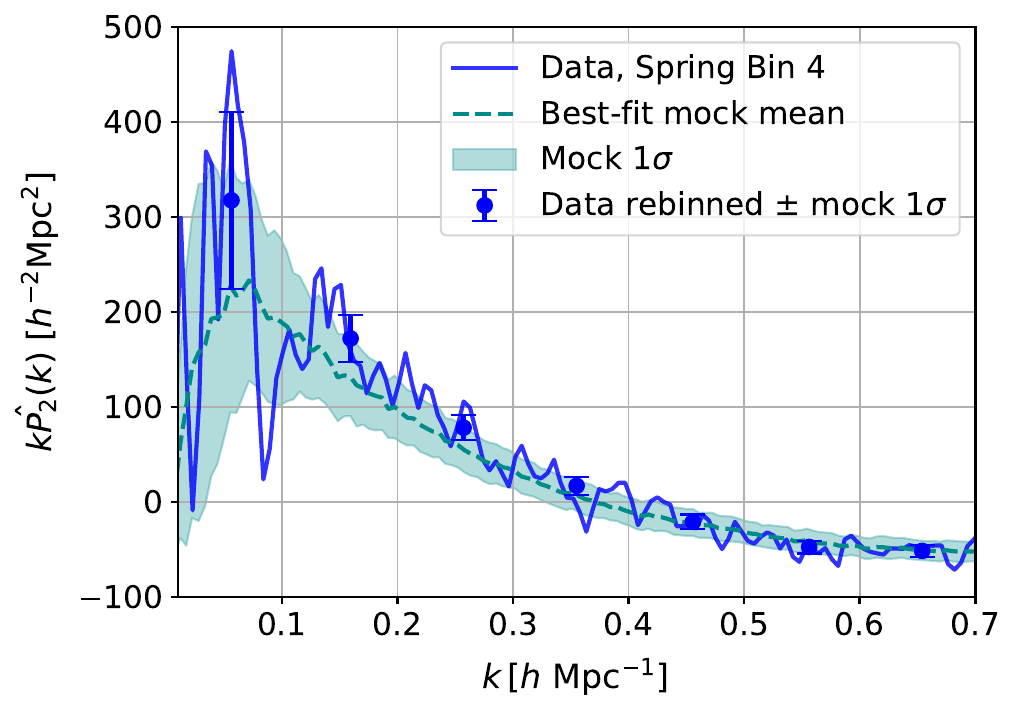}
    \caption{Monopole (upper panels) and quadrupole (lower panels) power spectrum measurements of the HETDEX [\ion{O}{2}] volume-limited samples for Bins 2 (left), 3 (middle), and 4 (right) in the Spring field, compared against MU1 mock measurements evaluated at the estimated best-fit $\boldsymbol{\alpha}$. The solid lines show the HETDEX data at a finer resolution of $\Delta k\simeq0.01~h~\mathrm{Mpc}^{-1}$, while the points with error bars show the same data rebinned to $\Delta k=0.1~h~\mathrm{Mpc}^{-1}$. The error bars represent $1\sigma$ uncertainties estimated from 400 mock realizations for Bins 2 and 3, and 200 for Bin 4. The dashed lines indicate the mean of the best-fit mocks, with the shaded regions denoting their $1\sigma$ uncertainties at the finer resolution. The data and mocks agree remarkably well across all wavenumbers up to $k_\mathrm{max}= 0.7\,h\,\mathrm{Mpc}^{-1}$.}
    \label{fig:plot_pk_best_spring}
\end{figure*}

\begin{figure*}
    \centering 
    \includegraphics[width=0.69\columnwidth]{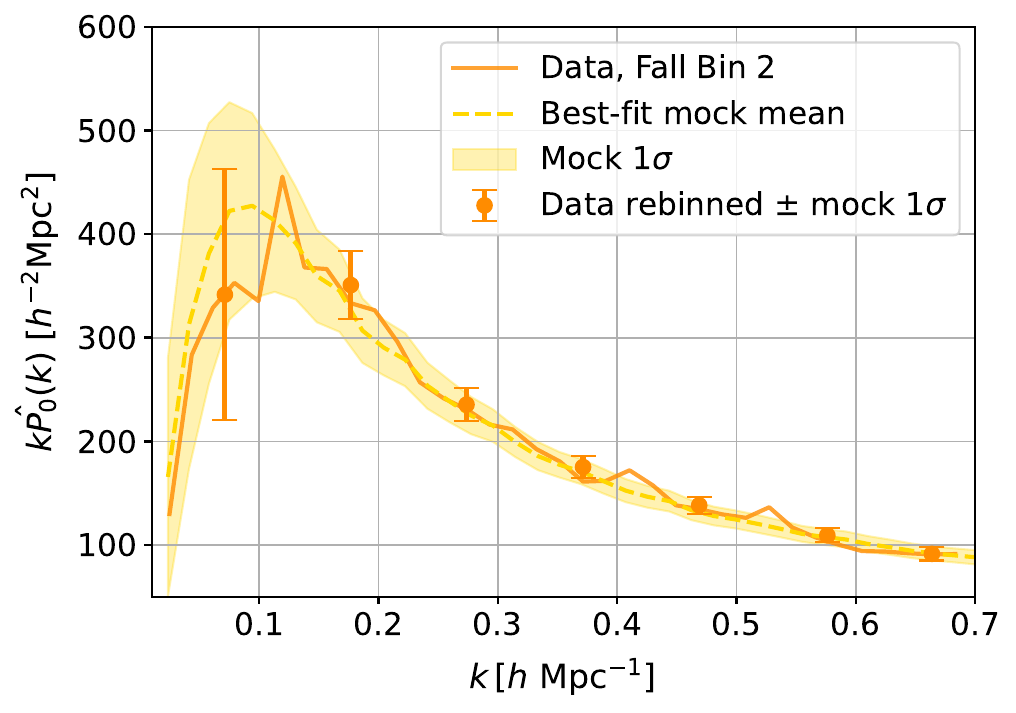}
    \includegraphics[width=0.69\columnwidth]{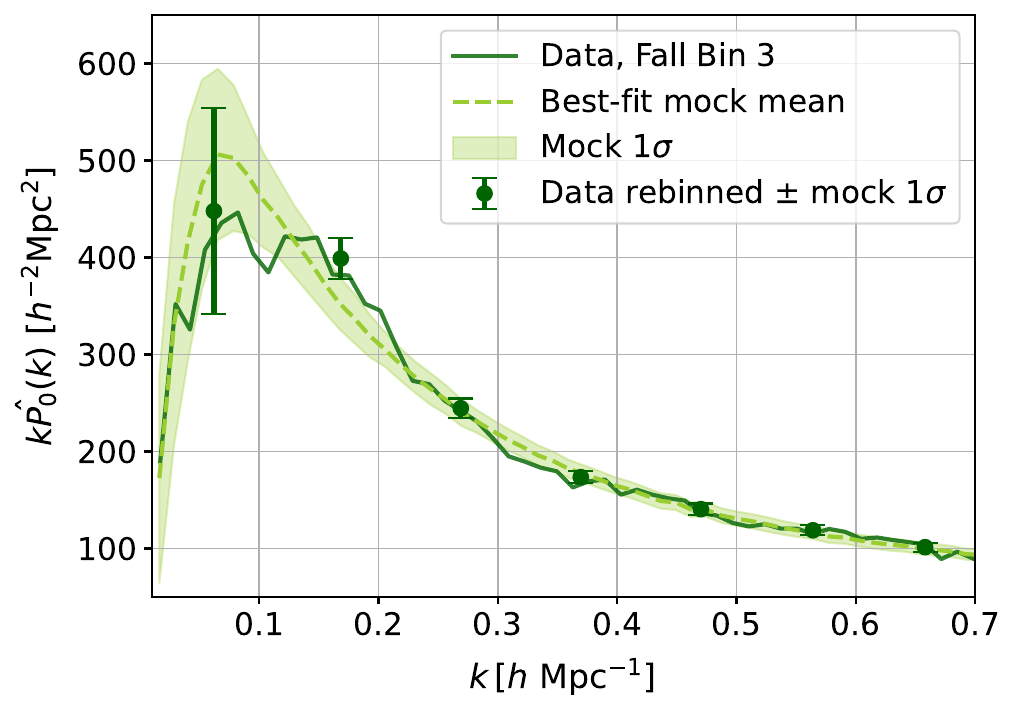}
    \includegraphics[width=0.69\columnwidth]{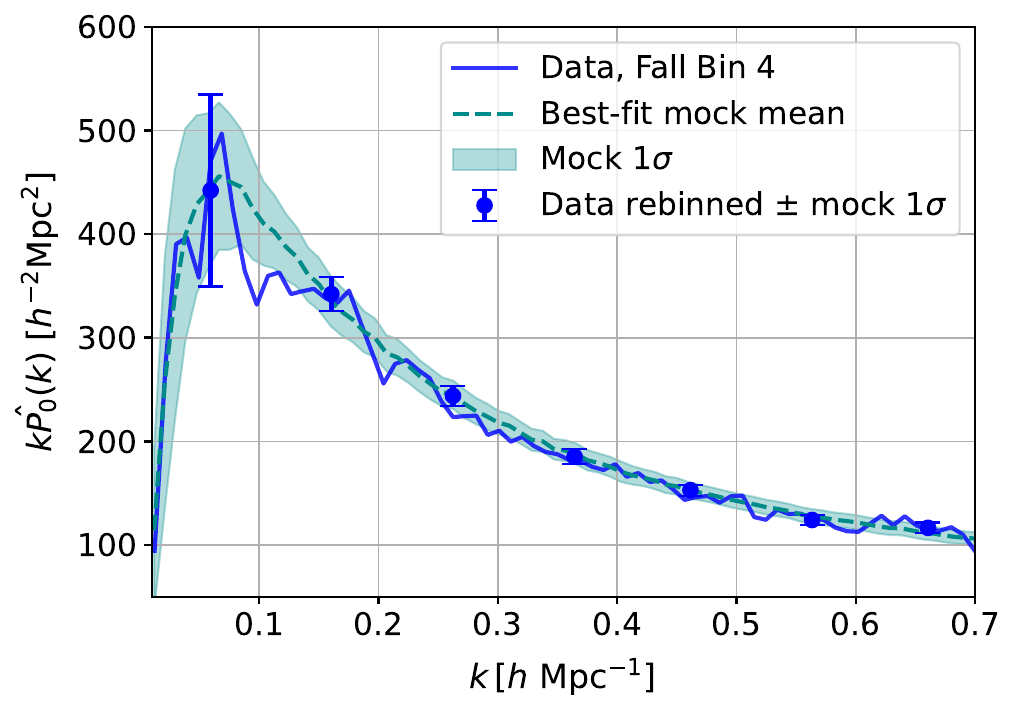}
    \includegraphics[width=0.695\columnwidth]{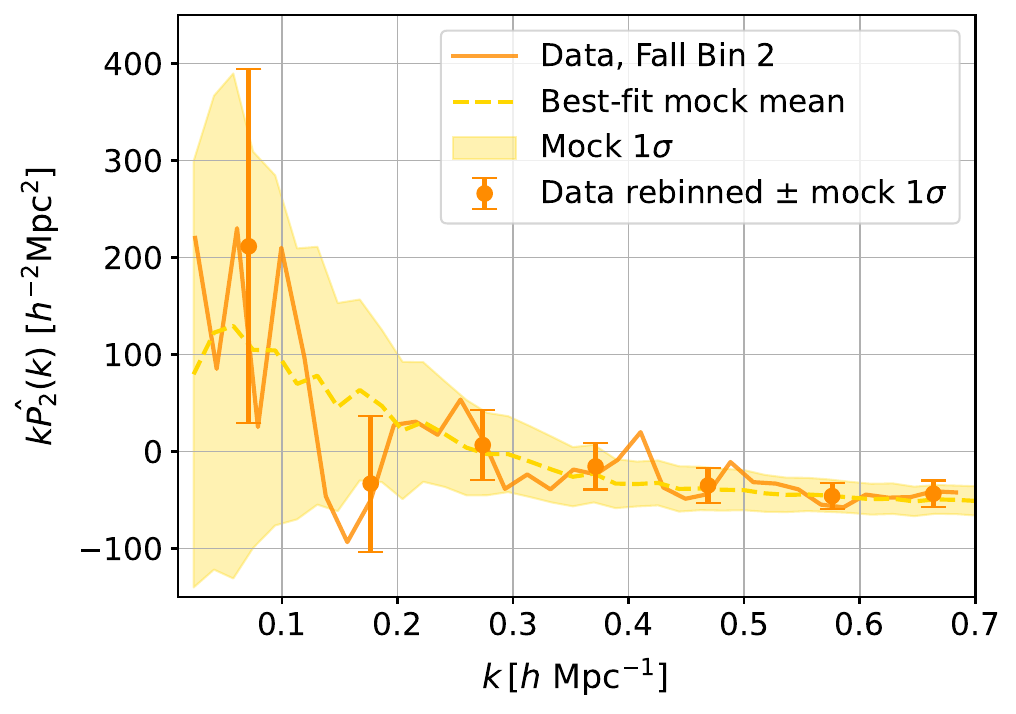}
    \includegraphics[width=0.695\columnwidth]{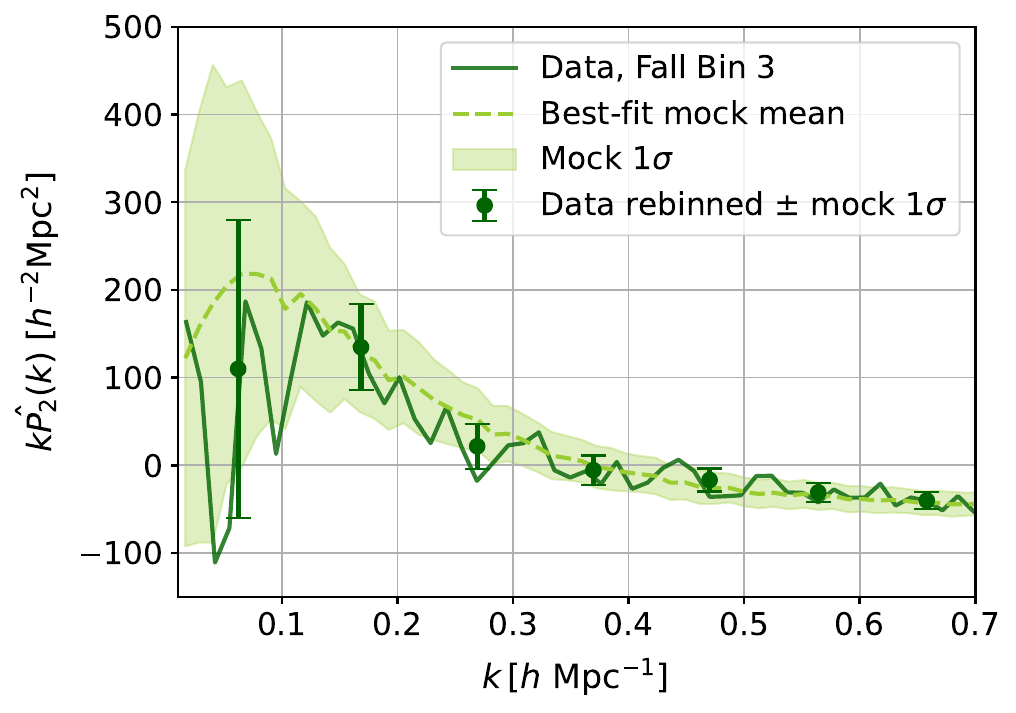}
    \includegraphics[width=0.695\columnwidth]{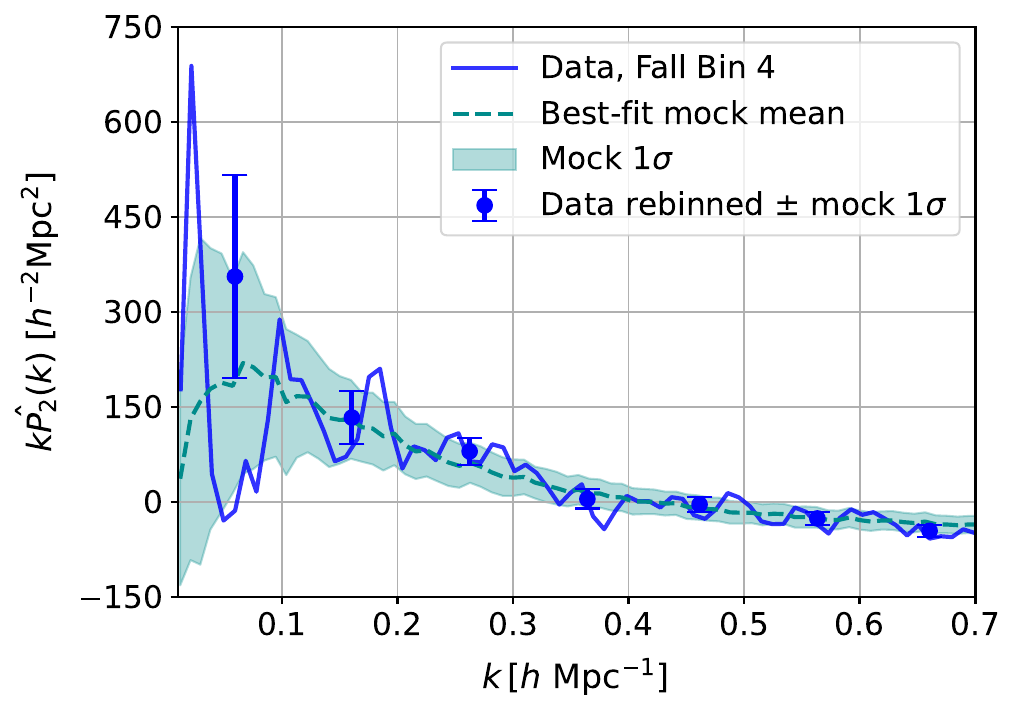}
    \caption{Same format as Figure~\ref{fig:plot_pk_best_spring}, but for the Fall field. As with the Spring field, the data and mocks agree remarkably well across all wavenumbers up to $k_\mathrm{max}= 0.7\,h\,\mathrm{Mpc}^{-1}$.}
    \label{fig:plot_pk_best_fall}
\end{figure*}

Finally, Figures~\ref{fig:plot_pk_best_spring} and~\ref{fig:plot_pk_best_fall} visualize the agreement between the data and the mocks by comparing the monopole and quadrupole power spectra of the HETDEX [\ion{O}{2}] volume-limited samples in the Spring and Fall fields, respectively, to those of the MU1 mocks with the best-fit $\boldsymbol{\alpha}$. They agree remarkably well for both the monopole and quadrupole power spectra, across all wavenumbers up to $k_\mathrm{max}= 0.7\,h\,\mathrm{Mpc}^{-1}$. We present the corresponding dipole and hexadecapole comparisons between the data and the mocks in Appendix~\ref{appendix:dipole_hexadecapole}.

\subsection{Constraints on the linear growth index}
\label{subsec:results}

\begin{figure}
    \centering 
    \includegraphics[width=0.95\columnwidth]{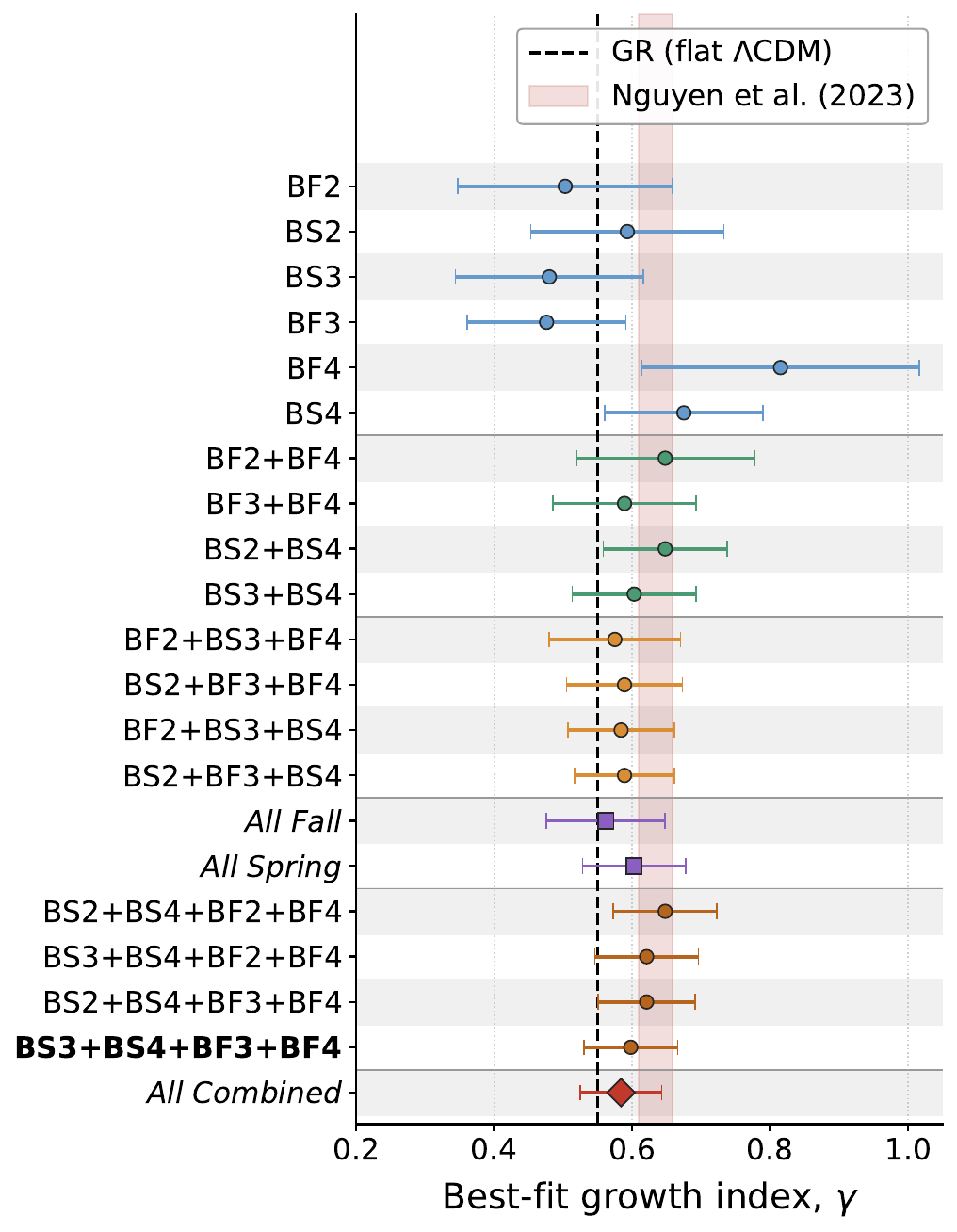}
    \caption{Best-fit $\gamma$ values estimated from individual bins (blue), pairs of bins (green), triples from different fields (orange), triples from the same field (purple), quadruples (brown), and the combination of all bins (red), each with $1\sigma$ error bars. For simplicity, bin combinations are abbreviated; for example, ``BS2'' denotes Spring Bin 2. Within each category, combinations are ordered by their combined comoving volume. The baseline result is $\gamma=0.60\pm 0.07$ from the largest volume, ``BS3+BS4+BF3+BF4,'' which is highlighted in bold. The combinations in italics use overlapping redshift bins and are shown only for illustrative purposes. The vertical dashed line indicates the value of $\gamma$ predicted by GR in a flat $\Lambda$CDM cosmology, while the shaded region indicates the best-fit $\gamma$ range reported by \citet{Nguyen_2023}. Most bin combinations are consistent with both predictions within their uncertainties.}
    \label{fig:plot_bestfit_gamma}
\end{figure}

\begin{figure}
    \centering 
    \includegraphics[width=0.95\columnwidth]{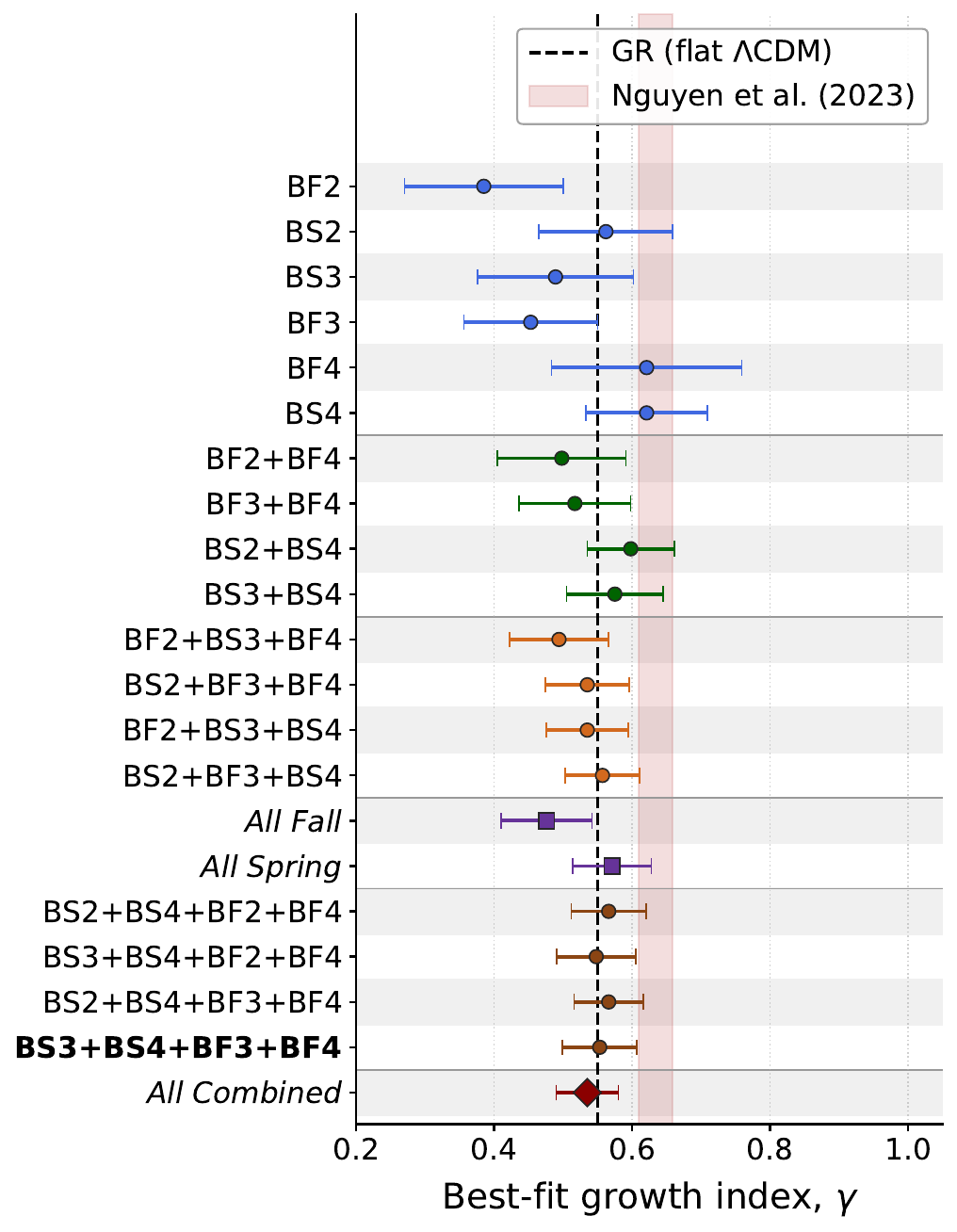}
    \caption{Same format as Figure~\ref{fig:plot_bestfit_gamma}, but with the AP parameter fixed to unity. Most combinations now favor the GR prediction. The baseline result is $\gamma=0.55\pm 0.05$.}
    \label{fig:plot_bestfit_gamma_AP_to_1}
\end{figure}

We derive the growth index, $\gamma$, of the HETDEX [\ion{O}{2}] volume-limited samples in this section. Given a choice of $\gamma$, we calculate $\alpha_v$ for each bin as
\begin{equation}
    \alpha_{v}(z_\mathrm{eff}) = \frac{\left[\Omega_{\mathrm{m},\mathrm{MU}}(z_\mathrm{eff})\right]^{\gamma}}{f_\mathrm{MU}(z_\mathrm{eff})},
\label{eq:gamma}
\end{equation}
where $z_\mathrm{eff}$ is the effective redshift of each bin (see Table~\ref{tab:table_oii_sub}), and $\Omega_{\mathrm{m},\mathrm{MU}}(z)$ and $f_\mathrm{MU}(z)$ are the matter density parameter and the linear growth rate predicted by the MU1 cosmology, respectively, at a given redshift.

We estimate the best-fit value of $\gamma$ for each bin, as well as for combinations of bins. Since we treat each bin as an independent measurement, we combine them into a joint constraint by summing the values of $\chi^2_p(\alpha_v)$ [Equation~\eqref{eq:profile_chi2}] across bins, i.e., $\chi^2(\gamma)=\sum_i\chi^2_p[\alpha_{v,i}(\gamma)]$, where $\chi^2_p(\alpha_{v,i})$ is $\chi^2_p$ from the $i$th bin. However, as shown in Table~\ref{tab:table_oii_sub} and Figure~\ref{fig:plot_bestfit_alpha_v_redshift}, Bins 2 and 3 overlap substantially in their redshift range for each field. Therefore, when computing the best-fit $\gamma$, we exclude combinations that include both Bins~2 and 3 from the same field.

Figure~\ref{fig:plot_bestfit_gamma} shows the resulting best-fit $\gamma$ values for different combinations of bins. 
For illustrative purposes, we also present in italics the combination of Bins~2, 3, and 4 from the same field and the combination of all bins, even though the substantial overlaps. All combinations are consistent with the general relativity (GR) prediction for a flat $\Lambda$CDM cosmology, $\gamma \simeq 0.55$ \citep[vertical dashed line;][]{peebles:1980,Lahav_1991,Bouchet_1995}. The baseline result is $\gamma=0.60\pm 0.07$ from ``BS3+BS4+BF3+BF4,'' a combination with the largest independent comoving volumes.

However, our results are also consistent with the $\gamma = 0.633^{+0.024}_{-0.025}$ value reported by \citet{Nguyen_2023}. At face value, \citet{Nguyen_2023}'s result suggests evidence for a suppressed growth rate of LSS during the dark-energy-dominated era at a statistical significance of 3.7$\sigma$. Our results do not rule out this possibility.

Conversely, the HETDEX data favor $\gamma\simeq 0.55$ when the AP parameters are fixed to their fiducial values. 
Figure~\ref{fig:plot_bestfit_gamma_AP_to_1} presents the same best-fit $\gamma$ estimation as Figure~\ref{fig:plot_bestfit_gamma}, but with the AP parameter set to unity. The resulting values have smaller error bars and are closer to the GR prediction for a flat $\Lambda$CDM cosmology. The baseline result is $\gamma=0.55\pm 0.05$.

The smaller uncertainties obtained when fixing the AP parameters to unity, compared to maximizing the likelihood with respect to them using a profile likelihood approach, reflect the correlations between the RSD and the AP parameters, as shown in Figure~\ref{fig:plot_mock_recovery_spring}.
Both effects imprint similar anisotropic signatures on the clustering multipoles in the linear regime (see Figures~\ref{fig:plot_pk_alpha_v_spring}, \ref{fig:plot_pk_alpha_perp_spring}, and \ref{fig:plot_pk_alpha_par_spring}). When the AP parameters are left free, statistical fluctuations in each bin can shift their best-fit values away from unity, which in turn propagates into the inferred values of $\alpha_v$ and, through Equation~\eqref{eq:gamma}, into $\gamma$. 
Setting the AP parameters to their fiducial values removes this additional source of scatter, yielding both tighter and less scattered estimates of $\gamma$. 
This finding motivates us to perform full cosmological inference, rather than using phenomenological parameters such as $\boldsymbol{\alpha}$. We leave this investigation for future work.

\subsection{Impact of modeling assumptions}

The constraints on the growth index presented in this work are obtained under several simplifying assumptions. 
We fix the galaxy--halo connection to the best-fit SHOD parameters derived in Paper I, rather than marginalizing over them simultaneously with the growth parameter. 
We also adopt a single velocity-rescaling parameter, $\alpha_v$, for all halos, including subhalos, and do not consider independent central- and satellite-velocity-bias parameters. 
Finally, the mock catalogs are generated at the \textit{Planck} $\Lambda$CDM cosmology of MU1, and the cosmological parameters controlling the shape and amplitude of the matter power spectrum are not varied in the inference.

Previous simulation-based emulator analyses have provided a useful estimate of the scale of these modeling uncertainties, although their galaxy samples and galaxy--halo connection parameterizations differ from ours. 
For example, \citet{Yuan_2022} has jointly varied cosmology and HOD parameters in an AbacusSummit-based analysis of the Baryon Oscillation Spectroscopic Survey Constant Mass (BOSS CMASS) sample~\citep{CMASS1_2016,CMASS2_2016} and obtained $f\sigma_8(z=0.52)=0.444\pm0.016$, corresponding to a $3.6\%$ statistical uncertainty. 
They have also found this constraint to be robust to the tested secondary-bias and alternative-HOD prescriptions. 
Taking a conservative $1$--$2\%$ fractional shift in the recovered growth rate as an indicative modeling scale, its impact on the growth index can be estimated from $f(z)=\left[\Omega_m(z)\right]^\gamma$,
\begin{equation}
    \Delta\gamma \simeq \frac{\Delta\ln f}{\ln\Omega_\mathrm{m}(z)}.
\end{equation}
For the effective redshift range of our samples, $z_{\rm eff}=0.17$--$0.39$, the fiducial \textit{Planck} cosmology gives $\Omega_m(z)\simeq0.42$--$0.54$. 
A $1\%$ shift in $f$ therefore corresponds to $|\Delta\gamma|\simeq0.012$--$0.016$, while a $2\%$ shift corresponds to $|\Delta\gamma|\simeq0.023$--$0.033$. 
Thus, if the sensitivity of our SHOD analysis to galaxy--halo modeling is comparable to that found in previous emulator studies, fixing the SHOD parameters would contribute an uncertainty of order $\sigma_{\gamma,\mathrm{SHOD}}\simeq 0.01$--$0.03$, which is subdominant to our statistical uncertainty, $\sigma_\gamma=0.05$--$0.07$.

Velocity bias is potentially more important because it directly affects the small-scale quadrupole power spectrum used in our analysis. 
Typical constraints from BOSS-like samples find central velocity-bias amplitudes consistent with unity to within a few percent and satellite velocity-bias parameters constrained at the $\simeq 10\%$ level \citep{Reid_2014,GAMA_2021}. 
Since our $\alpha_v$ parameter effectively rescales all halo velocities simultaneously, unmodeled velocity bias could be partially absorbed into $\alpha_v$. 
As a rough guide, the emulator results of \citet{Zhai_2019} imply that plausible velocity-bias variations generally shift the recovered growth amplitude by less than a few percent, corresponding to $\sigma_\gamma=0.02$--$0.03$ at the redshifts considered here.

The assumed cosmology introduces an additional limitation because changes in $\Omega_\mathrm{m,0}$, $\sigma_R$, $h$, and the spectral shape can alter both halo abundances and nonlinear redshift-space clustering. 
While the present AP analysis allows variations in geometric dilation parameters, it does not capture the full cosmological dependence of the real- and redshift-space power spectra.
A fully self-consistent analysis should therefore emulate the joint dependence on cosmology, SHOD parameters, and separate central and satellite velocity biases. 
Based on the estimates above, these modeling effects are likely subdominant to the current statistical error. However, they will need to be explicitly marginalized over in future analyses targeting percent-level constraints on $\gamma$.

\section{Conclusions}
\label{sec:conclusions}

We used the volume-limited samples of HETDEX [\ion{O}{2}]-emitting galaxies at $z\le0.48$, constructed in Paper I, to constrain the growth history of cosmic structures. Building on our previous analysis, which used only the monopole power spectrum, we incorporated the quadrupole power spectrum into our analysis to constrain the RSD parameter, $\alpha_v$, together with the AP parameters, $\alpha_\perp$ and $\alpha_\parallel$, for the Spring and Fall fields across different luminosity and redshift bins.

We interpreted the observed power spectrum multipoles using mock catalogs constructed from the MU1 $N$-body simulation suite, adopting the same SHOD galaxy--halo connection model as in~Paper I. 
We demonstrated that $\alpha_v$ and the AP parameters leave distinguishable imprints on the monopole and quadrupole power spectra across wavenumbers up to $k_\mathrm{max}=0.7~h~\mathrm{Mpc}^{-1}$.

Using a profile-likelihood approach, we obtained best-fit values of $\alpha_v$ and the AP parameters for each bin. The mock recovery tests confirm that our pipeline recovers unbiased estimates of these parameters. The best-fit values derived from the HETDEX data fall comfortably within the distribution of the mock catalogs. Goodness-of-fit tests further show that the resulting $\chi^2$ values are fully consistent with the theoretical Sellentin-Heavens distribution. The simulation-based best-fit model reproduces the observed monopole and quadrupole power spectra remarkably well, up to $k_\mathrm{max}=0.7~h~\mathrm{Mpc}^{-1}$. Together, these results confirm that the clustering of HETDEX [\ion{O}{2}] galaxies is consistent with the  \textit{Planck} flat $\Lambda$CDM cosmology adopted for MU1.

By translating the best-fit $\alpha_v$ values into the linear growth index, $\gamma$, essentially all combinations of bins are consistent with the GR prediction, $\gamma\simeq0.55$, when the AP parameters are held fixed at their fiducial values. Our baseline result is $\gamma=0.55\pm 0.05$, obtained from the largest volume and the independent combinations. However, when the AP parameters are allowed to vary, $\gamma=0.60\pm 0.07$, which is consistent with both the GR prediction and the suppressed growth rate, $\gamma=0.633^{+0.024}_{-0.025}$, reported by \citet{Nguyen_2023}.

We interpret this behavior as a consequence of the correlation between the RSD and AP parameters: they imprint similar anisotropic signatures on the clustering multipoles in the linear regime. When the AP parameters are allowed to vary freely, statistical fluctuations in their best-fit values propagate into $\alpha_v$ and, in turn, into $\gamma$. Fixing the AP parameters to their fiducial values removes this additional source of scatter, yielding a more precise constraint on $\gamma$, albeit under a more restrictive assumption.

These results demonstrate that the HETDEX [\ion{O}{2}] sample provides a competitive, independent low-redshift probe of the growth of cosmic structures, complementary to existing galaxy redshift surveys that rely on different target selection strategies and are subject to different systematics. While our current constraints on $\gamma$ cannot yet distinguish between the GR prediction and the suppressed growth rate reported by \citet{Nguyen_2023}, they establish the viability of HETDEX [\ion{O}{2}] galaxies for growth-of-structure studies.

This work motivates several directions for future investigations.
First, the correlation between $\alpha_v$ and the AP parameters highlights the need for a full cosmological inference that self-consistently constrains the growth rate together with the background expansion history, rather than relying on the phenomenological parameters such as $\boldsymbol{\alpha}$ adopted in this paper. 
Second, only ten percent of the anticipated MU1 was used. As more realizations become publicly available, we can incorporate additional statistics, such as the bispectrum, to reduce the correlations among the parameters and the uncertainty in $\gamma$.
Third, forthcoming HETDEX data releases from parallel observations, with larger sample sizes, will improve the statistical precision of the constraints, especially for the parameters related to the LOS, since the parallel observations are scattered across the sky. 
As we increase precision, we may need to vary $M_0$ simultaneously during inference, use a more flexible SHOD or SHAM, and account for a potential velocity bias for subhalos, although these were not necessary for the current level of precision.
Finally, combining the HETDEX [\ion{O}{2}] sample with other low-redshift LSS probes and CMB measurements, including ongoing projects such as the Vera C. Rubin Observatory~\citep{LSST_2019} and the Simons Observatory~\citep{Simons_2019}, will help further test the standard $\Lambda$CDM model and shed light on the origin of the tentative suppressed-growth signal and the tensions reported in the literature.

\section*{Acknowledgements}

HETDEX is led by the University of Texas at Austin McDonald Observatory and Department of Astronomy with participation from the Ludwig-Maximilians-Universität München, Max-Planck-Institut für Extraterrestrische Physik (MPE), Leibniz-Institut für Astrophysik Potsdam (AIP), Texas A\&M University, Pennsylvania State University, Institut für Astrophysik Göttingen, The University of Oxford, Max-Planck-Institut für Astrophysik (MPA), The University of Tokyo and Missouri University of Science and Technology.

Observations for HETDEX were obtained with the Hobby-Eberly Telescope (HET), which is a joint project of the University of Texas at Austin, the Pennsylvania State University, Ludwig-Maximilians-Universität München, and Georg-August-Universität Göttingen. The HET is named in honor of its principal benefactors, William P. Hobby and Robert E. Eberly. The Visible Integral-field Replicable Unit Spectrograph (VIRUS) was used for HETDEX observations. VIRUS is a joint project of the University of Texas at Austin, Leibniz-Institut für Astrophysik Potsdam (AIP), Texas A\&M University, Max-Planck-Institut fürExtraterrestrische Physik (MPE), Ludwig-Maximilians-Universität München, Pennsylvania State University, Institut für Astrophysik Göttingen, University of Oxford, and the Max-Planck-Institut fur Astrophysik (MPA).

The authors acknowledge the Texas Advanced Computing Center (TACC) at The University of Texas at Austin for providing high performance computing, visualization, and storage resources that have contributed to the research results reported within this paper. URL: http://www.tacc.utexas.edu

Funding for HETDEX has been provided by the partner institutions, the National Science Foundation, the State of Texas, the US Air Force, and by generous support from private individuals and foundations.

This work was also supported in part by the Deutsche Forschungsgemeinschaft (DFG, German Research Foundation) under Germany's Excellence Strategy – EXC-2094/2 – 390783311.

C.G. and C.G. acknowledge support from the National Science Foundation under grant AST-2408358.

D.J. was supported by NSF grant AST-2407298 at PSU.

\appendix
\section{Comparison of AP parameter application in the catalog level and P(\texorpdfstring{$k_\perp$}{kperp},\texorpdfstring{$k_{\parallel}$}{kpara}) level}
\label{appendix:AP_param_test}

\begin{figure*}
    \centering
    \includegraphics[width=\textwidth]{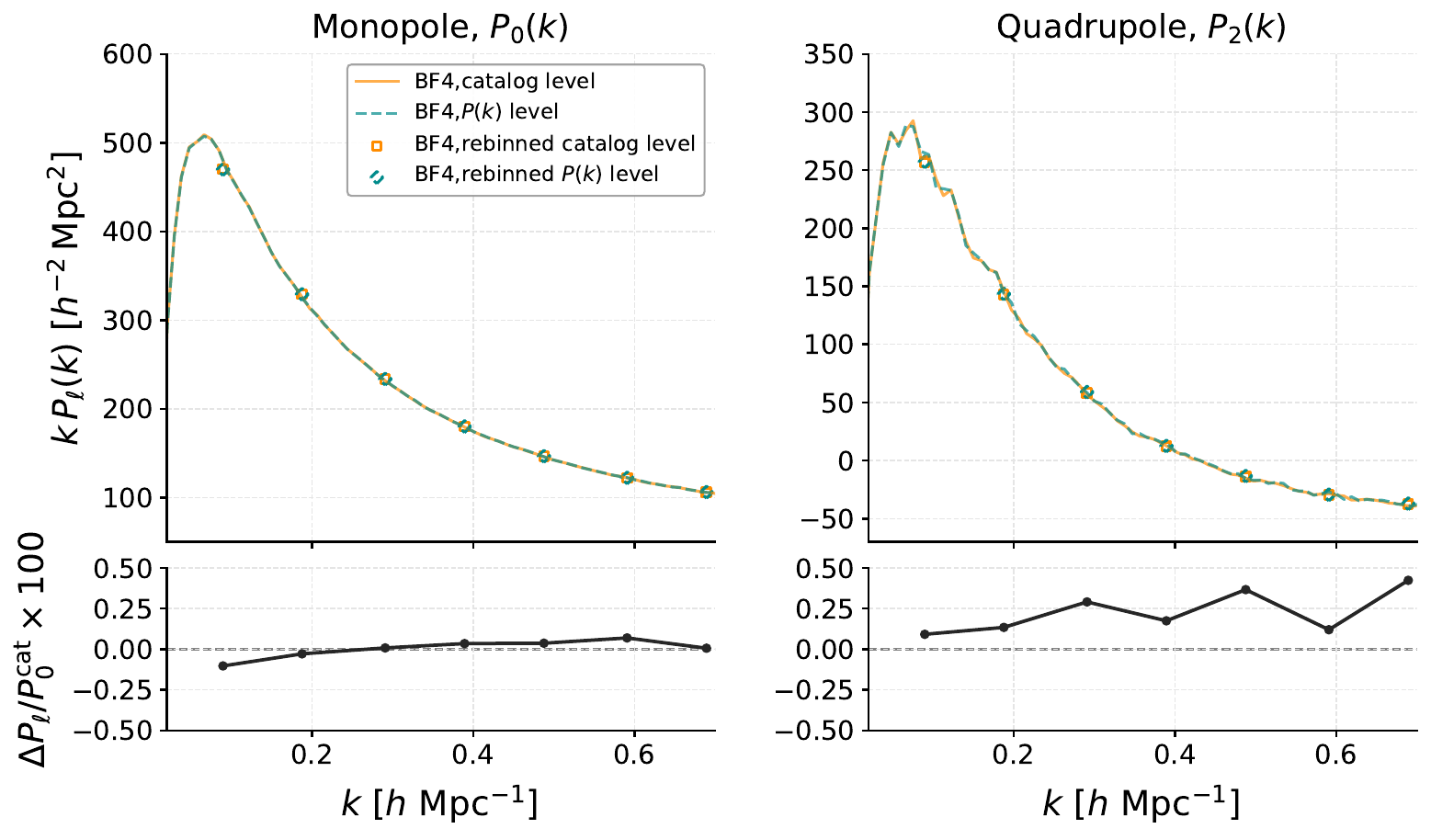}
    \caption{Top panels: monopole (left) and quadrupole (right) power spectra of 400 realizations for Fall Bin 4, comparing two methods of applying the AP parameters, evaluated at $\alpha_\perp=1.017$ and $\alpha_\parallel=1.035$. The first method rescales the comoving positions of galaxies directly in the mock catalog before measuring the power spectrum (``catalog level'', orange). The second method measures the power spectrum at the fiducial AP values and then maps it analytically (``$P(k)$ level'', blue). For each method, the line shows the power spectrum with $\Delta k=0.01~h~\mathrm{Mpc}^{-1}$, while the points show the same measurement rebinned to $\Delta k=0.1~h~\mathrm{Mpc}^{-1}$.
    Bottom panels: fractional difference (in percent) between the two methods computed from the rebinned power spectra, $\Delta P_\ell \equiv P_\ell^{P(k)}(k) - P_\ell^\mathrm{cat}(k)$, normalized by the rebinned catalog-level monopole, $P_0^\mathrm{cat}(k)$.
    The quadrupole difference is also normalized by $P_0^\mathrm{cat}(k)$ to avoid artificial divergences near the quadrupole's zero-crossing.}
    \label{fig:pk_catalog_vs_pk_level}
\end{figure*}

Evaluating $\chi^2(\boldsymbol{\alpha})$ over a three-dimensional grid of $\boldsymbol{\alpha}=(\alpha_v, \alpha_\perp, \alpha_\parallel)$ (Section~\ref{subsec:likelihood}) would, in principle, require rescaling the comoving positions of every galaxy in each mock catalog according to the assumed $\alpha_v$ and AP parameters and re-measuring the power spectrum multipoles at each grid point. 
However, since this ``catalog-level'' rescaling is computationally expensive when repeated over a fine grid and many mock realizations, we instead measure the power spectrum multipoles, first at the fiducial AP values, then rescale them analytically by mapping the observed $(k,\mu)$ onto the true cosmology using the \citet{Ballinger_1996} formulae described in Section~\ref{sec:methodology}. We refer to this computationally efficient approach as ``$P(k)$-level'' rescaling.

In this appendix, we validate that the $P(k)$-level AP rescaling yields power spectrum multipoles that are consistent with those from the full catalog-level rescaling. Figure~\ref{fig:pk_catalog_vs_pk_level} compares the monopole and quadrupole power spectra for Fall Bin 4, as a representative of all bins, computed using both methods at $\alpha_\perp=1.017$ and $\alpha_\parallel=1.035$. To quantify their agreement, we compute the fractional difference between the two methods, $\Delta P_\ell/P_0^\mathrm{cat}(k)$, where $\Delta P_\ell \equiv P_\ell^{P(k)}(k)-P_\ell^\mathrm{cat}(k)$. 
We normalize the differences in both the monopole and quadrupole by the catalog-level monopole, $P_0^\mathrm{cat}(k)$, rather than by each multipole's own amplitude, since the quadrupole crosses zero and a self-normalized fractional difference would artificially diverge near that crossing. The two approaches agree to within $0.25\%$ and $0.5\%$ of $P_0^\mathrm{cat}(k)$ for the monopole and quadrupole power spectra, respectively, across all wave numbers up to $k_\mathrm{max}=0.7\,h\,\mathrm{Mpc}^{-1}$, confirming that the $P(k)$-level rescaling is an accurate and computationally efficient substitute for catalog-level rescaling of the AP parameters. We therefore adopt the $P(k)$-level approach in our likelihood analysis (Section~\ref{sec:interpretation}).

\section{Comparison between Uchuu and MU1}
\label{appendix:sim_comparison}

In this appendix, we assess whether populating SHOD galaxies onto MU1 halo catalogs, using the same best-fit $M_0$, reported for Uchuu in Paper I, remains a reasonable approximation for this analysis. As discussed in Section~\ref{subsec:mock}, we adopt MU1 in place of the Uchuu simulation primarily to take advantage of its 100 publicly available independent realizations of a cubic box of $1~h^{-1}~\mathrm{Gpc}$. However, this comes at the cost of coarser mass resolution, a smaller box size, and a slightly different cosmology relative to Uchuu's single $2~h^{-1}~\mathrm{Gpc}$ realization. 

\begin{figure*}
    \centering
    \includegraphics[width=\textwidth]{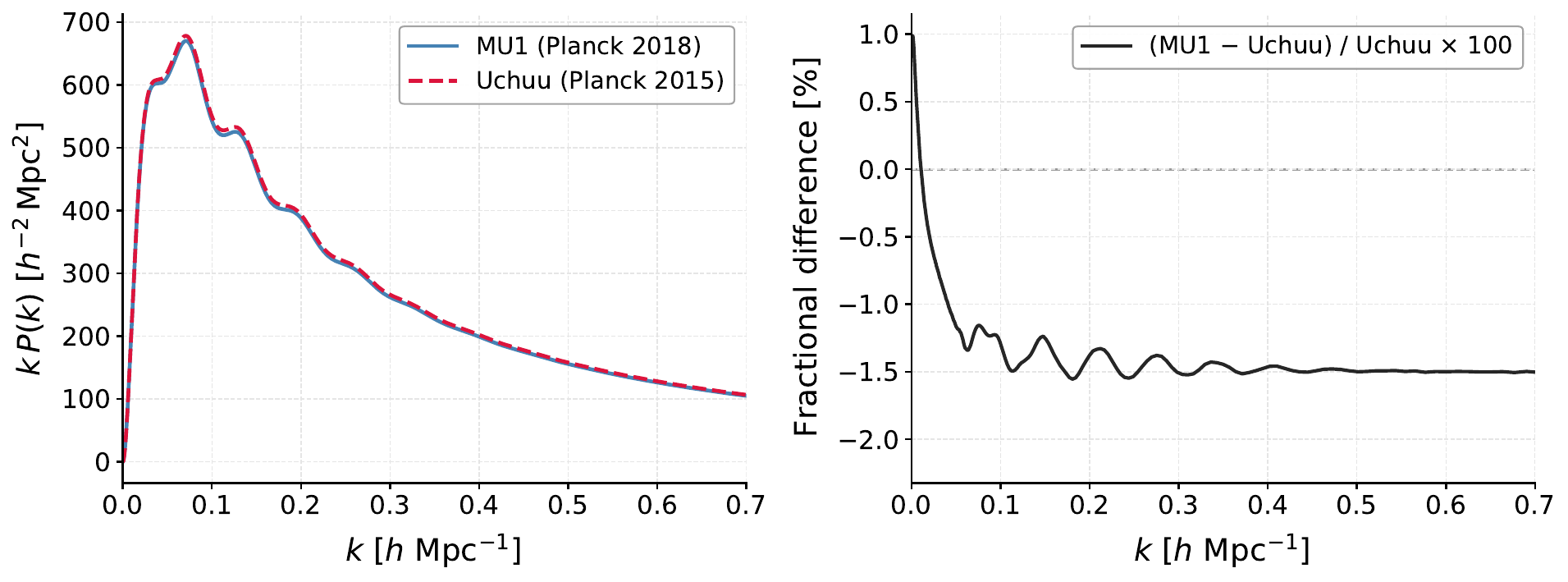}
    \caption{Left: Input linear matter power spectra for MU1 (blue, \textit{Planck} 2018 cosmology) and Uchuu (red dashed, \textit{Planck} 2015 cosmology). Right: Fractional difference (in percent) between the two power spectra, $(P_\mathrm{MU}-P_\mathrm{U})/P_\mathrm{U}\times100$. The power spectra of the two cosmologies agree to within 1.5\% over the range of scales relevant to this analysis, $k\le0.7\,h\,\mathrm{Mpc}^{-1}$.}
    \label{fig:pk_uchuu_muchouchuu_comparison}
\end{figure*}

\begin{figure*}
    \centering
    \includegraphics[width=\textwidth]{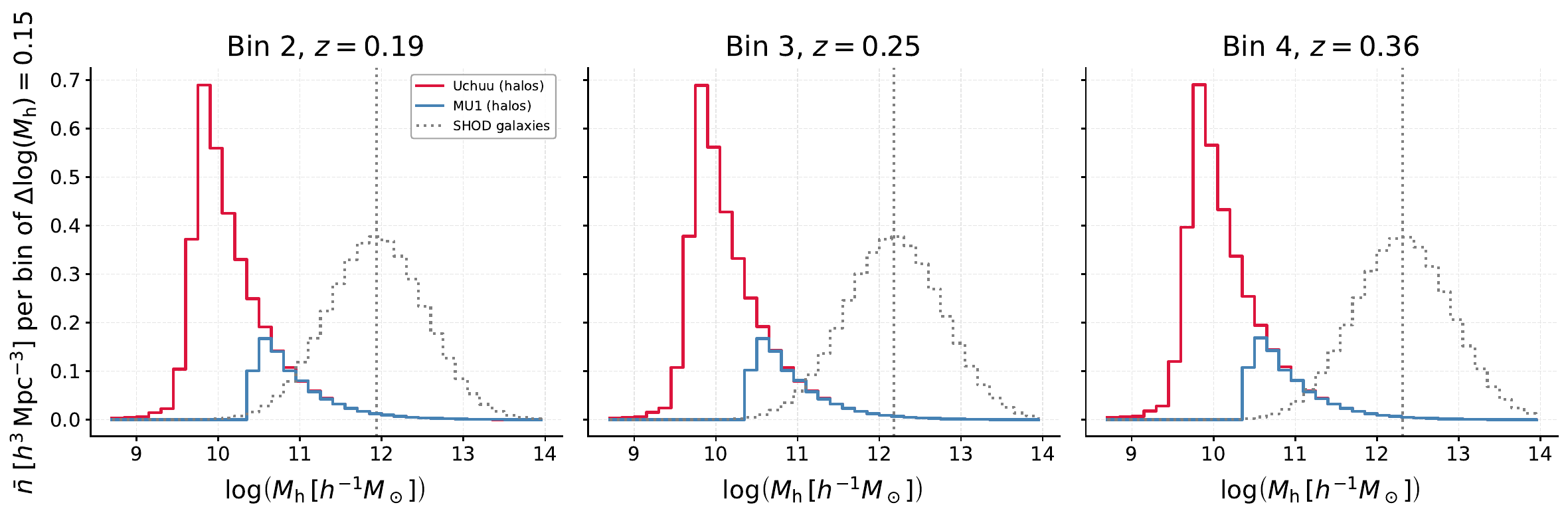}
    \caption{Halo number density per bin of $\Delta\log(M_\mathrm{h})=0.15$ for Uchuu (red) and MU1 (blue), for Bins 2 (left), 3 (middle), and 4 (right), shown as solid lines. The vertical dotted lines mark the best-fit characteristic mass, $M_0$, obtained for the Spring field using the Uchuu simulation (see Paper I). The dotted curves show the shape of the SHOD occupation function following Equation~\eqref{eq:lognormal}.  The curves, which are arbitrarily rescaled for visibility, illustrate the range of halo masses that contribute to the galaxy sample relative to each simulation.}
    \label{fig:halo_number_density_with_shod}
\end{figure*}

\begin{figure*}
    \centering
    \begin{minipage}{0.32\textwidth}
        \centering
        \includegraphics[width=\textwidth]{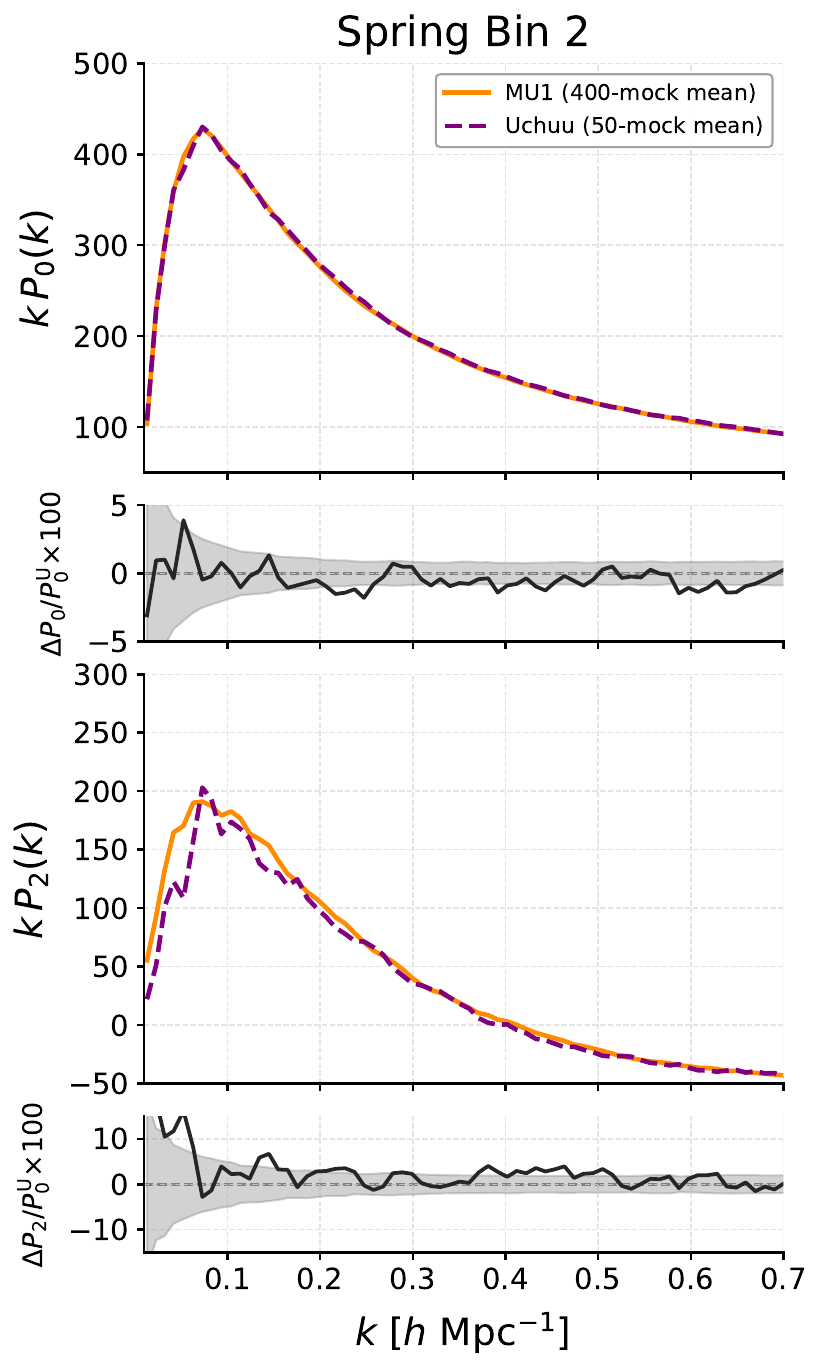}
    \end{minipage}
    \begin{minipage}{0.32\textwidth}
        \centering
        \includegraphics[width=\textwidth]{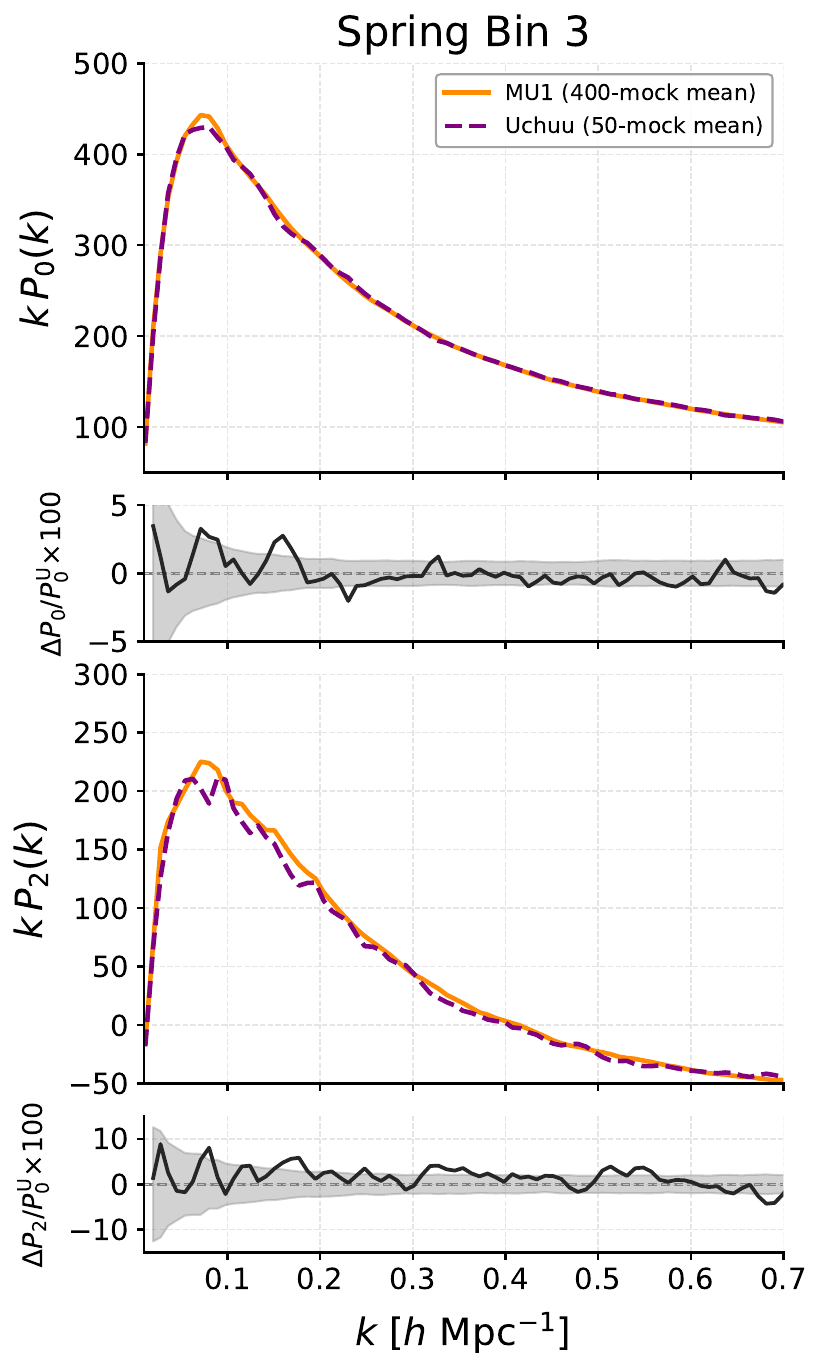}
    \end{minipage}
    \begin{minipage}{0.32\textwidth}
        \centering
        \includegraphics[width=\textwidth]{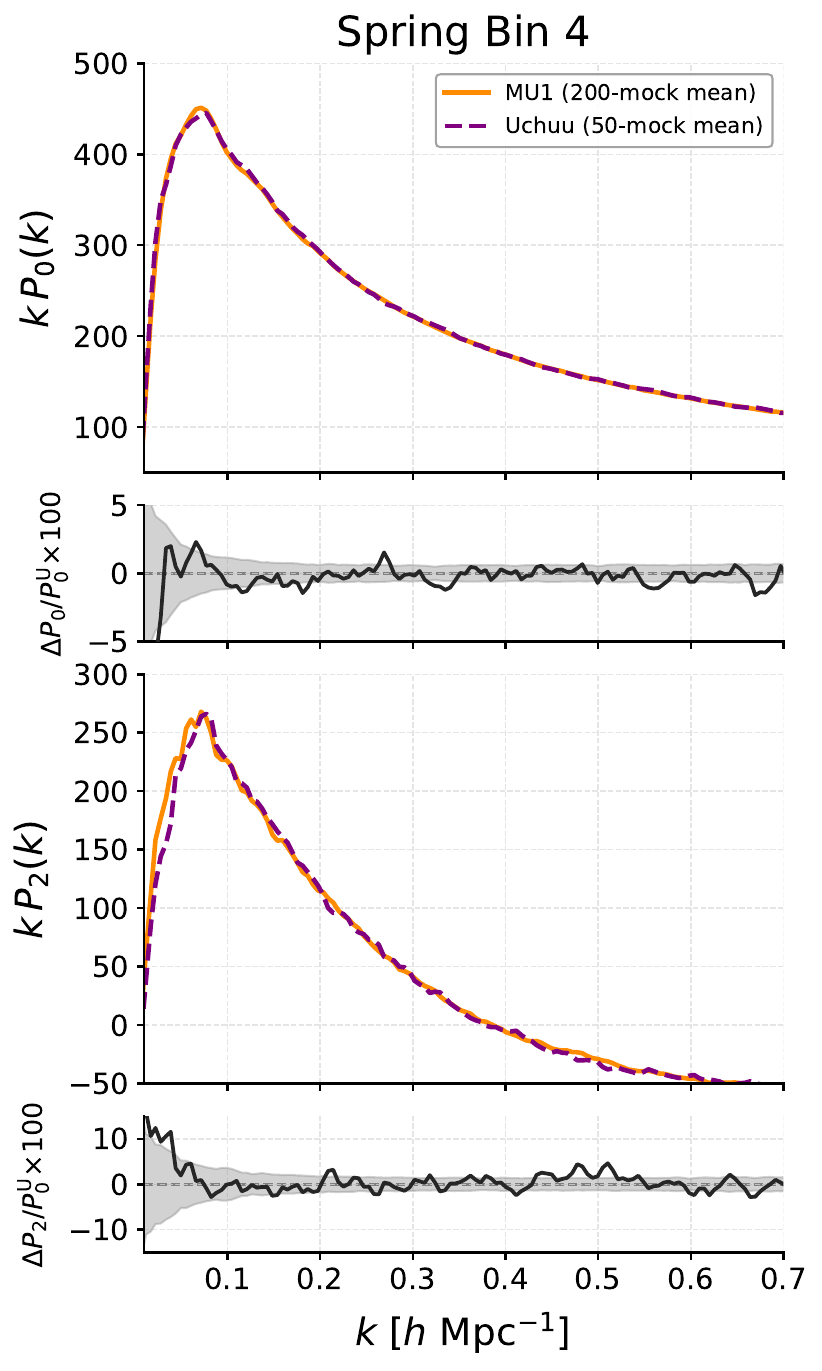}
    \end{minipage}
    \caption{Monopole (top) and quadrupole (bottom) power spectra in the Spring field, comparing the MU1 mean averaged over 400 (left and middle) and 200 (right) realizations (orange) with the Uchuu mean averaged over 50 realizations (purple, dashed). Since the MU1 mean is averaged over many more realizations, its own sampling noise is negligible, so the remaining scatter  can be attributed primarily to the 50 realizations in the Uchuu sample. The bottom of each panel displays the fractional difference (in percent) between the two simulations, $\Delta P_\ell \equiv P_\ell^\mathrm{MU}(k) - P_\ell^\mathrm{U}(k)$, normalized by the Uchuu monopole power spectrum, $P_0^\mathrm{U}(k)$. The quadrupole difference is normalized by $P_0^\mathrm{U}(k)$ to avoid artificial divergences near the quadrupole's zero-crossing. The shaded bands represent the expected $1\sigma$ scatter of a 50-realization average, estimated by dividing the MU1 scatter by $\sqrt{50}$. The two simulations show consistent power spectrum amplitudes and shapes within the scatter of the mean of 50 realizations.}
    \label{fig:plot_pkg_uchuu_muchouchuu_comparison_spring}
\end{figure*}

\begin{figure*}
    \centering
    \begin{minipage}{0.32\textwidth}
        \centering
        \includegraphics[width=\textwidth]{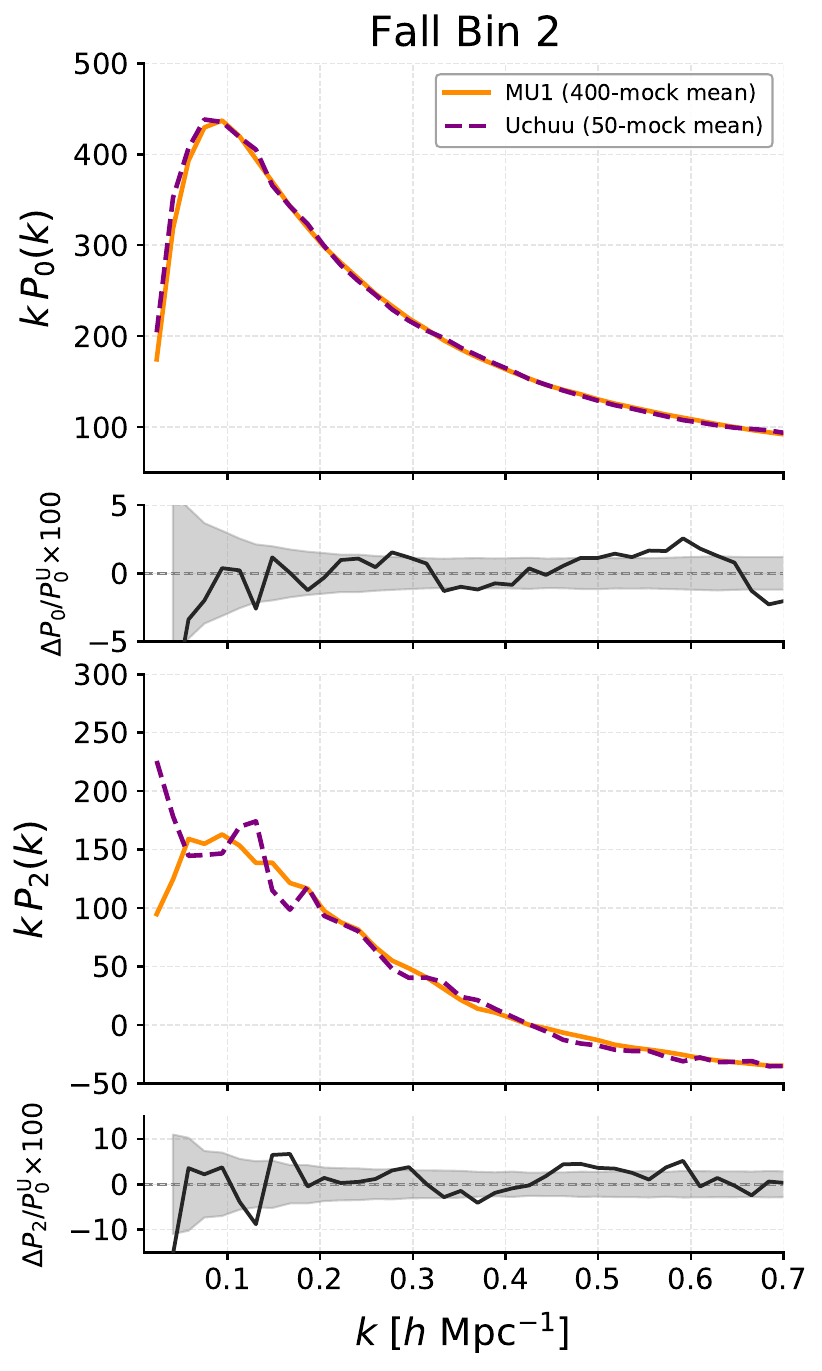}
    \end{minipage}
    \begin{minipage}{0.32\textwidth}
        \centering
        \includegraphics[width=\textwidth]{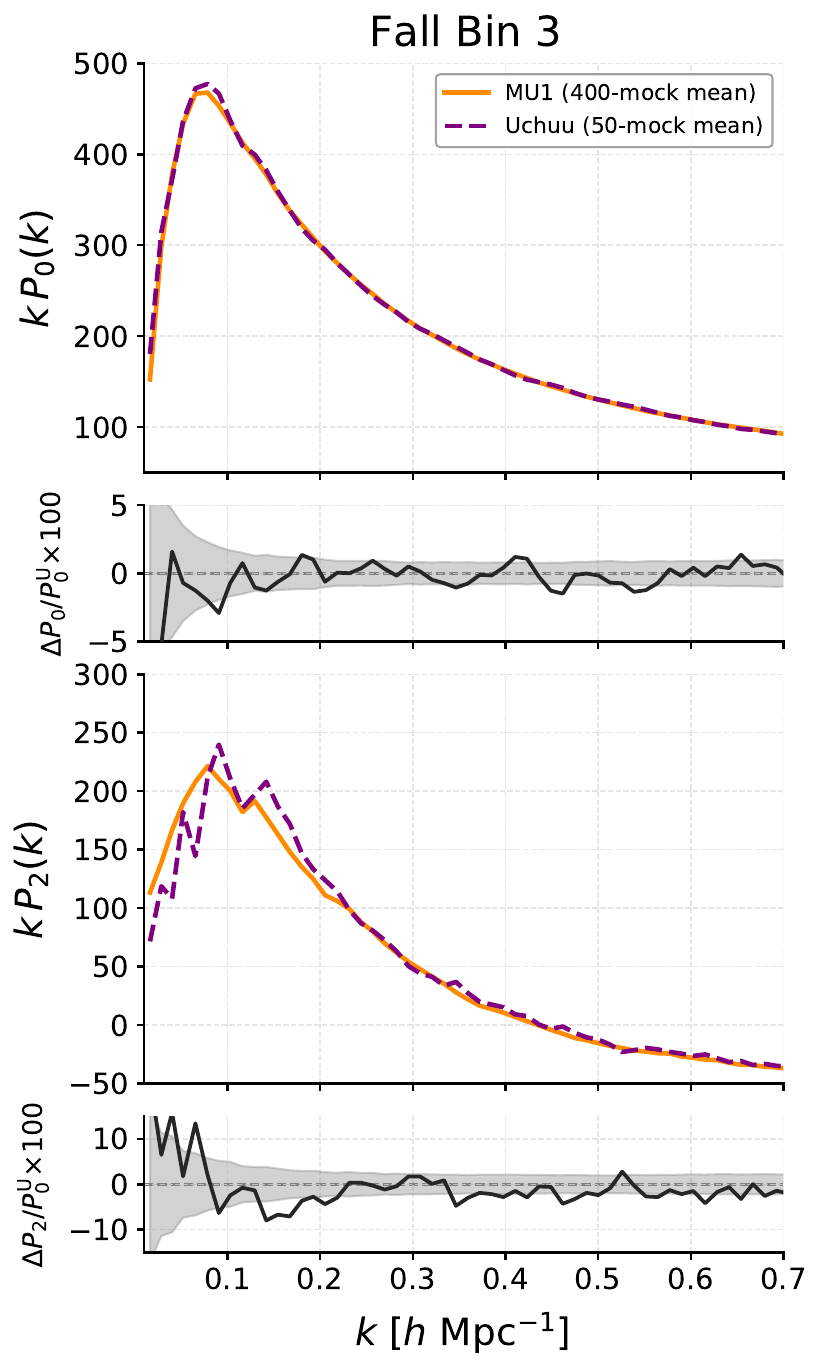}
    \end{minipage}
    \begin{minipage}{0.32\textwidth}
        \centering
        \includegraphics[width=\textwidth]{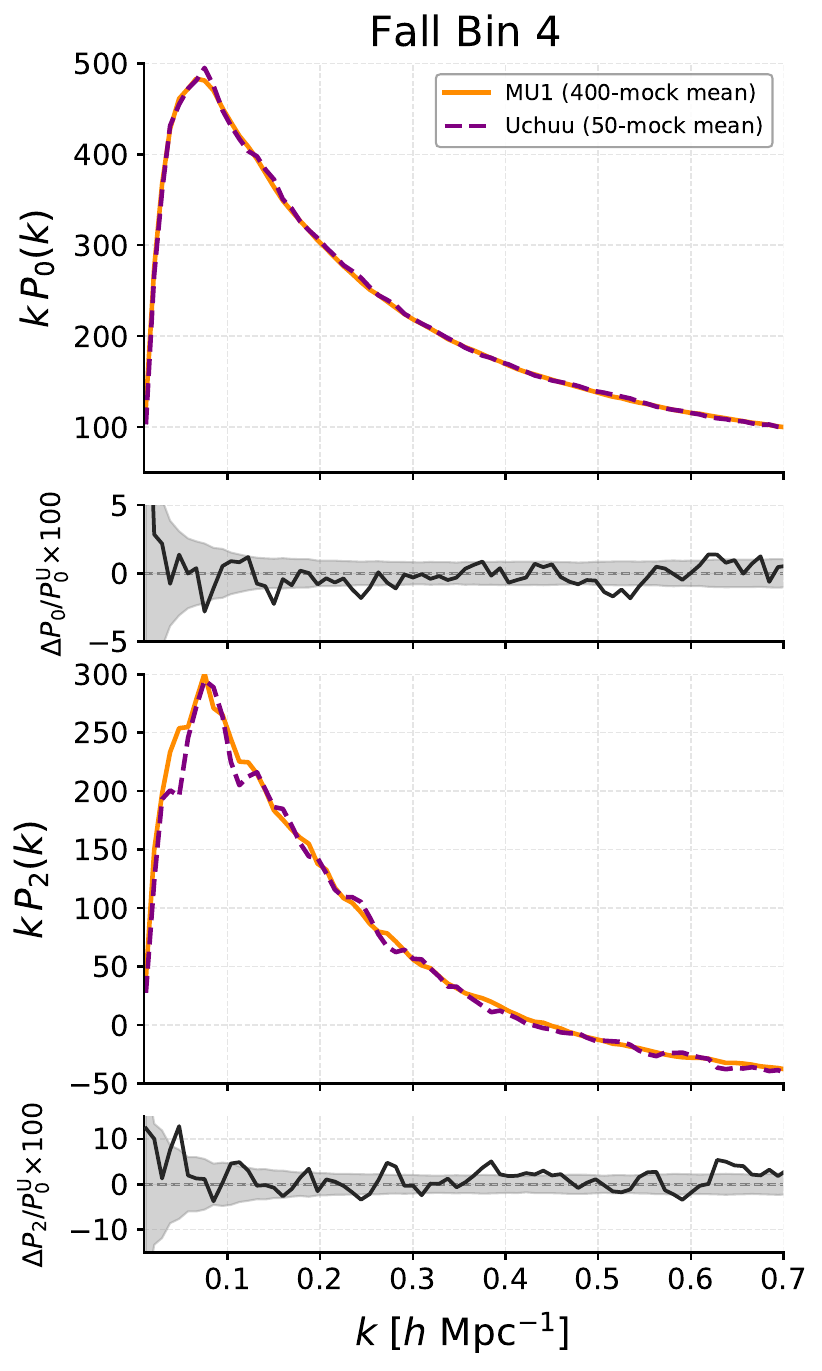}
    \end{minipage}
    \caption{Same format as Figure~\ref{fig:plot_pkg_uchuu_muchouchuu_comparison_spring}, but for the Fall field. As in the Spring field, the two simulations show consistent power spectra within the scatter of the mean of 50 realizations.}
    \label{fig:plot_pkg_uchuu_muchouchuu_comparison_fall}
\end{figure*}

Uchuu and MU1 are based on \textit{Planck} flat $\Lambda$CDM cosmology from the 2015~\citep{Planck15,Uchuu_dr1} and 2018~\citep{Planck18,Fernandez_2026,MuchoUchuu} data releases, respectively. Figure~\ref{fig:pk_uchuu_muchouchuu_comparison} compares the input linear matter power spectra of the two cosmologies together with their fractional difference. The two agree to within 1.5\% over $k\le0.7\,h\,\mathrm{Mpc}^{-1}$, with MU1 showing a slightly lower amplitude than Uchuu.

The slightly lower amplitude of the input power spectrum is compensated for by MU1's coarser mass resolution, approximately four times coarser than that of Uchuu, which results in a correspondingly higher minimum resolved halo mass at all three redshifts. This is shown in Figure~\ref{fig:halo_number_density_with_shod}. The tail of the best-fit lognormal function for Uchuu found in Paper I in the small mass region is slightly below the minimum mass resolved in MU1. Since more massive halos are more strongly biased, the resulting galaxy power spectrum amplitude in MU1 is slightly enhanced, compensating for the lower linear matter power spectrum. As a result, Uchuu and MU1 yield consistent amplitudes and shapes for both the monopole and quadrupole power spectra within the scatter of the mean of 50 realizations; see Figures~\ref{fig:plot_pkg_uchuu_muchouchuu_comparison_spring} and~\ref{fig:plot_pkg_uchuu_muchouchuu_comparison_fall}.
Only 50 mocks are currently available from the single Uchuu simulation~(see Paper I), compared to the 400 (or 200 for Spring Bin 4) mocks available from MU1. 
In conclusion, the effects of differences in cosmology, halo mass resolution, and box size are all small and do not significantly impact the cosmological inferences presented in this paper.

\section{Results from Fall field}

We present the results for the Fall field in this appendix, following the same analysis as for the Spring field. 
Section~\ref{appendix:pk_fall} describes the effect of $\alpha_v$, $\alpha_\perp$, and $\alpha_\parallel$ on the monopole and quadrupole power spectra in the Fall-field bins, while Section~\ref{appendix:mock_recovery_fall} presents the mock recovery test of the $\alpha_v$ and AP parameters for the Fall-field bins.

\subsection{Power spectrum multipole measurements of Fall field}
\label{appendix:pk_fall}

\begin{figure*}
    \centering
    \begin{minipage}{0.32\textwidth}
        \centering
        \includegraphics[width=\textwidth]{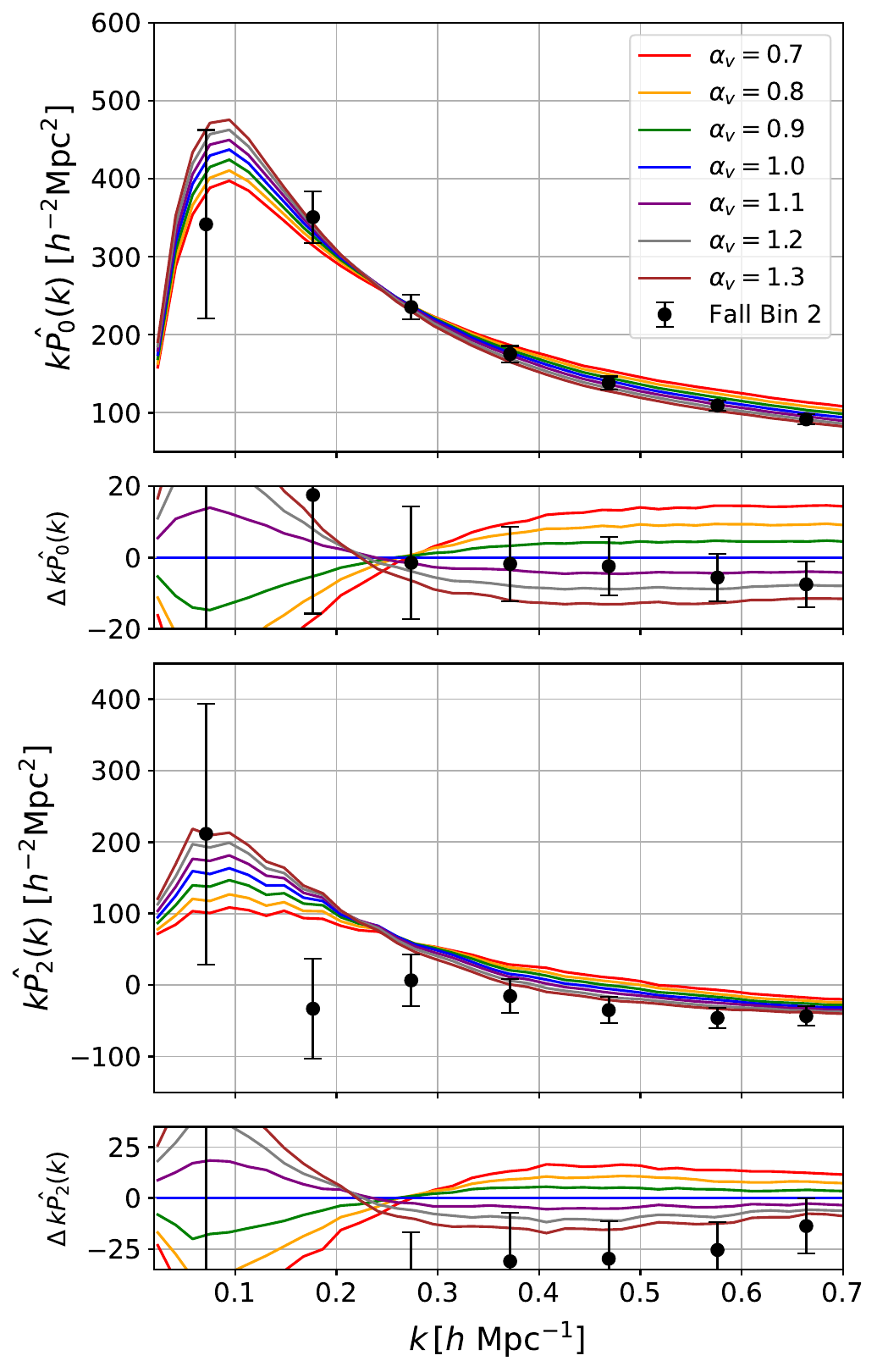}
    \end{minipage}
    \begin{minipage}{0.32\textwidth}
        \centering
        \includegraphics[width=\textwidth]{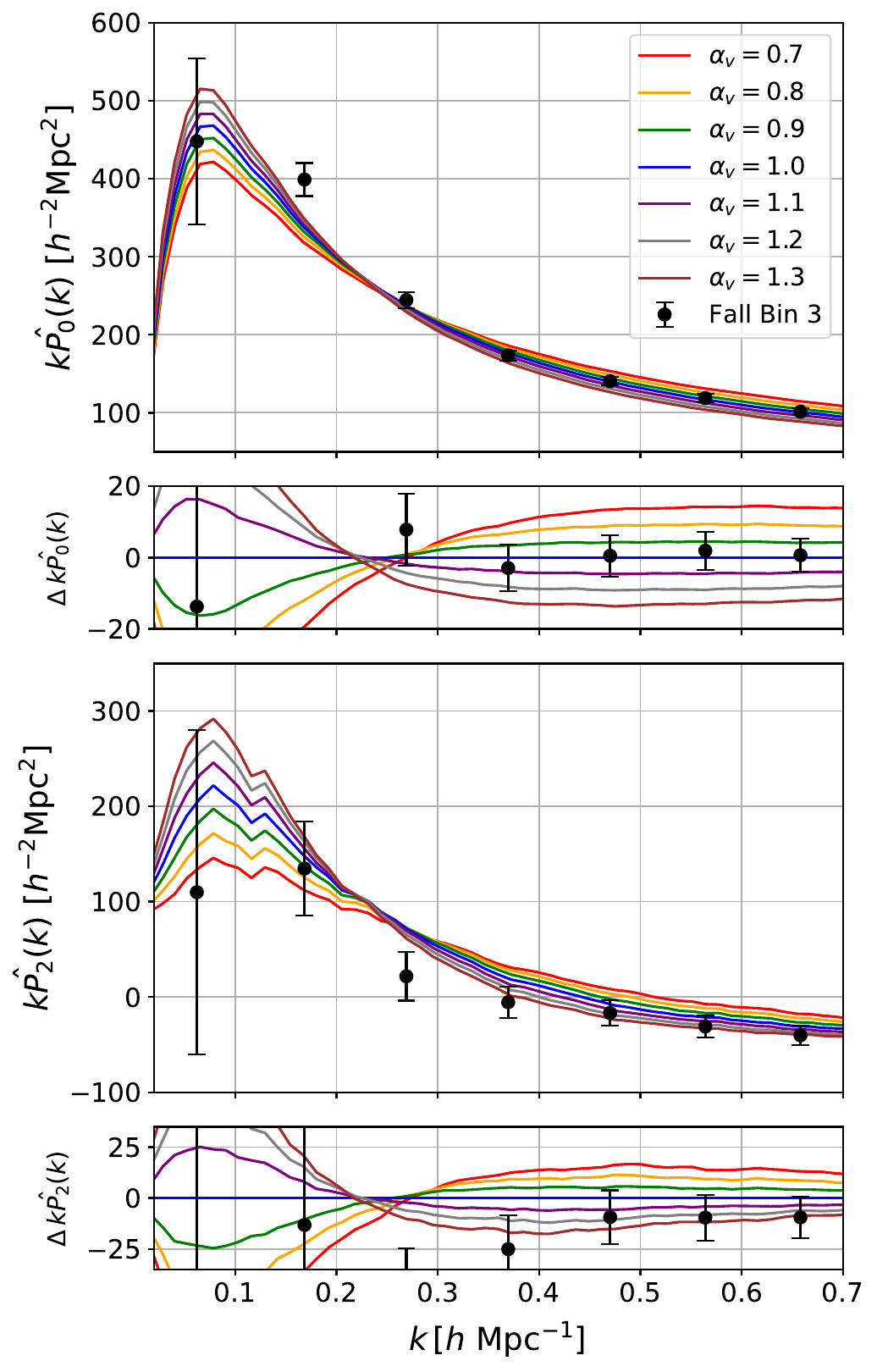}
    \end{minipage}
    \begin{minipage}{0.32\textwidth}
        \centering
        \includegraphics[width=\textwidth]{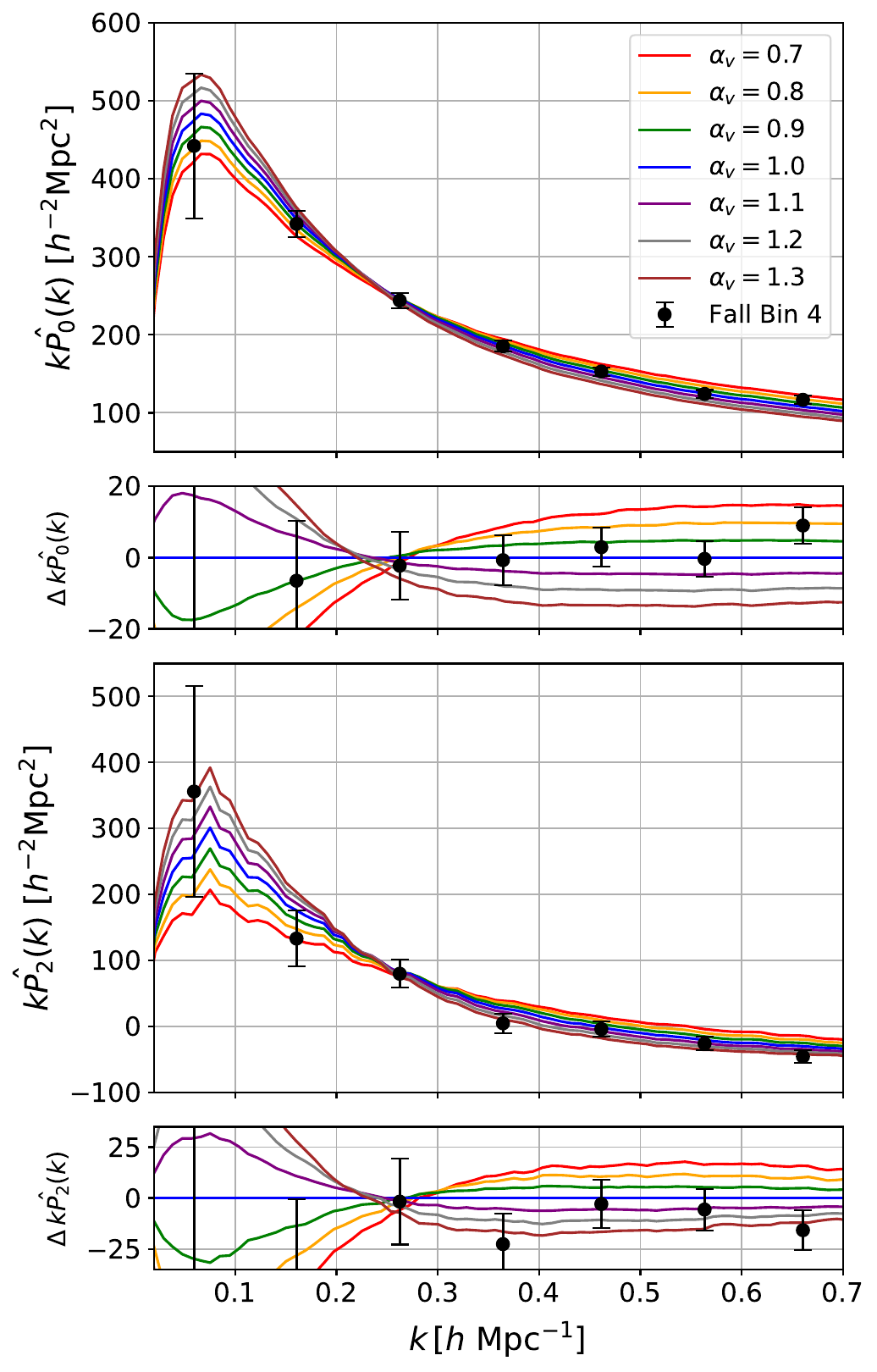}
    \end{minipage}
    \caption{Same format as Figure~\ref{fig:plot_pk_alpha_v_spring}, but for the Fall field.}
    \label{fig:plot_pk_alpha_v_fall}
\end{figure*}

\begin{figure*}
    \centering
    \begin{minipage}{0.32\textwidth}
        \centering
        \includegraphics[width=\textwidth]{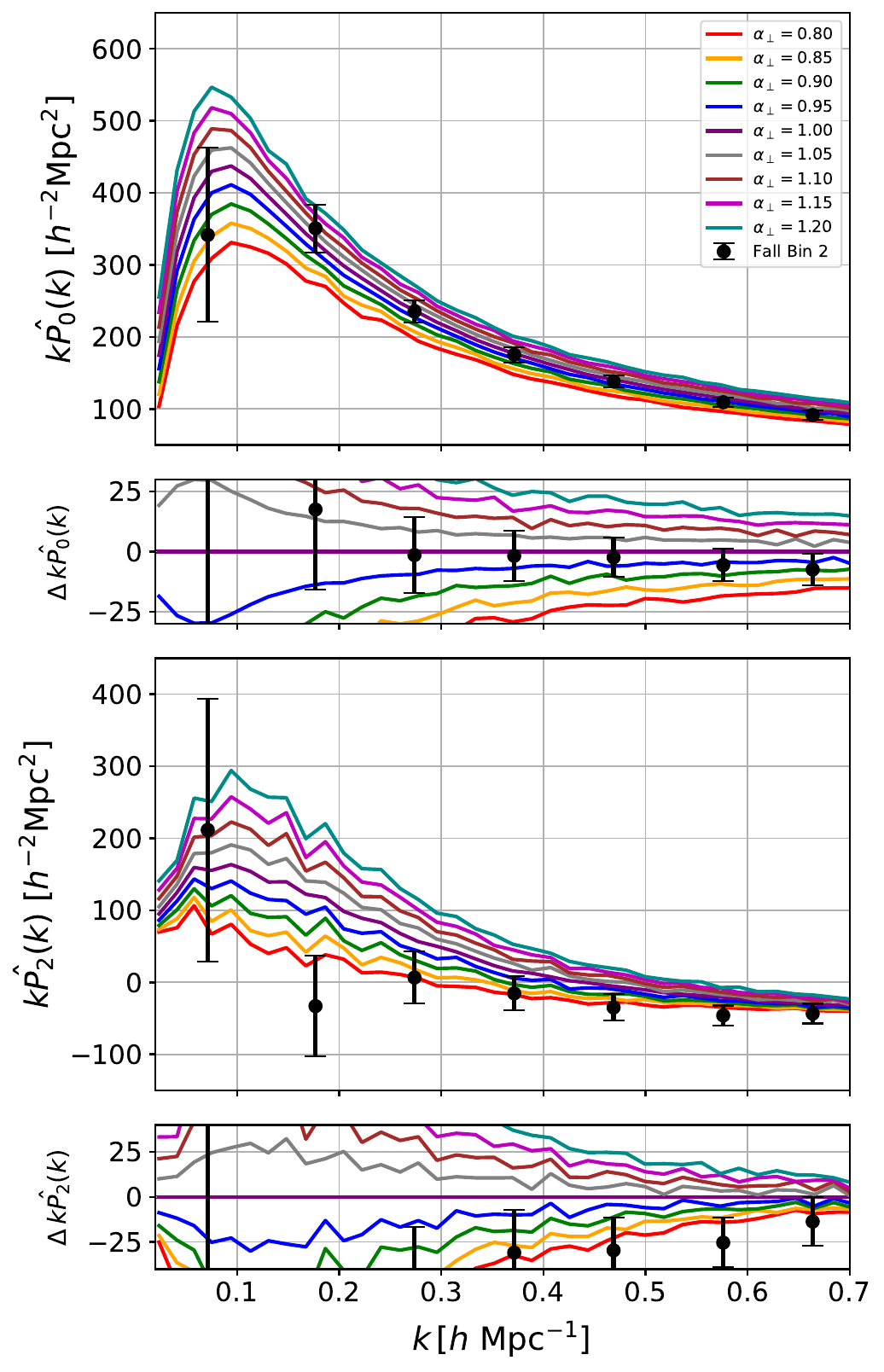}
    \end{minipage}
    \begin{minipage}{0.32\textwidth}
        \centering
        \includegraphics[width=\textwidth]{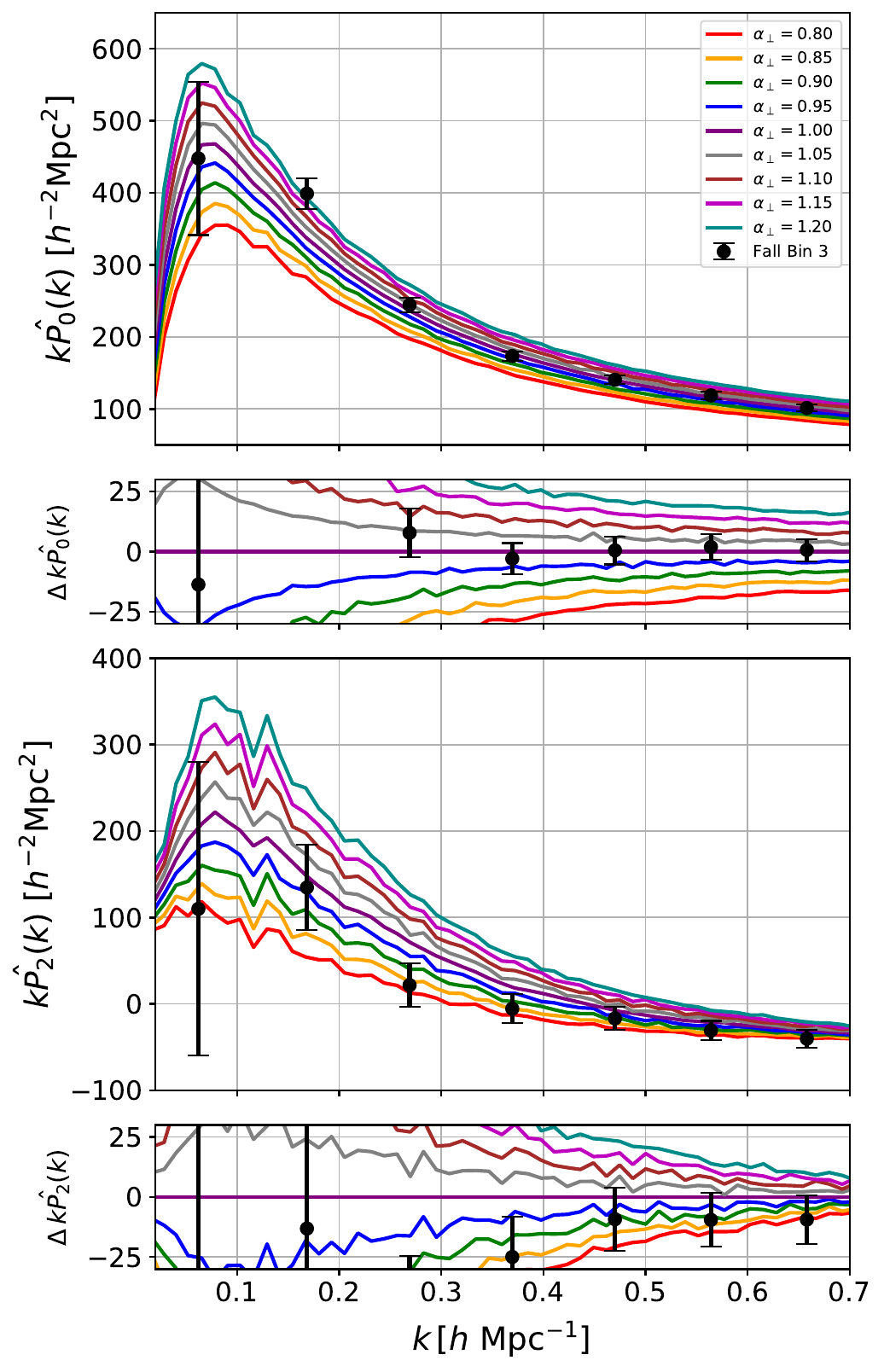}
    \end{minipage}
    \begin{minipage}{0.32\textwidth}
        \centering
        \includegraphics[width=\textwidth]{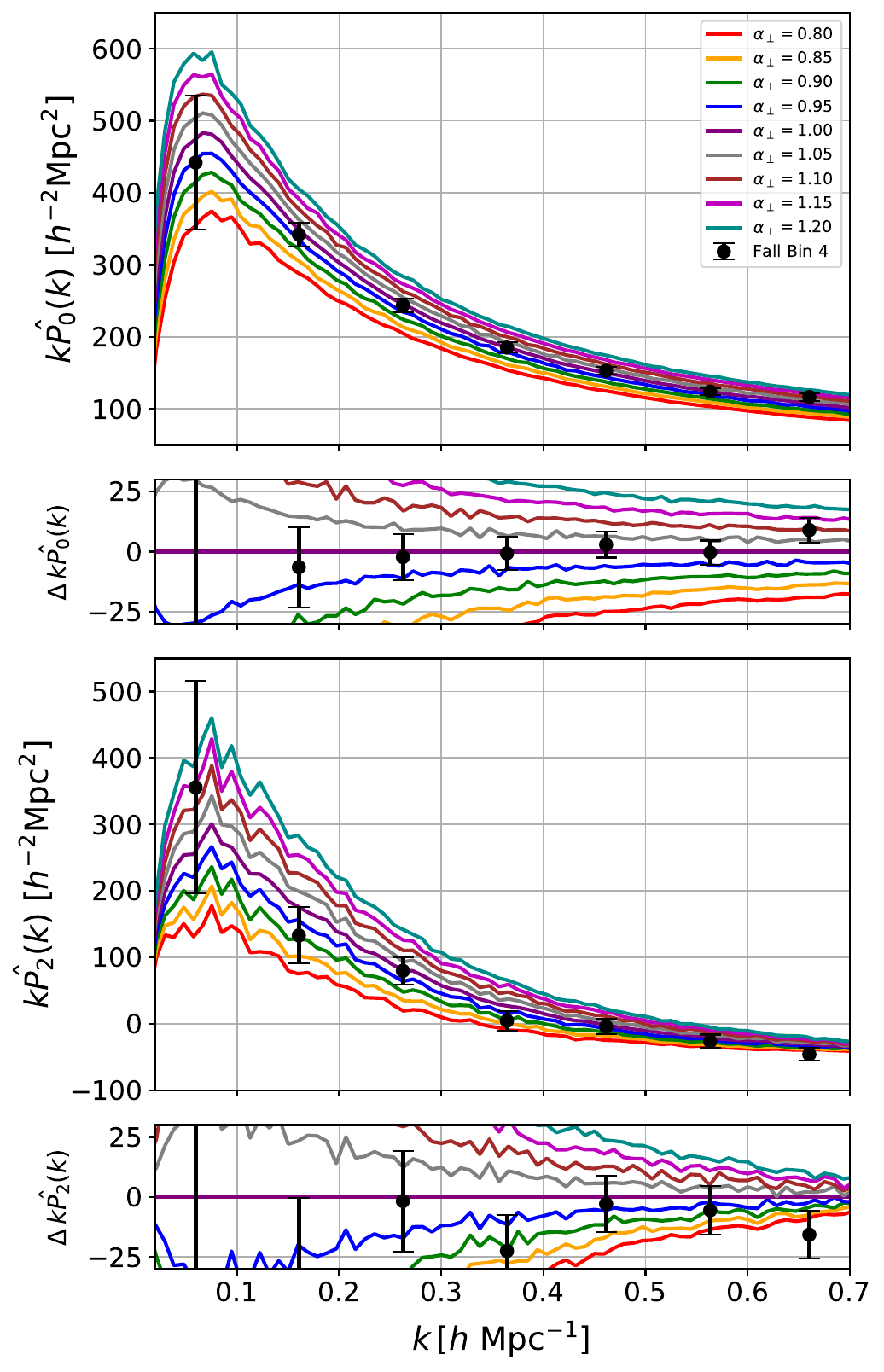}
    \end{minipage}
    \caption{Same format as Figure~\ref{fig:plot_pk_alpha_perp_spring}, but for the Fall field.}
    \label{fig:plot_pk_alpha_perp_fall}
\end{figure*}

\begin{figure*}
    \centering
    \begin{minipage}{0.32\textwidth}
        \centering
        \includegraphics[width=\textwidth]{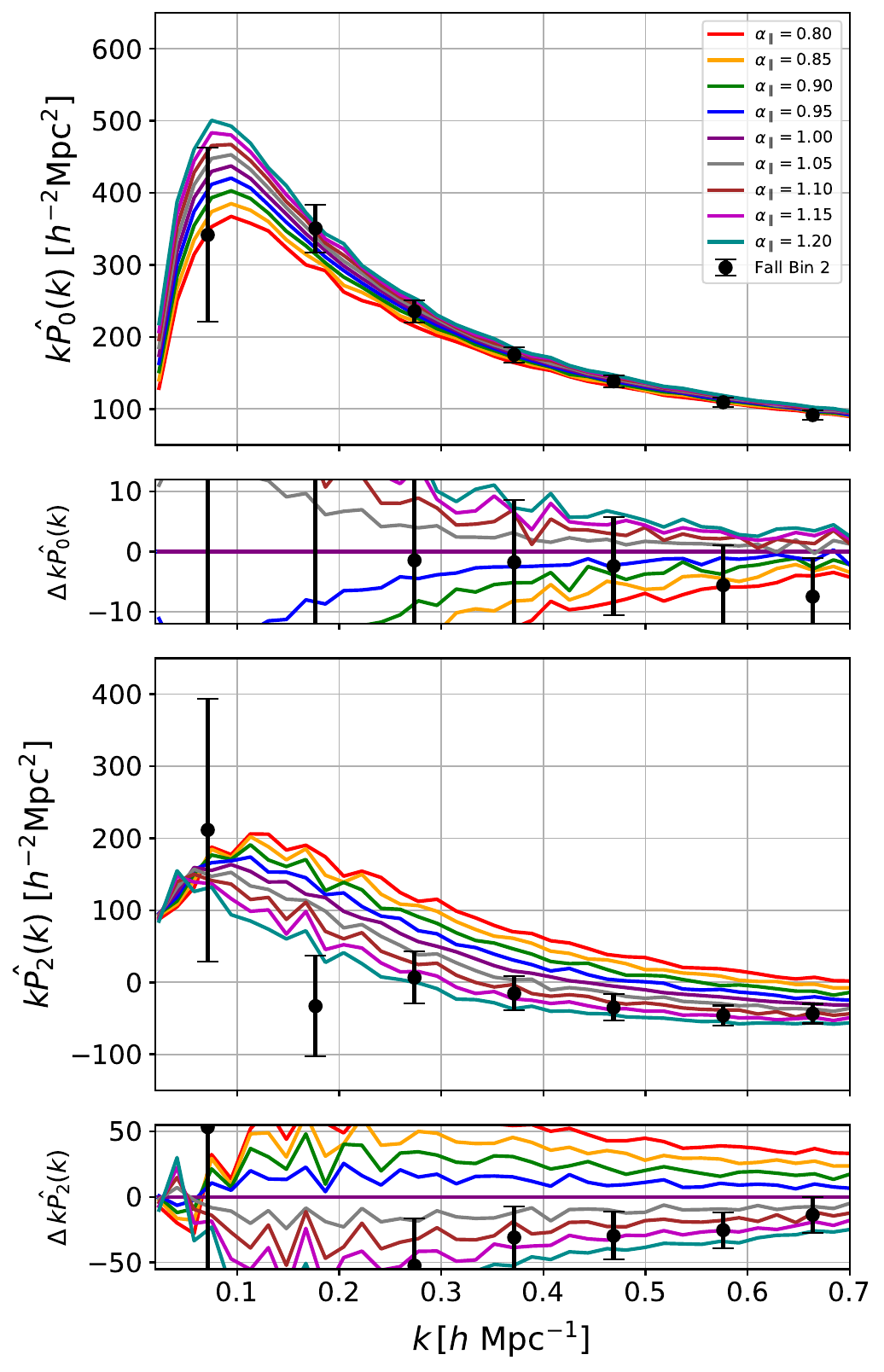}
    \end{minipage}
    \begin{minipage}{0.32\textwidth}
        \centering
        \includegraphics[width=\textwidth]{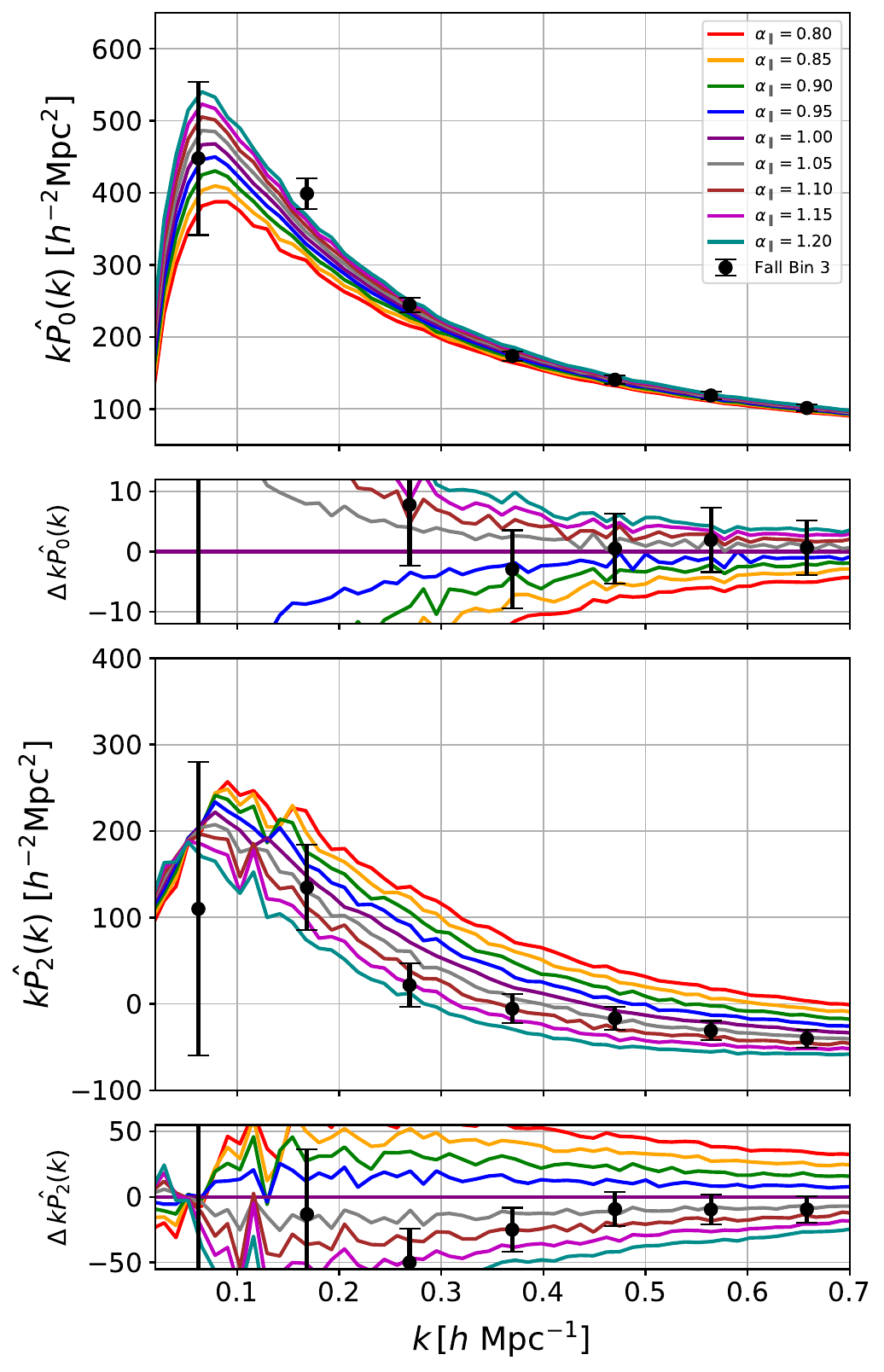}
    \end{minipage}
    \begin{minipage}{0.32\textwidth}
        \centering
        \includegraphics[width=\textwidth]{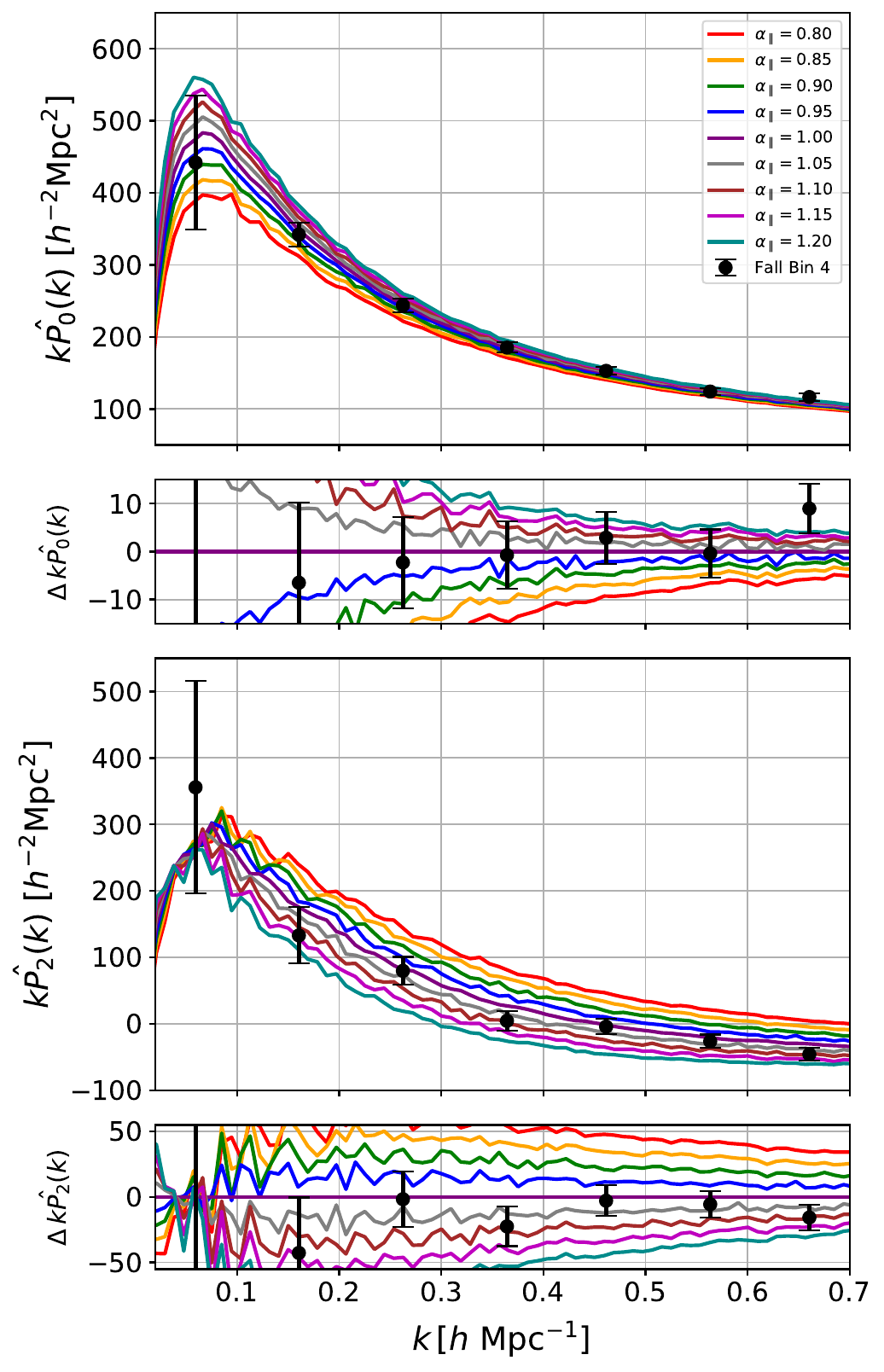}
    \end{minipage}
    \caption{Same format as Figure~\ref{fig:plot_pk_alpha_par_spring}, but for the Fall field.}
    \label{fig:plot_pk_alpha_par_fall}
\end{figure*}

Figure~\ref{fig:plot_pk_alpha_v_fall} presents the corresponding monopole and quadrupole power spectra for the Fall field, following the same comparison with mocks for varying $\alpha_v$ as described for the Spring field in Section~\ref{subsec:results} (see Figure~\ref{fig:plot_pk_alpha_v_spring}). The $M_0$ values are again taken from the best-fit results of Paper I. The Fall field exhibits the same qualitative trends with $\alpha_v$ and $k$ as the Spring field and remains distinguishable from the $M_0$ and AP parameter dependencies shown in Figures~\ref{fig:plot_pk_alpha_perp_fall} and~\ref{fig:plot_pk_alpha_par_fall}.

Figures~\ref{fig:plot_pk_alpha_perp_fall} and~\ref{fig:plot_pk_alpha_par_fall} show the analogous comparison for $\alpha_{\perp}$ and $\alpha_{\parallel}$ in the Fall field. As in the Spring field, $\alpha_\perp$ and $\alpha_\parallel$ leave distinct imprints on the monopole and quadrupole power spectra, in terms of both amplitude and dependence on $k$.

As in the Spring field, the distinct responses of the power spectrum multipoles to $\alpha_v$, $\alpha_\perp$, and $\alpha_\parallel$ allow us to constrain each parameter from the Fall-field bins up to $k_\mathrm{max} = 0.7\,h\,\mathrm{Mpc}^{-1}$, further supporting our choice of a lognormal SHOD model.

\subsection{Mock recovery tests of Fall field}
\label{appendix:mock_recovery_fall}

\begin{figure*}
    \centering 
    \includegraphics[width=0.95\textwidth]{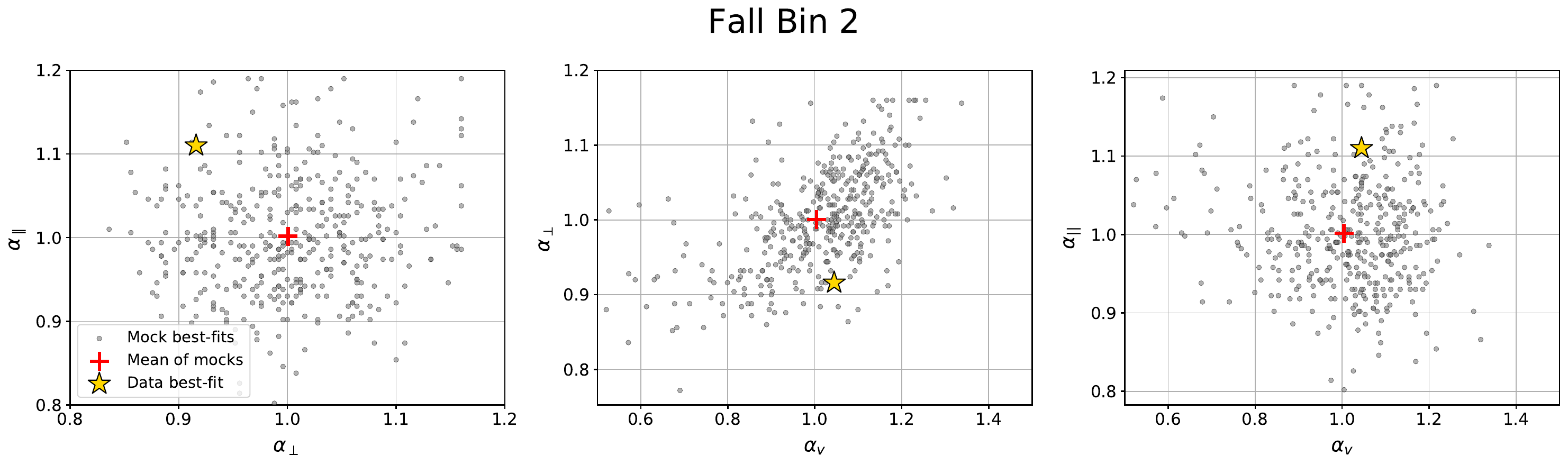}
    \includegraphics[width=0.95\textwidth]{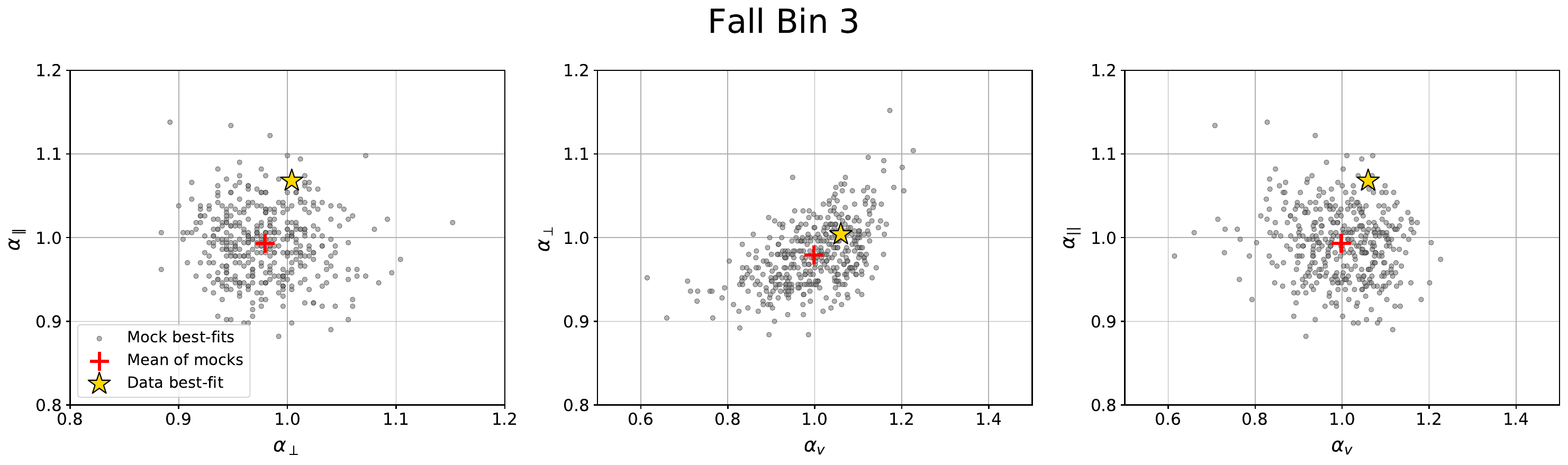}
    \includegraphics[width=0.95\textwidth]{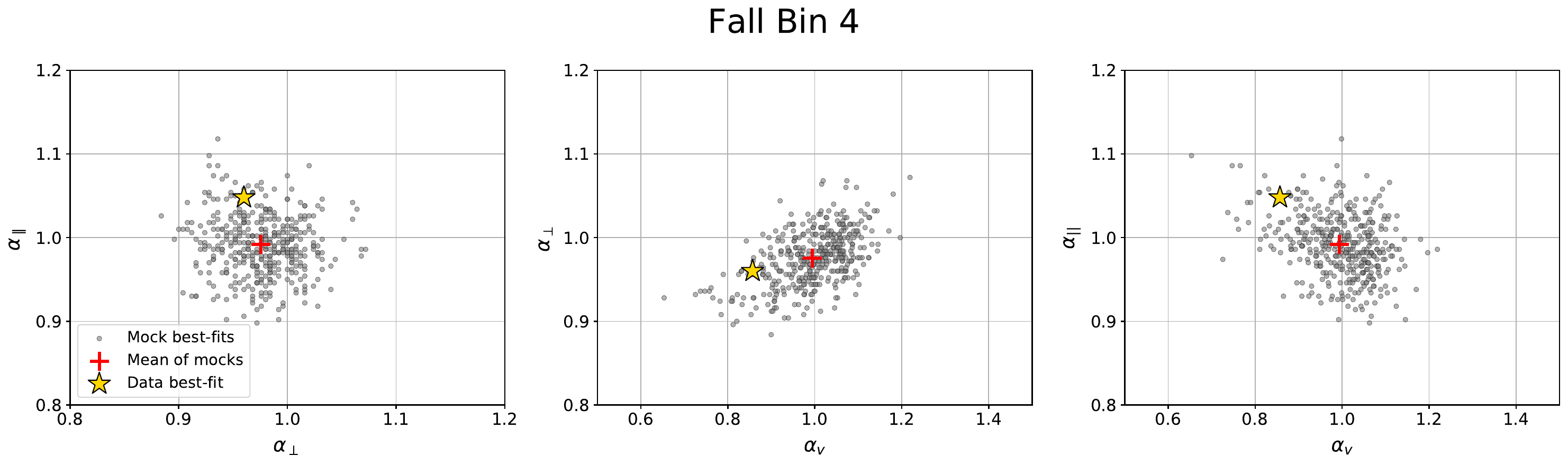}
    \caption{Same format as Figure~\ref{fig:plot_mock_recovery_spring}, but for the Fall field. As with the Spring field, the mean best-fit $\boldsymbol{\alpha}$ (red cross) is well recovered.}
    \label{fig:plot_mock_recovery_fall}
\end{figure*}

Figure~\ref{fig:plot_mock_recovery_fall} presents the corresponding mock recovery test for the Fall-field bins, following the same procedure described for the Spring field in Section~\ref{subsec:likelihood} (see Figure~\ref{fig:plot_mock_recovery_spring}). As in the Spring field, the mean best-fit, $\bar{\boldsymbol{\alpha}} \simeq \mathbf{1}$, is well recovered within the scatter of the individual mock realizations. This result confirms that our pipeline yields unbiased estimates of $\boldsymbol{\alpha}$ for the Fall field as well as the Spring field.

\section{Hexadecapole and dipole power spectra}
\label{appendix:dipole_hexadecapole}

This appendix presents the hexadecapole and dipole power spectrum measurements of the HETDEX [\ion{O}{2}] volume-limited samples to further verify the best-fit mocks obtained in Section~\ref{sec:interpretation} and check the impact of wide-angle effects on the dipole power spectrum.

\subsection{Hexadecapole}
\label{appendix:hexadecapole}

\begin{figure*}
    \centering
    \begin{minipage}{0.32\textwidth}
        \centering
        \includegraphics[width=\textwidth]{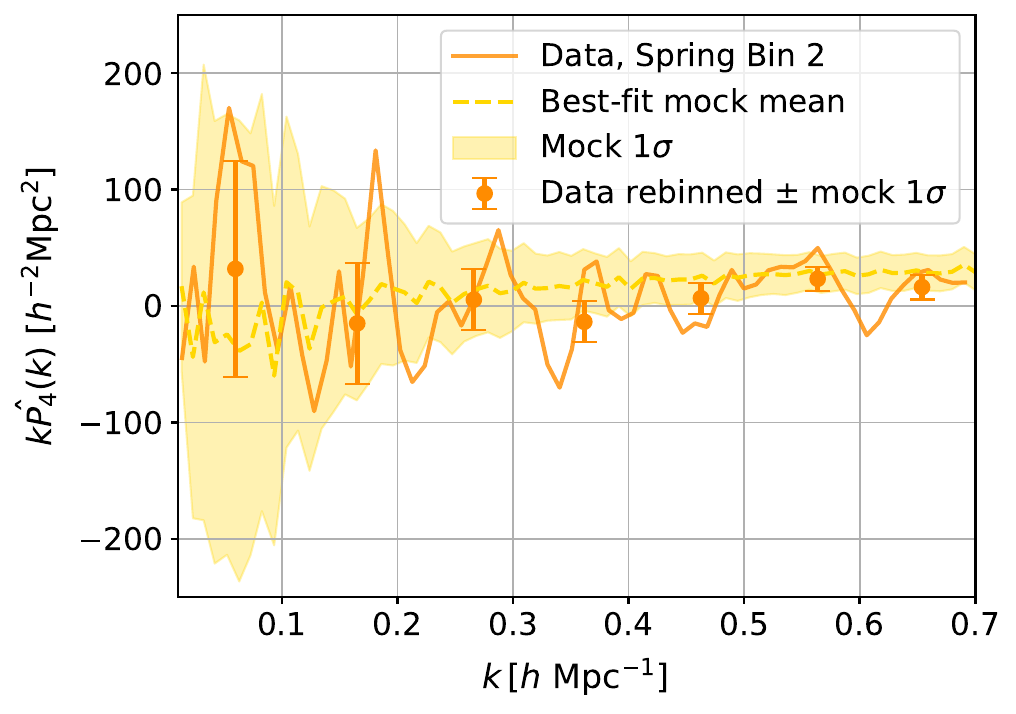}
    \end{minipage}
    \begin{minipage}{0.32\textwidth}
        \centering
        \includegraphics[width=\textwidth]{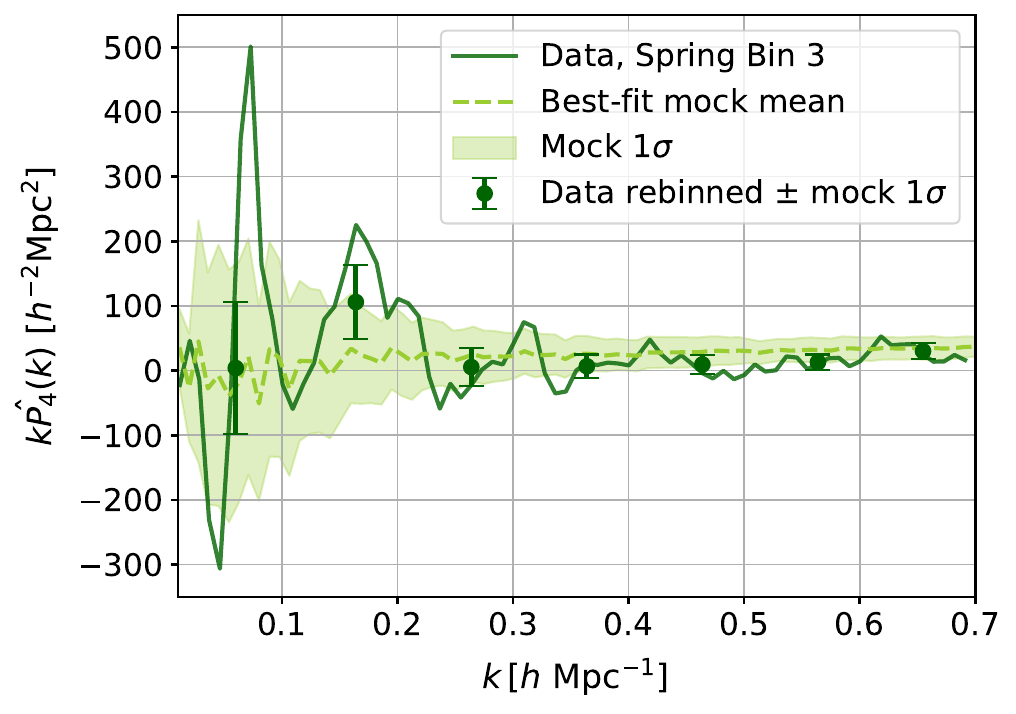}
    \end{minipage}
    \begin{minipage}{0.32\textwidth}
        \centering
        \includegraphics[width=\textwidth]{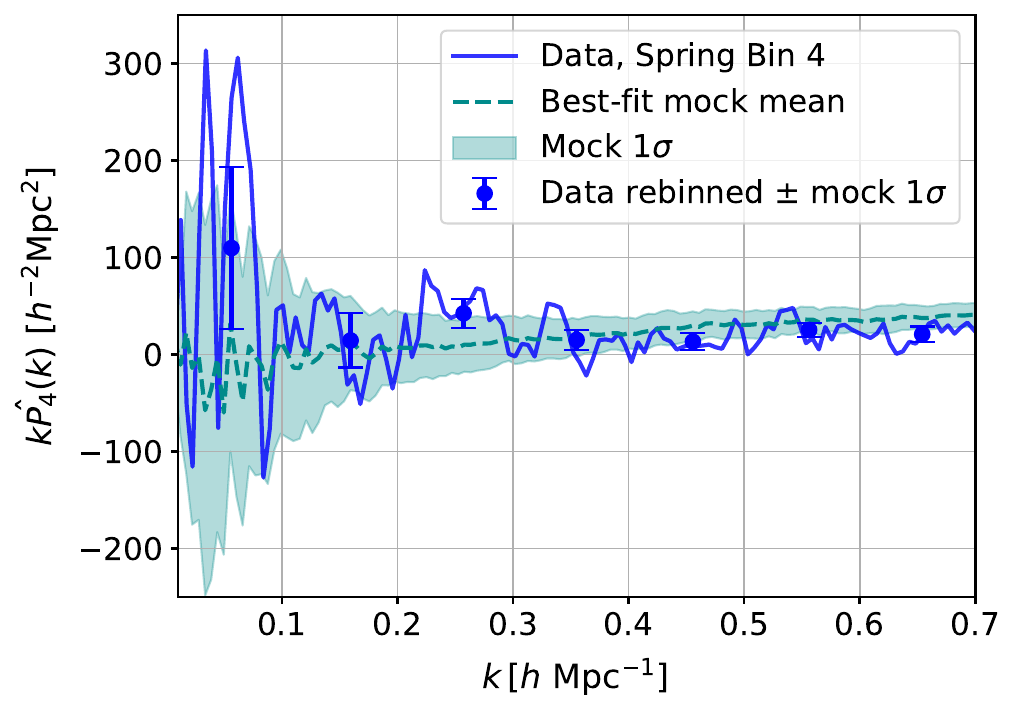}
    \end{minipage}
    \begin{minipage}{0.32\textwidth}
        \centering
        \includegraphics[width=\textwidth]{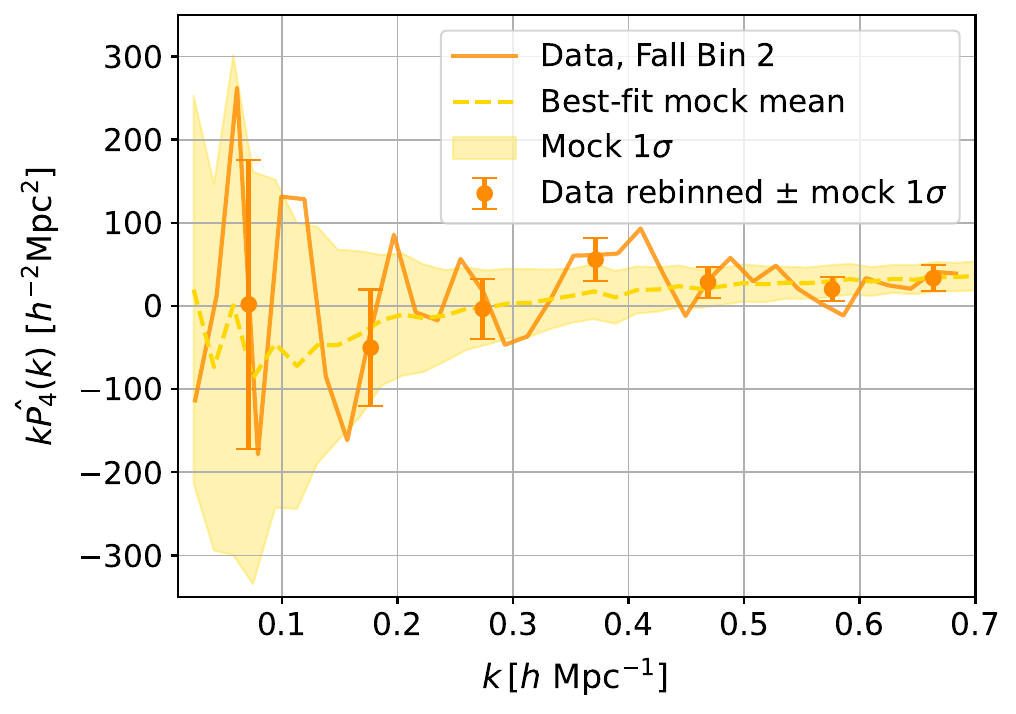}
    \end{minipage}
    \begin{minipage}{0.32\textwidth}
        \centering
        \includegraphics[width=\textwidth]{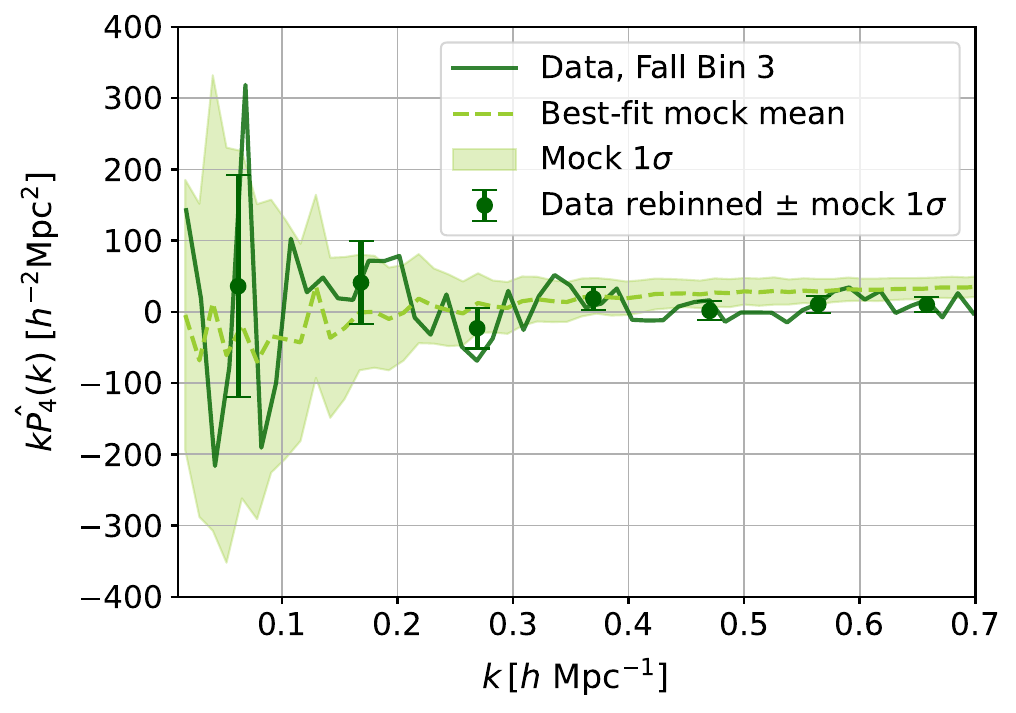}
    \end{minipage}
    \begin{minipage}{0.32\textwidth}
        \centering
        \includegraphics[width=\textwidth]{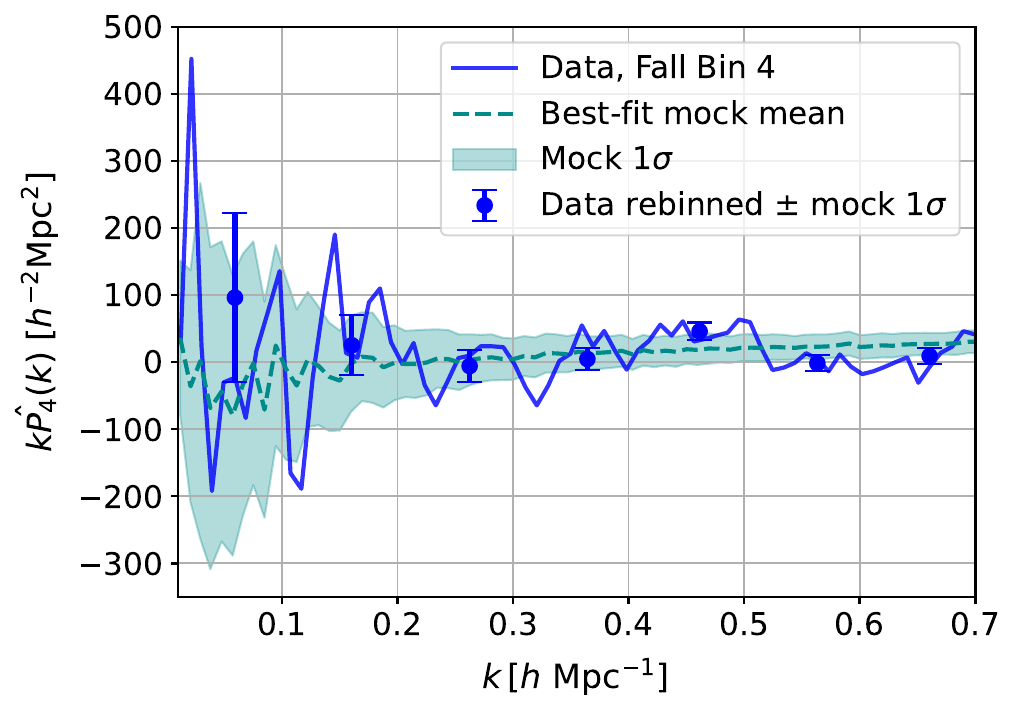}
    \end{minipage}
    \caption{Same format as Figure~\ref{fig:plot_pk_best_spring}, but for the hexadecapole power spectrum measurements.
    The data and mocks agree remarkably well across all wavenumbers up to $k_\mathrm{max}=0.7\,h\,\mathrm{Mpc}^{-1}$.}
    \label{fig:plot_p4_best}
\end{figure*}

We did not use the hexadecapole power spectra for cosmological inference, but they provide an additional, independent, higher-order test of the best-fit model derived from the monopole and quadrupole power spectra. 

Figure~\ref{fig:plot_p4_best} compares the hexadecapole power spectra of the HETDEX [\ion{O}{2}] volume-limited samples to those of the MU1 mocks evaluated at the best-fit $\boldsymbol{\alpha}$, for the Spring and Fall fields, respectively. The data and mocks agree well across all wavenumbers up to $k_\mathrm{max}=0.7\,h\,\mathrm{Mpc}^{-1}$, although with larger relative scatter than the monopole and quadrupole.  This is consistent with the hexadecapole's smaller signal-to-noise ratio. This agreement provides an additional consistency check on our SHOD model and the best-fit $\boldsymbol{\alpha}$ values.

\subsection{Dipole}
\label{appendix:dipole}

\begin{figure*}
    \centering
    \begin{minipage}{0.32\textwidth}
        \centering
        \includegraphics[width=\textwidth]{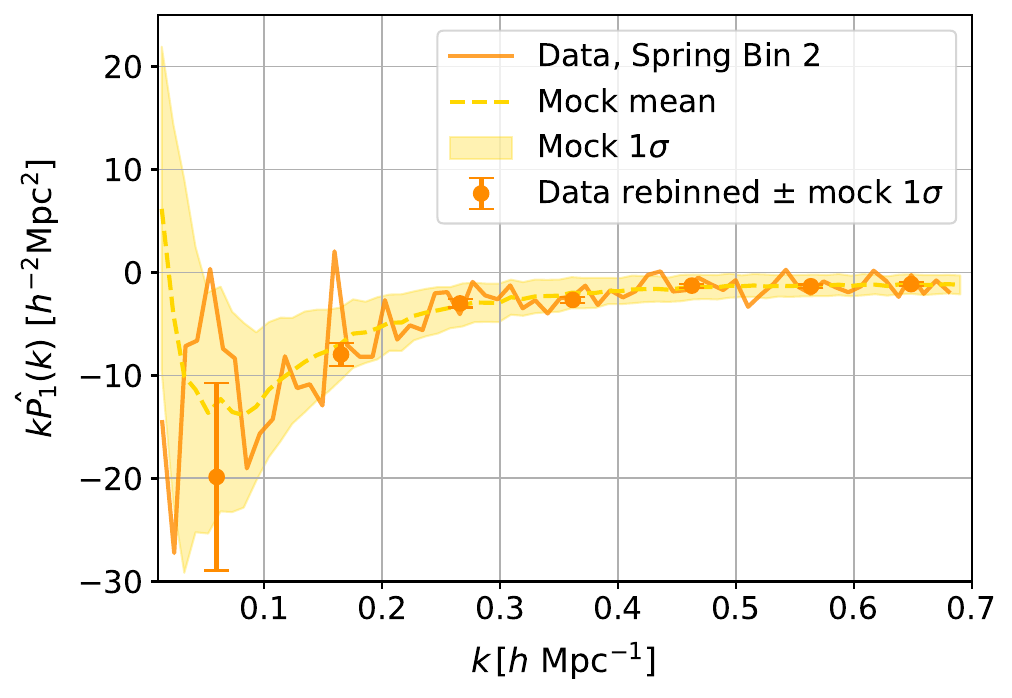}
    \end{minipage}
    \begin{minipage}{0.32\textwidth}
        \centering
        \includegraphics[width=\textwidth]{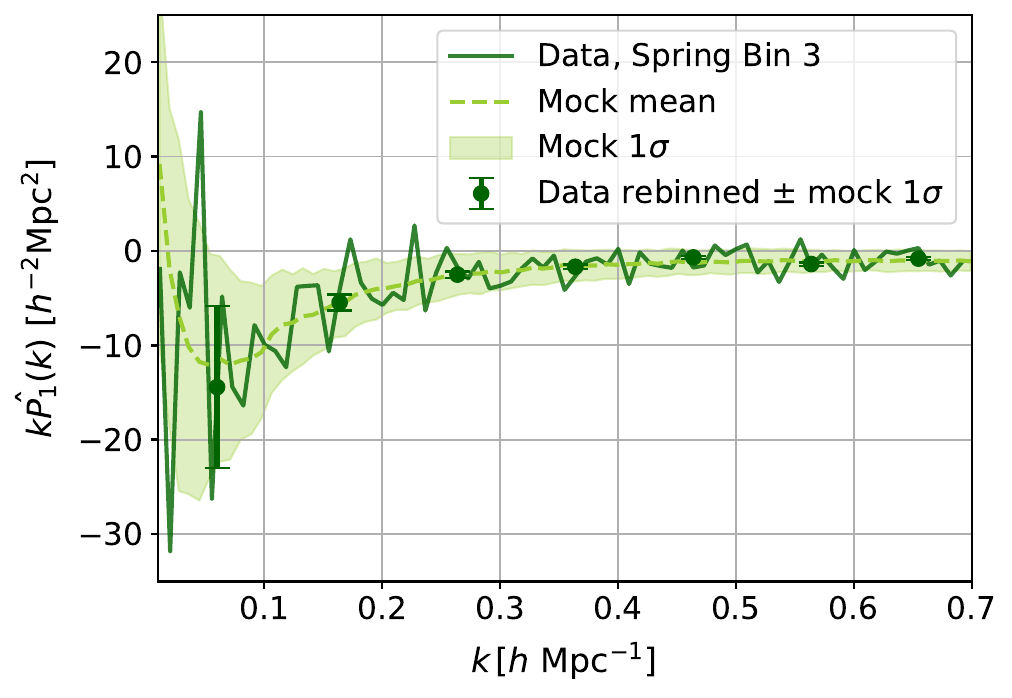}
    \end{minipage}
    \begin{minipage}{0.32\textwidth}
        \centering
        \includegraphics[width=\textwidth]{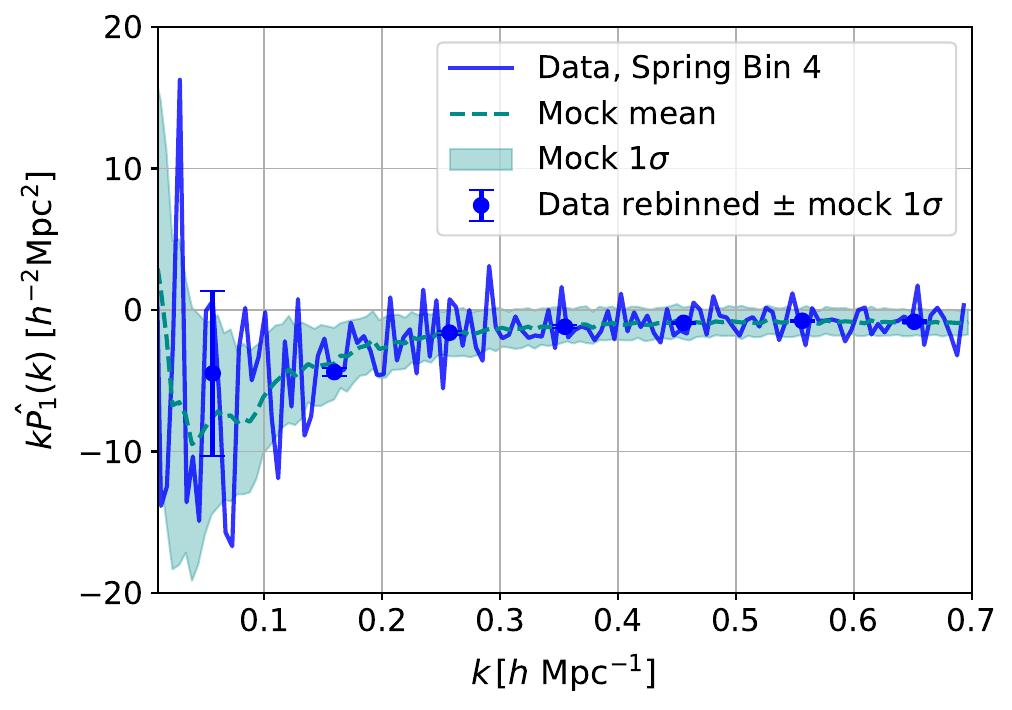}
    \end{minipage}
    \begin{minipage}{0.32\textwidth}
        \centering
        \includegraphics[width=\textwidth]{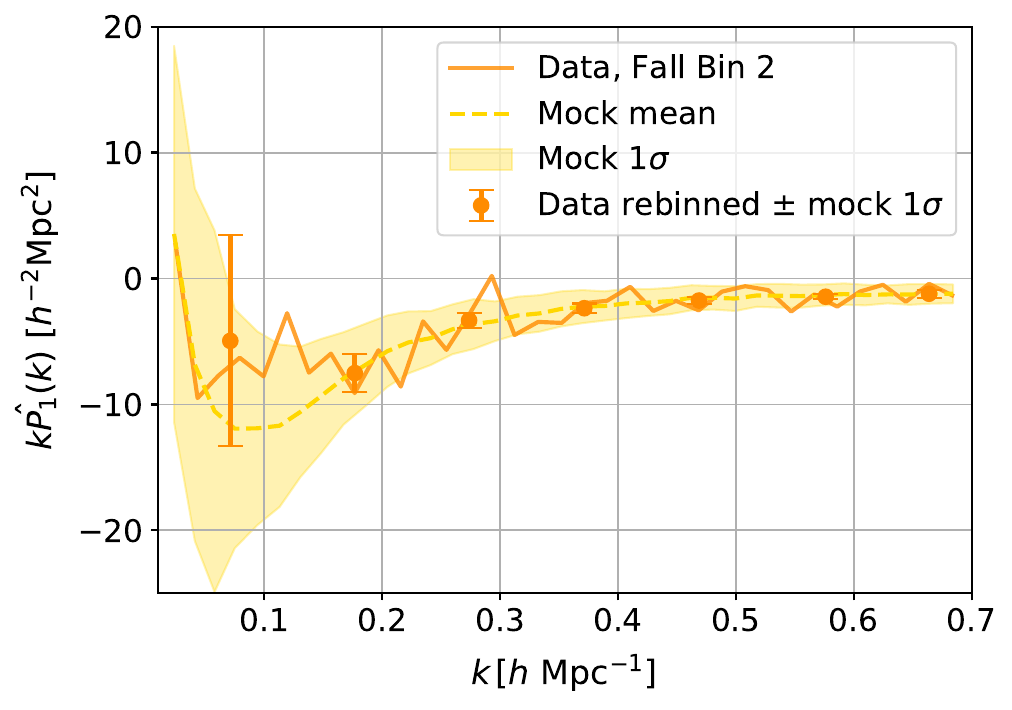}
    \end{minipage}
    \begin{minipage}{0.32\textwidth}
        \centering
        \includegraphics[width=\textwidth]{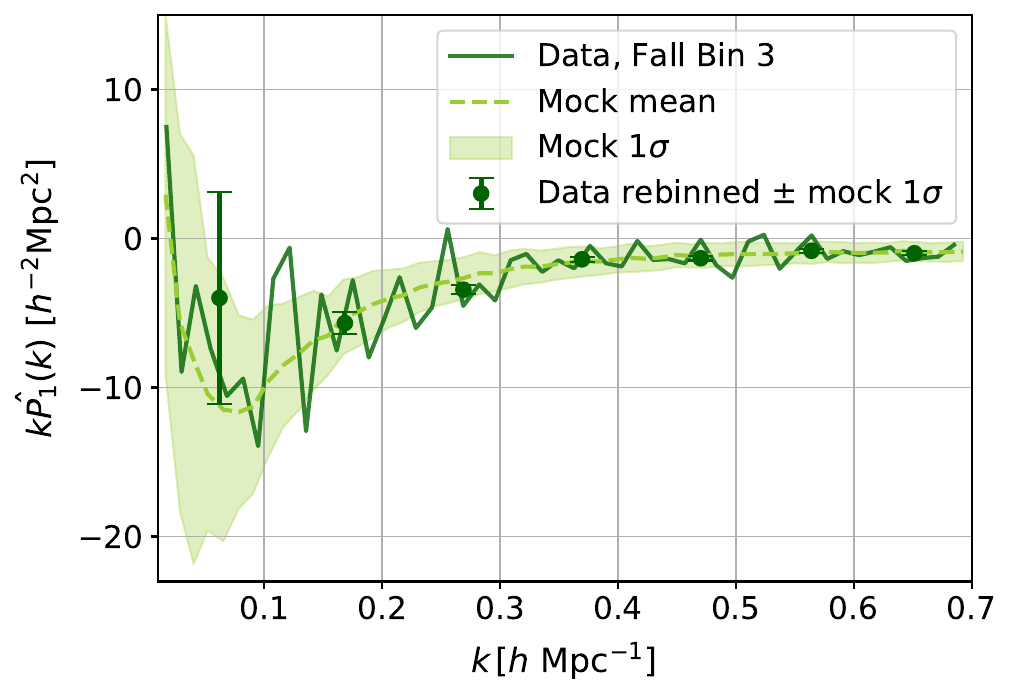}
    \end{minipage}
    \begin{minipage}{0.32\textwidth}
        \centering
        \includegraphics[width=\textwidth]{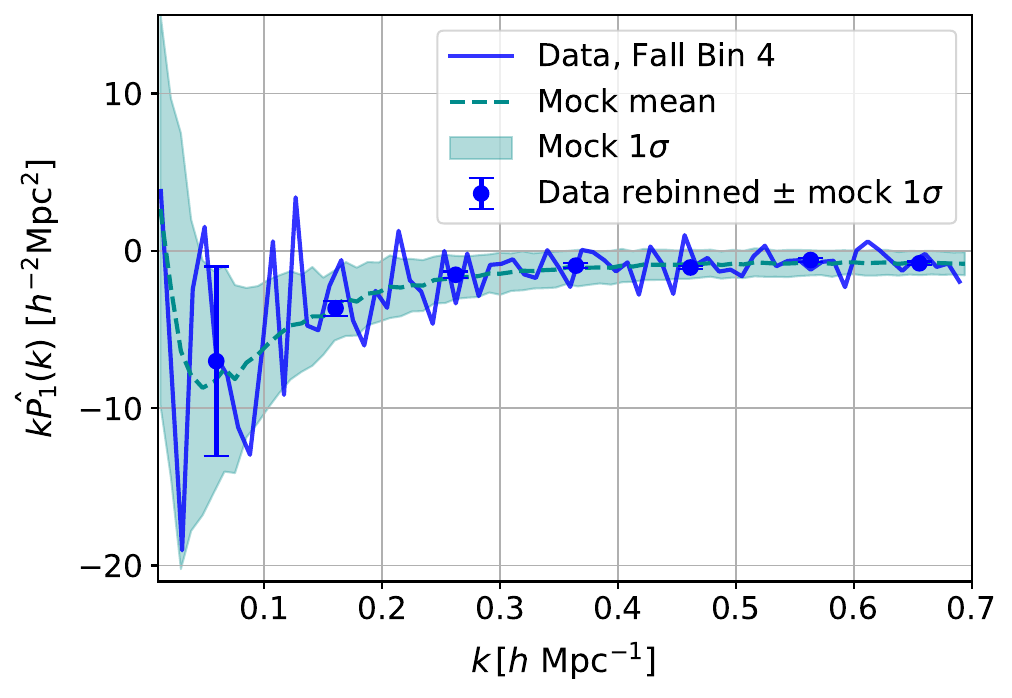}
    \end{minipage}
    \caption{Same format as Figure~\ref{fig:plot_pk_best_spring}, but for the dipole power spectrum measurements. The data and mocks agree remarkably well across all wavenumbers up to $k_\mathrm{max}=0.7\,h\,\mathrm{Mpc}^{-1}$.}
    \label{fig:plot_p1_best}
\end{figure*}

The power spectrum of a single tracer is expected to be an even function of $\mu$ in the global plane-parallel approximation, since both linear RSD and the Newtonian galaxy density field depend only on $\mu^2$. 
As a result, all odd-$\ell$ multipoles, including the dipole, vanish by symmetry, as mentioned in Section~\ref{sec:methodology}.
Nonetheless, a nonzero dipole can arise from the so-called wide-angle effects, since our estimator adopts a local, moving LOS defined by the end-point convention rather than a single, globally fixed LOS~\citep{Gaztanaga_2017,Beutler_2019}. 
Although HETDEX's sky coverage is moderate, the [\ion{O}{2}] galaxies are located relatively close to us; thus, we examine the impact of the wide-angle effect on the HETDEX dipole power spectra.

Figure~\ref{fig:plot_p1_best} compares the dipole power spectrum of the HETDEX [\ion{O}{2}] volume-limited samples to that of the MU1 mocks evaluated at the best-fit $\boldsymbol{\alpha}$. Note that odd-multipole power spectra contain signals in the imaginary part of the estimator. The data and mocks agree well across all wavenumbers up to $k_\mathrm{max}=0.7\,h\,\mathrm{Mpc}^{-1}$. A non-zero dipole signal is detected, particularly at low $k$ and low $z$. The signal is more significant in the Spring field bins, which have larger sky coverage than the Fall field bins. These results suggest that these dipole power spectra originate from the wide-angle effect. Our forward-model approach automatically accounts for this effect.

\bibliographystyle{aasjournalv7}
\bibliography{sample701}
\end{document}